# The Polana-Eulalia Complex with JWST NIRSpec: Connections to Bennu, Ryugu, & Outer Solar System Bodies

Lucas T. McClure[a], Joshua P. Emery[a], Driss Takir[b], Julia de León[c,d], Mário de Prá[e], Brittany Harvison[f], Bryan Holler[g], Javier Licandro[c,d], Tania Le Pivert-Jolivet[c,d], Joe Masiero[h], Noemi Pinilla-Alonso[i,j]

[a]*Northern Arizona University, Department of Astronomy and Planetary Science, Flagstaff, AZ 86011, United States*
[b]*Space Science Institute, Boulder, CO, USA*
[c]*Instituto de Astrofísica de Canarias, C/Vía Láctea s/n, E-38205, La Laguna, Tenerife, Spain*
[d]*Departamento de Astrofísica, Universidad de La Laguna, C/Astrofísico Francisco Sánchez s/n, E-38200, La Laguna, Tenerife, Spain*
[e]*Observatório Nacional, Rio de Janeiro, Brazil*
[f]University of Central Florida, Department of Physics, Orlando, FL, USA
[g]*Space Telescope Science Institute, Baltimore, MD, USA*
[h]*IPAC, Caltech, 1200 E California Blvd, Pasadena, MC 100-22, CA, USA*
[i]*Instituto de Ciencias y Tecnologías Espaciales de Asturias (ICTEA): Oviedo, Asturias, Spain, ES*
[j]*Department of Physics, Universidad de Oviedo, Oviedo, Spain, ES*

Corresponding E-mail:
Lucas McClure: ltm87@nau.edu

Keywords – asteroids, JWST, spectroscopy

## Abstract

Carbon-rich (C-complex) asteroids provide insight into the early Solar System, and families of C-complex asteroids serve as robust samples for learning about the internal structure of planetesimals and the materials of the protoplanetary disk. This work investigates the compositions of asteroids within the Polana-Eulalia Complex (PEC) through analyses of JWST NIRSpec. Obtained as part of the **S**pectral **A**nalysis of **M**ain **B**elt **A**steroids in the 3-µm Region (SAMBA-3) project. NIRSpec observations were conducted for the namesake asteroid of the Eulalia family, (495) Eulalia, as well as three members from the Polana family: (2441) Hibbs, (6712) Hornstein, and (6769) Brokoff. Analyses of the 2.7-µm region demonstrate that PEC asteroids exhibit absorptions associated with Mg-rich phyllosilicates, in turn underscoring high – though incomplete – degrees of aqueous alteration in the past for both PEC families. Although consistent in regard to composition, (142) Polana and (495) Eulalia show deeper 2.7-µm bands than the Polana family asteroids, suggesting slight differences in the degrees of aqueous alteration between large and small PEC asteroids. Minor absorptions from 3.0 to 4.0 µm indicate that (495) Eulalia's surface may contain more carbonates than (142) Polana, which potentially highlights slight differences between each family's formation and evolution. Additionally, PEC asteroids also show spectral consistency with Bennu and Ryugu, reinforcing their shared origin in the PEC. Minor absorptions in PEC asteroid spectra also connect to small bodies within and beyond the Main Belt, providing further evidence of outer Solar System formation for the PEC parent bodies.

## Introduction

### 1.1 Background

Carbon-rich (C-complex), primitive asteroids provide key insights into the formation and evolution of the Solar System. As the predominant class of asteroids in the Main Belt, C-complex asteroids collectively hold the majority of the Solar System's un-melted (or mostly un-melted) materials from the time of formation (e.g., DeMeo & Carry 2013, 2014; Keil, 2000). C-complex asteroids also have spectral connections to carbonaceous chondrites (CCs) (e.g., Weisburg & Righter et al. 2014; Takir et al. 2013). The link between C-complex asteroids and CCs provides some constraint on the compositional distribution throughout and beyond the Main Belt.

Populations of C-complex asteroids serve as statistically-robust samples that can be used to trace the history of materials from the early Solar System. Families of carbon-rich asteroids are populations that exhibit similar dynamical (Nesvorny et al. 2015) and spectroscopic properties (*e.g.,* spectral slope, absorptions, albedo; *see* DeMeo & Carry 2013; Masiero et al. 2015). Carbon-rich planetesimals likely originated in the outer Solar System and were then implanted into the Main Belt through Gas Giant migration (Walsh et al. 2011; Tsganis et al. 2005; Gomes et al. 2005; Morbidelli et al. 2005). Subsequent collisional evolution in the Main Belt generated numerous asteroid families centered around large, carbon-rich "parent bodies" (Masiero et al. 2015). Main Belt families of C-complex asteroids are samples of great interest for probing key materials and processes from both the inner and outer Solar System, with the parent bodies of the Inner Main Belt (IMB; 2.1 – 2.5 AU) families being the likely origin for many carbonaceous near-earth

Asteroids (NEAs; e.g., Arredondo et al. 2020; Morate et al. 2016, 2018, 2019; de Leon et al. 2016; Pinilla-Alonso et al. 2016;. Harvison et al. 2024).

Discovered by Walsh et al. 2013), the Polana and Eulalia families are similar, though independent, IMB families of C-complex asteroids. Both families are comprised of low-albedo, low-inclination asteroids that are dynamically-centered around (142) Polana and (495) Eulalia – the largest and namesake members of the respective families (Fig. 1; Walsh et al. 2013). The families are often discussed together as the Polana-Eulalia Complex (PEC) due to similar orbital and spectral characteristics in near-ultraviolet (NUV; 0.35 – 0.46 µm; Tatsumi et al. 2022), visible (VIS; 0.46—0.9 µm; de Leon et al. 2016), and near-infrared (NIR; 0.9—2.5 µm; Pinilla-Alonso et al. 2016) wavelengths. The PEC has a close proximity to the $\nu_6$ secular resonance has also likely played a role in the delivery of its (preferentially small) members to NEA-space, which is the plausible scenario for spacecraft mission targets Bennu and Ryugu (see Section 1.2 for more details; Walsh et al 2013; Campins et al. 2010,2013; Bottke et al. 2015). PEC asteroids also display featureless spectra in VISNIR wavelengths (de Leon et al. 2016; Pinilla-Alonso et al. 2016). No member of either family exhibits a 0.7-µm band associated with Fe-rich phyllosilicates (de Leon et al. 2016). Although these two families exhibit similar orbital properties, the Polana family's "V"-shaped drift is wider than that of the Eulalia family. The relative widths of the envelopes underscore that that the Polana family is older (>2000 Myr) than the Eulalia family (900-1500 Myr) (Walsh et al. 2013). Furthermore, archival photometry has shown some non-negligible difference in NUV-VIS wavelength spectrophotometric slope distributions between Polana and Eulalia family asteroids, with Polana and Eulalia family asteroids exhibiting more positively-sloped (i.e., redder) and negatively-sloped (i.e., bluer) spectra, respectively (Delbo et al. 2025; McClure et al. 2025).

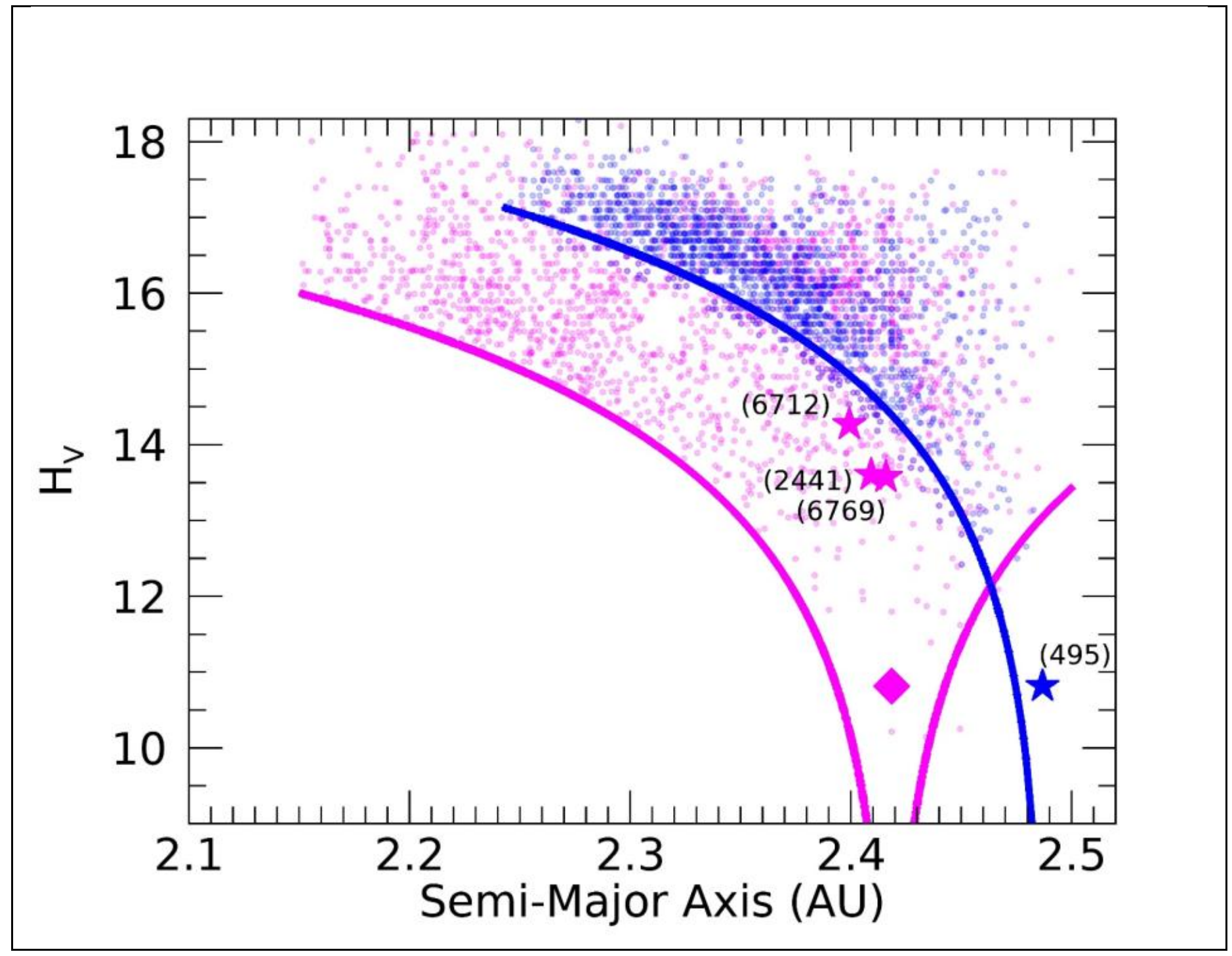

**Figure 1**: Envelopes in semi-major axis of the Polana and Eulalia families over the full magnitude range. The "V"-shapes are due to size-dependent drift from the Yarkovsky effect (see Vokrouhlický et al. 2006). Objects in this work are represented by star shapes with colors corresponding to the family envelope outline, with the three Polana family asteroids: (2441) Hibbs, (6712) Hornstein, and (6769) Brokoff (shown in pink) and (495) Eulalia (in blue). (142) Polana is highlighted as a pink diamond for reference, with Polana and Eulalia family members in light pink and dark blue dots, respectively. (For interpretation of the references to color in this figure legend, the reader is referred to the web version of this article).

### 1.2 Motivation

This work focuses on JWST NIRSpec (0.7 – 5.2 μm) observations of PEC asteroids. The PEC asteroids in this work include (495) Eulalia and three Polana family asteroids (PFAs): (2441) Hibbs, (6712) Hornstein, and (6769) Brokoff (see **Figure 1**). As shown in **Table 1**, the PFAs are roughly 5 km in radius, which places them within the mid-sized class of members in the family. Each of these targets is part of a larger campaign from JWST Cycle 3 GO time, named the **S**pectral **A**nalysis of **M**ain **B**elt **A**steroids in the **3**-μm region (**SAMBA-3**; ID #: 6384). SAMBA-3 has the broad goal of measuring and analyzing spectra of the largest bodies of numerous IMB families. However, the spectrum of (142) Polana was already measured during Cycle 2 (Program #:3760; PI: A. Arredondo; see also Arredondo et al. 2025), so we instead observed three members from the Polana family of asteroids, listed in the Polana family list used in Walsh et al. 2013. Results from SAMBA-3 (*e.g.,* Le Pivert-Jolivet et al. 2025a; Harvison et al. 2025; McClure et al. 2025; Takir et al. 2025) and other JWST NIRSpec observations (*e.g.,* Arredondo et al. 2025) demonstrate a range of surface compositions that are in turn reflective of different alteration histories. Thus, JWST-based spectral analysis of PEC asteroids will (1) further our understanding of primitive small-bodies as well as (2) refine the PEC's specific role as a source region for NEAs.

Many questions pertaining to the PEC's role in the formation and evolution of the Solar System can be addressed in the era of JWST. The wavelengths accessible by NIRSpec can provide insight into numerous surface minerals that are potentially present on the surfaces of this work's targets. NIRSpec coverage includes the 2.7-μm band, caused by the OH stretch within Fe- or Mg-rich phyllosilicates (*e.g.,* Rivkin et al. 2002). Shorter-wavelength observations (de Leon et al. 2016) have suggested a lack of Fe-rich minerals on the surfaces of PEC asteroids. Thus, this work's observations will test the hypothesis that PEC asteroids lack 2.7-μm features that are consistent with surfaces containing Fe-rich phyllosilicates. A previous observation of (142) Polana shows a 2.7-m band associated with Mg-rich phyllosilicates (Arredondo et al. 2025), so given the strong spectral similarities between the PEC families, we particularly expect broad consistency with respect to the metal within the predominant surface phyllosilicate, if present. Building upon previous JWST-based works that were centered on primitive asteroids (*e.g.,* Le Pivert-Jolivet et al. 2025a; Arredondo et al. 2025), this work also seeks to test the hypothesis that PEC asteroid spectra from 3.0 – 4.0 μm include features associated with salts, organics, and carbonates, which are typically found in primitive carbonaceous chondrites and Ryugu and Bennu samples (e.g., King et al. 1992; Dartois et al. 2004; Huang & Kerr, 1960; Pilogret et al. 2022; Lauretta et al. 2024 and refs. therein). In total, compositional characterization of the NIRSpec wavelength range can provide insights into an asteroid family's formation and evolution (e.g., Takir et al. 2013). Observations of multiple PEC members comprehensively probe the range of composition(s).

JWST-based observations of PEC asteroids could elucidate key questions regarding the origin of NEAs Bennu and Ryugu, the targets of sample-return spacecraft missions *OSIRIS-REx (NASA)* and *Hayabusa2 (JAXA).* Namely, from which family do Bennu and Ryugu originate? Dynamical modeling and spectroscopic observations strongly suggest a connection between both of these NEAs and the PEC (e.g., Campins et al. 2010, 2013; Bottke et al. 2015; de Leon et al. 2016; Pinilla-Alonso et al. 2016). However, unambiguous linking of Bennu and/or Ryugu to the PEC has been limited due to the probabilistic nature of dynamical modeling and to the featureless nature of PEC spectra in VISNIR wavelengths. In NUV-VIS wavelengths, the Polana family exhibits slightly redder average slopes compared to the Eulalia family, which has tentatively linked the Polana and Eulalia families to (redder) Ryugu and (bluer) Bennu, respectively (e.g., Delbo et al. 2023; McClure et al. 2025). NUV-VIS wavelength slope differences between the PEC families could be influenced by bulk differences in Fe-rich phyllosilicates, hydration, and/or organic matter on surfaces throughout the families (Feierberg, et al. 1985; Applin et al. 2018; Trang et al. 2021). Regardless of the tentative compositional connections, bulk differences in the NUV-VIS range could imply differences in longer wavelengths with more diagnostic spectral features, providing significant constraints on the origin and evolution of these very similar families. Thus, the detection and characterization of the NIRSpec wavelength range for PEC asteroids could provide strong constraints on the precise family lineage of both Bennu and Ryugu. Thus, our observations will test the hypothesis that Bennu and Ryugu are spectroscopic matches to the Eulalia and Polana families, respectively. We will leverage the diagnostic wavelengths of NIRSpec and compare our observations with data from spacecraft and returned-samples.

**Table 1**: Orbital and physical parameters of each target in this work.

| Object | *a* (au)[1] | *e*[1] | *i* (°)[1] | R (km) | Geometric albedo ($p_v$) | Absolute Magnitude (H)[2] |
|---|---|---|---|---|---|---|
| 495 | 2.49 | 0.13 | 2.28 | 18.64 ± 0.02[3] | 0.062 ± 0.008[3] | 11.34 |
| 2441 | 2.41 | 0.19 | 3.75 | 5.17 ± 1.60[4] | 0.055 ± 0.036[4] | 14.49 |
| 6712 | 2.40 | 0.12 | 0.97 | 4.64 ± 0.02[3] | 0.040 ± 0.007[3] | 14.64 |
| 6769 | 2.42 | 0.12 | 3.89 | 4.90 ± 0.10[3] | 0.068 ± 0.010[3] | 14.33 |

[1]Values are collected from Minor Planet Center ( https://www.minorplanetcenter.net/db_search).
[2]Reported H values were collected from the Minor Planet Center (Tedesco et al. 2004) and subsequently corrected to WISE-derived H values using the formulation from Pravec et al. (2012). This was done to maintain consistency with the R and $p_v$ values, which are also reported from WISE and used in reduction (i.e., see footnotes 3-4 & **Section 2.2**)
[3]Values are collected from Mainzer et al. (2019).

[4]Values are collected from Mainzer et al. (2016).

## 2. Methods

### 2.1 Observations & Data Reduction

This work used JWST's NIRSpec instrument in integral field unit (IFU) mode. We used the medium resolution gratings for (495) Eulalia, namely G140M/F100LP (0.97–1.84 µm), G235M/F170LP (1.66–3.07 µm), and G395M/F290LP (2.87–5.10 µm). SAMBA-3 observations of the three PFAs invoked the PRISM grating with the CLEAR filter (0.7 – 5.1 µm). All four targets were observed in a 4-point-dither and readout pattern of NRSIRS2RAPID.

This work's observations were mostly conducted over Cycle 3. During 2024 (495) Eulalia, (2441) Hibbs, and (6712) Hornstein were observed on November 20$^{th}$, October 24$^{th}$, and September 10$^{th}$, respectively. In November 2024, the acquisition of the guide star for (6769) Brokoff failed, leading to inadequate signal-to-noise. A remedial observation of (6769) Brokoff occurred on October 12, 2025 (see **Table 2**).

Data reduction and subsequent spectral extraction followed a similar path to prior JWST-based works revolving around primitive asteroids (e.g., Le Pivert-Jolivet et al. 2025a; Arredondo et al. 2025). Due to the different timing of the observations, we used version 1.15.1 of the JWST data calibration pipeline to spectrally reduce the uncalibrated data files from (495) Eulalia, (2441) Hibbs, and (6712) Hornstein, whereas version 1.20.1 was employed for (6769) Brokoff's uncalibrated data files (see Bushouse et al. 2024; 2025). Although separate versions, the calibration pipeline differences are negligible particularly with regard to the effect on fully-reduced spectra. Moreover, we followed the default, standard procedures to produce calibrated spectral data cubes that contain wavelength slices. The context file jwst_1293.pmap was used for (495) Eulalia and (2441) Hibbs, whereas jwst_1290.pmap and jwst_1462.pmap were the context files for (6712) Hornstein and (6769) Brokoff, respectively. Like in Le Pivert-Jolivet et al. 2025, we used the NSClean algorithm, which removes the 1/f pattern noise from the second stage of the pipeline. To construct local point-spread functions (PSFs) at each wavelength, the median of the 10 adjacent wavelength slices on both sides of each wavelength slice was calculated. Then, the background emission was subtracted to isolate the PSFs, which were then normalized to a unit sum. Then, a scaling factor and a constant background level were fit at each original wavelength. This entire process was repeated for each of the 4 separate dithers, with spectral values ≥3σ being removed via use of a 21-pixel-wide running-boxcar. Errors were then calculated as the standard deviation of each of the 4 dithers. Each asteroid target had a corresponding, IFU-based standard star observation. (495) Eulalia's standard star was P330E, whereas the PFAs used SNAP-2. All stars' spectra in this work underwent the same process as outlined for the asteroid spectra.

### 2.2 Removal of Thermal Contribution

The measured spectra include contributions from thermally-emitted light, particularly at longer wavelengths. An example of the thermal contribution can be seen in the irradiance spectrum of (495) Eulalia (**Fig. 2a**), which produced a "tail" longward of 2.8 µm. To model and remove thermal emission and thereby isolate the reflectance in this region, we use the Near-Earth Asteroid Thermal Model (NEATM; Harris 1998).

The removal of thermal contributions to each asteroid spectrum involves a series of steps described below. First, the reflected component is estimated by scaling the irradiance spectrum of the standard star to that of the associated asteroid in the 3.5 – 5 µm range. The scaled stellar flux is subtracted from the asteroid flux, leaving just the thermal component. The thermal flux spectrum is then modeled within the NEATM framework (*see* **Fig. 2b**), incorporating the observational details (**Table 2**) in order to calculate the temperature distribution across the visible surface. The model results in estimates of radius (*r*), visible geometric albedo ($p_v$), and beaming parameter ($\eta$).

In this work, we use a $\chi^2$ minimization to find the best-fit NEATM model from 3.5 – 5 µm. For each asteroid we use a range of values for η and $p_v$ starting with an initial guess of η = 1. Additionally, we set the Solar Constant (S) to be S = 1366 W/m$^2$ and the emissivity (ε) to be ε = 0.95. The best-fit model is subtracted from the original asteroid irradiance spectrum to isolate the reflected component, which is then divided by the scaled star spectrum to yield an asteroid reflectance spectrum. All NEATM-derived values are provided in **Table 3**. Uncertainties are computed through 1000 Monte Carlo-based spectral re-samplings using the reflectance errors. Uncertainties incorporate JWST's NIRSpec IFU Calibration Status[1] (±3%), H-magnitude (extracted from Pravec et al. 2012; ±0.2-0.3) and light curve amplitude (Oey et al. 2017 for (495) Eulalia; Waszczak et al. 2015 for (2441) Hibbs; Hanuš et al. 2011 for (6712) Hornstein & (6769) Brokoff; ± 0.3-0.4) are also included in the total uncertainties of NEATM-derived parameters, shown in **Table 3**. We note that the albedos and radii listed in **Table 1** and **Table 3** are slightly different from each other. We attribute these small discrepancies to the NEATM fits being calculated for a specific wavelength range and as best-fits for all three parameters ($p_v$, η, r). Additionally, rotational lightcurves for the PFAs are incomplete, which further limits precise matches in diameter and albedo.

[1] https://jwst-docs.stsci.edu/jwst-calibration-status/nirspec-calibration-status/nirspec-ifu-calibration-status#gsc.tab=0

**Table 2**: Observational details for each target in this work.

| Object | Grating/Filter | Start Time (UT) | Exposure Time (s) | r (au) | Δ (au) | α (°) |
|---|---|---|---|---|---|---|
| 495 | G140M/F100LP | 16:23:59 | 1050 | 2.293 | 1.578 | 21.09 |
| | G235M/F170LP | 16:54:32 | 642 | | | |
| | G395M/F290LP | 17:25:05 | 700 | | | |
| 2441 | PRISM/CLEAR | 12:06:18 | 175 | 1.949 | 1.385 | 29.24 |
| 6712 | PRISM/CLEAR | 23:21:02 | 759 | 2.229 | 1.936 | 27.04 |
| 6769 | PRISM/CLEAR | 05:09:56 | 233 | 2.191 | 1.717 | 26.46 |

**Table 3: V**alues derived from NEATM-based correction.

| **Object** | **NEATM-derived values** | | |
|---|---|---|---|
| | $\mathbf{p_v}$ | **η** | **r (km)** |
| (495) Eulalia | 0.07 ± 0.03 | 0.87 ± 0.01 | 13.4 ± 0.4 |
| (2441) Hibbs | 0.04 ± 0.01 | 0.93 ± 0.01 | 4.1 ± 0.1 |
| (6712) Hornstein | 0.03 ± 0.02 | 0.85 ± 0.01 | 4.4 ± 0.1 |
| (6769) Brokoff | 0.04 ± 0.02 | 0.90 ± 0.01 | 4.5 ± 0.1 |

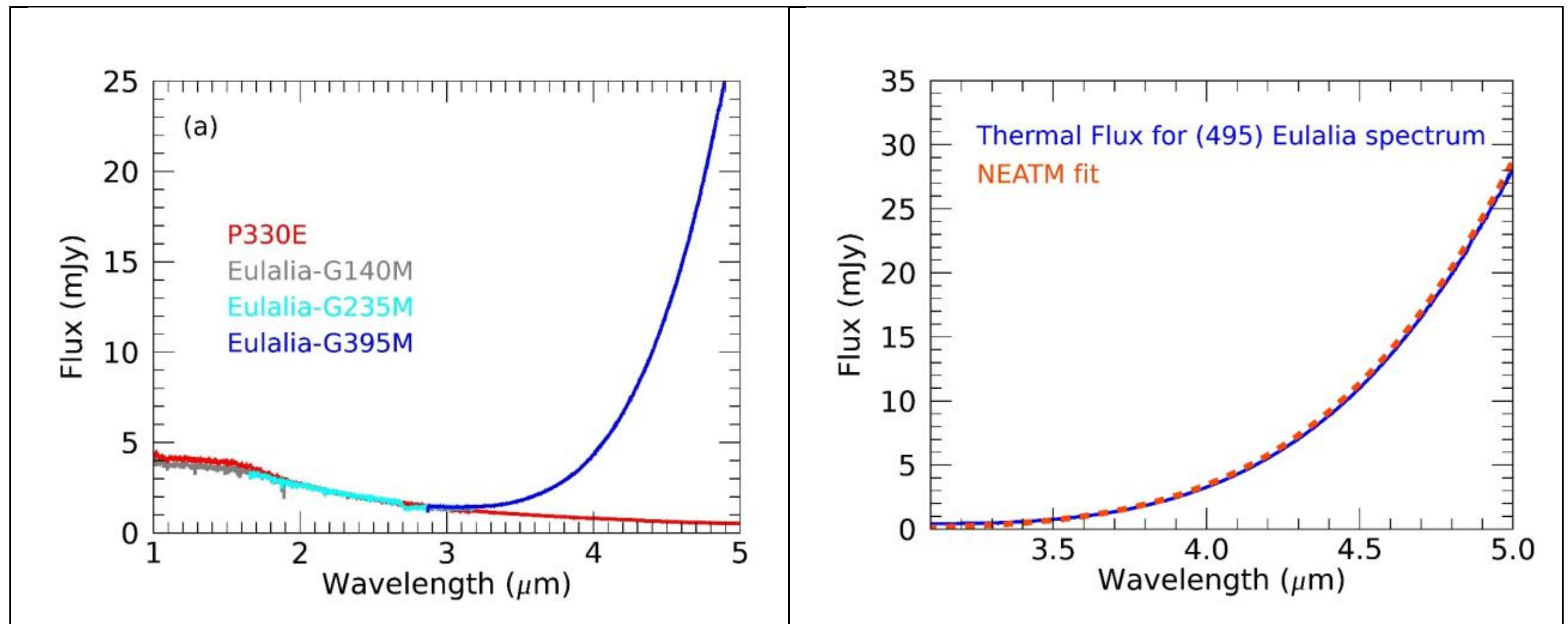

**Figure 2**: **(a)**: The irradiance spectrum of Eulalia in each of the gratings, shown in shades of blue and gray, overlaying the standard star, P330, spectrum in red. **(b)**: The NEATM-derived model (dashed dark orange line) overlaid with the isolated thermal contribution from the (495) Eulalia spectrum (blue solid line). (For interpretation of the references to color in this figure legend, the reader is referred to the web version of this article).

## 2.3 Band Parameter Analysis

To perform the band parameter analyses of the 2.7-µm region, we either use multi-Gaussian or polynomial fitting to determine the band center ($B_C$) and depth ($B_D$). Before either method, the thermally-corrected spectrum was normalized to unity at 2.67 µm (*see* **Appendix**). A linear continuum was constructed and divided from the spectra so as to normalize the top of the band to 1. The linear continuum range for each object is provided in **Appendix.** Multi-Gaussian fitting was used in cases where bands were too complex for a single polynomial fit, based on visual inspection of the spectral band. In cases when multi-Gaussian fitting was used, we invoked 5 Gaussian curves from 2.665 – 3.6 µm to calculate a best-fit model that matches the shape of the band. Due to assumptions inherent to multi-Gaussian fitting, we performed visual inspections to ensure the shape of the best-fit model provided an adequate basis for subsequent depth and center calculations. In cases when the polynomial fitting was used, we invoked 3$^{rd}$ order polynomials that were fit to the bottom 1/3 to ½ of the band (see **Appendix** for more details), which was determined through visual inspection.

We also parameterized the relatively minor absorptions from 3 to 4 µm when present, namely features around 3.1 µm, 3.4 µm and 3.9 µm. Linear continua were constructed for each feature (see **Appendix** for more detail). Each band in this range was divided by the linear continuum, which in turn encapsulated both band edges. We then fit 2$^{nd}$ order polynomials to the visually-determined bottom of the 3.1 µm, 3.4 µm, and 3.9 µm bands, since multi-Gaussian fitting was not necessary for these band shapes (see **Appendix** for more detail).

We estimated uncertainties for both band parameters by conducting a Monte Carlo analysis. For each target and band, the spectral reflectance values were resampled at each wavelength in the band ranges for the multi-Gaussian/polynomial fits mentioned in the **Appendix**. The resampling bounds were determined by the uncertainty at each respective wavelength. Upon each resampling, either the multi-Gaussian or polynomial fitting method was performed, again depending on the method invoked initially. Resampling was repeated 1000 times to in turn determine each band parameter's mean and standard deviation. We find an additional uncertainty in band center of ≤0.005 µm due to the chosen polynomial fit range (see **Appendix**), so we include this value in calculating the total uncertainty. For each band, $B_C$ was determined as the mean minimum point in the 1000 curve fits, whereas $B_D$ was calculated as the mean of the values corresponding to 1 – the reflectance at the band center.

## 2.4 Euclidean Distance

The Euclidean distance, which was used to characterize the spectrum of (84) Klio in a complementary SAMBA-3 work (see Le Pivert-Jolivet et al 2025a), can quantitatively assess the spectral similarity between the targets of this work and other primitive bodies, including carbonaceous chondrites, NEAs, and the JWST spectrum of (142) Polana. To perform this analysis, we normalize all spectra to 2.67 µm, fit a linear continuum from 2.67 to 3.1 µm, and divided the linear continuum from the spectral reflectance. We then calculate the Euclidean distance according to Equation 1.

$d = \sqrt{\sum_{i=1}^{N}(P_i - S_i)^2}$ Eq. (1),

where P is the normalized spectrum of each SAMBA-3 PEC target and S is the normalized spectrum of each comparison sample (*i.e.,* meteorite, NEA, & (142) Polana). To calculate d, the top and bottom of the band are normalized to a reflectance of 1 and 0.5, respectively, to primarily accentuate shape. This process was done for a spectral suite of carbonaceous chondrites (Takir et al. 2013), the NIRSpec spectrum of (142) Polana from Arredondo et al. (2025), two spectra of Bennu (OVIRS spectrum from Hamilton et al. 2019 & averaged sample spectrum from Lauretta et al. 2024), and two spectra of Ryugu (NIRS3 spectrum from Kitazato et al. 2019; A0026/TD1 sample from Pilogret et al. 2022; *see also* Yada et al. 2022; Nakamura et al. 2023). We note that the carbonaceous chondrites samples likely experienced some terrestrial alteration, so their spectra were measured in heated environments to remove any absorbed water. The representative samples from OSIRIS-REx and Hayabusa2 are unaffected by terrestrial contamination/weathering. The samples from Bennu and Ryugu are unaffected by terrestrial contamination/weathering, as they were preserved upon returning to Earth. We restrict this analysis to the 2.67 – 3.1 µm to solely probe the 2.7-µm band and not substantially overlap with the analysis of the potential bands at longer wavelengths (see **Section 2.3**).

## 3. Results

### 3.1 Spectrum of (495) Eulalia

The short-wavelength (0.9 – 2.65 µm) spectrum of (495) Eulalia is characterized by a few distinct properties. As shown in **Fig. 3**, the spectrum exhibits a slightly blue slope from 0.9 – 1.6 µm, and a steeper red slope from 1.6 – 2.65 µm. In total, the short-wavelength portion of (495) Eulalia's spectrum shows a broad feature that can be described by a concave-up shape, typical of many C-complex asteroids (**Fig. 3a**; DeMeo & Carry 2013).

Distinct absorptions are present longward of 2.65 µm (**Fig. 3b-c**). Firstly, we used visual inspection/comparison with the broader asteroid spectrum literature to determine that the spectrum of (495) Eulalia exhibits a sheer drop-off in reflectance that culminates in a "sharp"-type absorption with maximum depth around 2.7 µm (Rivkin et al. 2022). Additional, weaker absorptions are potentially present around 3.1 µm, 3.4 µm, and 3.9 µm. The potential bands at 3.1 µm and 3.4 µm may overlap with the 2.7-µm band, whereas the possible 3.9-µm absorption is more separated from the rest of the features. Beyond 4.1-µm, the noise leftover after the removal of thermal contribution is high relative to the reflectance, limiting meaningful reflectance interpretation in this range. The full spectrum is provided in the **Appendix**.

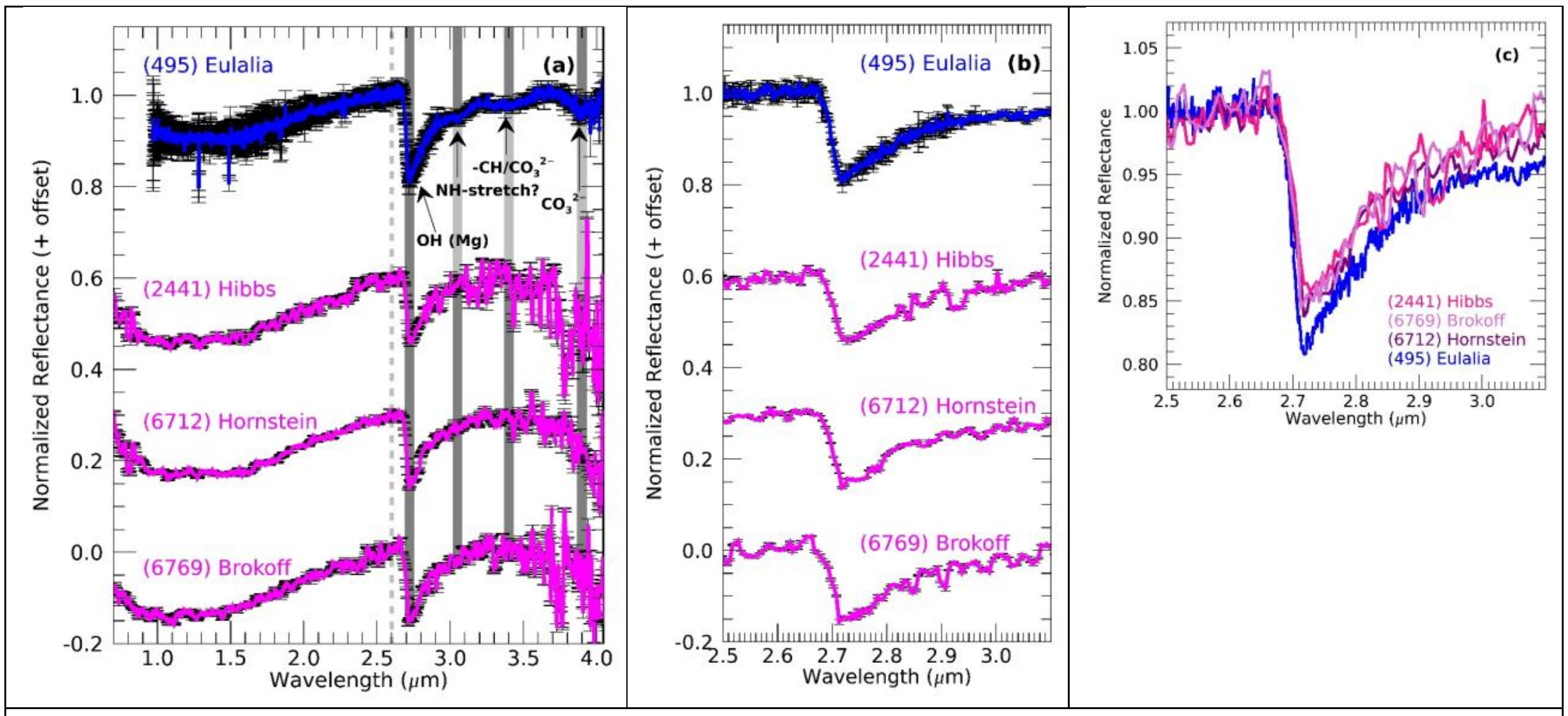

**Figure 3**: **(a)** The reflectance spectra of each PEC target in this work normalized to 2.6 µm (gray dashed vertical line). Spectra are offset in reflectance, by 0.4 for (2441) Hibbs and by increments of 0.3 for the remaining PFAs. All pink objects are members of the Polana asteroid family to distinguish them from the namesake asteroid of the Eulalia Family, (495) Eulalia (in blue). Gray vertical bars locate the broad centers of the identified/tentative absorption bands. **(b)** The 2.7-µm region for each of the targets to highlight the slight shape differences within the sample. Normalization, color-coding, and offsets are the same as the adjacent figure. **(c)** The 2.7 µm region for all 4 PEC targets in this work, with their spectra overlapping to show slight differences in band depth and shape. All spectra are normalized to 2.6 µm and each corresponds to a specific color: (495) Eulalia is in blue, and the three PFAs are in shades of pink/purple. Error bars are removed for clarity. (For interpretation of the references to color in this figure legend, the reader is referred to the web version of this article).

### 3.2 Spectra of (2441) Hibbs, (6712) Hornstein, & (6769) Brokoff

The spectra of the three PFAs in this work (**Fig. 3**) display similar short-wavelength characteristics. All three asteroids exhibit a broad spectral feature that presents as a "concave-up" shape, characterized by a change of slope direction around 1.6 µm (**Fig. 3a**). Namely, blue slopes are observed until ~1.0 µm, whereas red slopes are exhibited from 1.6 – 2.65 µm.

The suite of PFAs in this work exhibits key spectral similarities as well as distinctions. All three PFAs show steep drops in reflectance longward of 2.65 µm with maximum depths that are centered around 2.7 µm (**Fig. 3b-c**). However, the shape of this band is variable among the three PFAs. All four PEC targets in this work exhibit "sharp" absorption minima determined through visual comparisons with asteroid other spectra (e.g., Rivkin et al. 2022), though we note (2441) Hibbs shows a relatively smoother band-bottom than the other two PFAs.

Longward of 3.0 µm, (2441) Hibbs and (6712) Hornstein may show minor absorptions around 3.4 µm and longward 3.8 µm, where (6769) Brokoff may not show the 3.4 µm feature. We stress that low signal-to-noise ratios (SNR) are observed even when the data are binned and that high relative thermal contributions make this wavelength range challenging to fully characterize any of the three PFAs. However, the 3.4-µm regions for (2441) Hibbs and (6712) Hornstein show slight reflectance drops, which are only slightly greater than the local noise of the spectrum. Longward of 3.8-µm, all three PFAs exhibit negative slopes until around 4-µm, potentially underscoring the presence of a band in this region for both objects. Longward of 4-µm, low SNR restricts the analysis and discussion of this range for these objects (see **Appendix** for full-range spectra).

### 3.3 Comparison with Ground-Based Spectra

Ground-based, NIR spectra of all four asteroids in this work are roughly consistent with their JWST NIRSpec counterparts (**Fig. 4**). The Primitive Asteroid Spectroscopic Survey (PRIMASS; Pinilla-Alonso et al. 2024) used the SpeX instrument (prism mode) on the Infrared Telescope Facility (IRTF; 0.7 – 2.5 µm) as well as the NICS instrument on the Telescopio Nazionale Galileo (TNG; 0.9 – 2.5 µm) to observe each object in wavelengths that partially overlap with the JWST data (Pinilla-Alonso et al. 2016). Additionally, a prior observation of (495) Eulalia from Fieber-Beyer et al. 2008 showed consistency with the PRIMASS spectrum. Each JWST spectrum shows general agreement with its PRIMASS counterpart in terms of slope and concavity. When slopes are calculated from 1.0 – 2.3 µm for each spectrum (i.e., wavelength range of consistent coverage for all 8 spectra in Fig. 4), we find that that slope differences between the ground-based and JWST-based spectra are < 0.05 $\mu m^{-1}$ (0.5%/1000 Angstrom), which is within the slope uncertainty in Pinilla-Alonso et al. 2016 (~0.06 $\mu m^{-1}$ or 0.6%/1000 Angstrom). We also note that phase angle can affect parameters like slope, but the effect is usually an order of magnitude smaller (~ 0.001 $\mu m^{-1}$; Lantz et al. 2018) than the observed difference. Thus, we can potentially attribute the slight spectral difference to the use of different solar analogs and/or potential surface variabilities in composition/grain size, though more observations are required to test potential surface variability.

Each of the PEC spectra presents a 2.7-µm band consistent with a distinct subset of Main Belt asteroids. Although this work's targets have not been included in previous works' ground-based observations of the 3-µm region (2.5 – 4.0 µm), each of the asteroid spectra presented in this work exhibit "sharp"-type 2.7-µm bands (Takir & Emery 2012; Rivkin 2022). Sharp-type bands have been correlated with asteroids that have potentially formed interior to the ammonia snow-line (compared to "not-sharp-types"; Rivkin et al. 2022), and they are predominantly located currently in the Inner (2.1 – 2.5 AU) and Middle (2.5 – 2.82 AU) Main Belt.

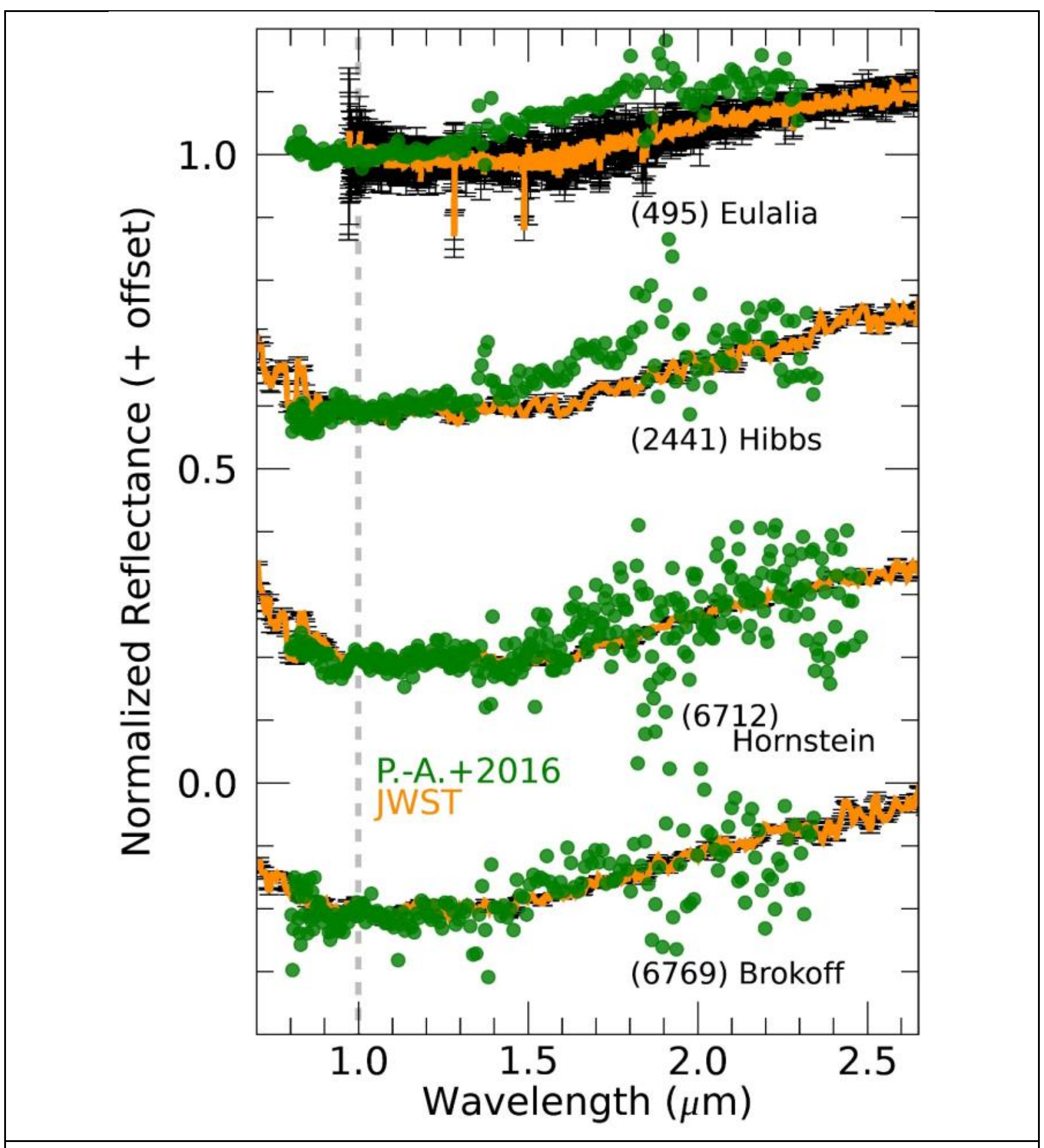

**Figure 4:** Comparison between ground-based (green; P.-A.+2016 = Pinilla-Alonso et al. 2016; *see also* Pinilla-Alonso et al. 2024) and JWST NIRSpec (orange) spectra for the targets in this work, normalized to 1 at 1 µm (labeled as the dashed gray vertical line). Reflectance values for each spectrum are offset by increments of 0.4. Errors are not included in the PRIMASS library and therefore are not included in this figure, though as mentioned in PRIMASS-related work, the error is roughly the point-to-point spread in the data (e.g., de Leon et al. 2016; Pinilla-Alonso et al. 2016; also see Pinilla-Alonso et al. 2024 for more details). (For interpretation of the references to color in this figure legend, the reader is referred to the web version of this article).

## 4. Analysis

The PEC asteroids in this work have a variety of spectral similarities and differences with each other and with the broader spectral suite of carbon-rich bodies in the literature. JWST spectra in this work exhibit NIR slopes (1.0 – 2.2 µm) that are consistent with previous ground-based observations (**Fig. 4;** Pinilla-Alonso et al. 2016). A 2.7-µm band is observed for each of the four PEC asteroids of this work. The 2.7-µm bands are roughly similar in shape and exhibit a "sharp"- type band (e.g., Takir & Emery 2012; Takir et al 2013; Rivkin et al. 2022). (495) Eulalia's 2.7-µm band is quantitatively deeper than that of each of the three PFAs, and its spectrum is consistent with a smattering of additional absorptions at longer wavelengths. In this section, we parameterize, analyze, and report $B_C$ and $B_D$ for all four 2.7-µm bands as well as the longer-wavelength absorptions observed for (495) Eulalia. We also make estimates of the tentative absorptions for the PFAs.

### 4.1 Band Parameter Analysis of (495) Eulalia

The parameterization of the 2.7-µm band for (495) Eulalia invoked multi-Gaussian fitting, similar to that which was done for (84) Klio in a complementary SAMBA-3 work (*see* Le Pivert-Jolivet et al. 2025a). The linear continuum for (495) Eulalia's band was constructed from the band edges, which are around 2.60 µm and 3.67 µm. The actual band range is defined as 2.665 – 3.6 µm (see **Appendix**). The resulting model produced for the spectrum of (495) Eulalia is provided in **Figure 5**, and band parameters are provided in **Table 4**.

We also parameterized (495) Eulalia's relatively minor absorptions from 3.0 – 4.0 µm. As mentioned in **Section 2.3**, we used 2$^{nd}$ order polynomial fitting for bands around 3.1 µm, 3.4 µm, and 3.9 µm (see the **Appendix** for ranges). Ultimately, we determined the band parameters for each absorption, which are reported in **Table 4**. In order to make direct band parameter comparisons with (495) Eulalia, we also re-determined band parameters for (142) Polana (**Table 4; Appendix**). Some differences in band depth and center are found between this work and Arredondo et al. 2025, but we attribute these small differences to the specific definitions used to determine linear continua and polynomial fit ranges (see **Appendix** for full details on wavelength ranges).

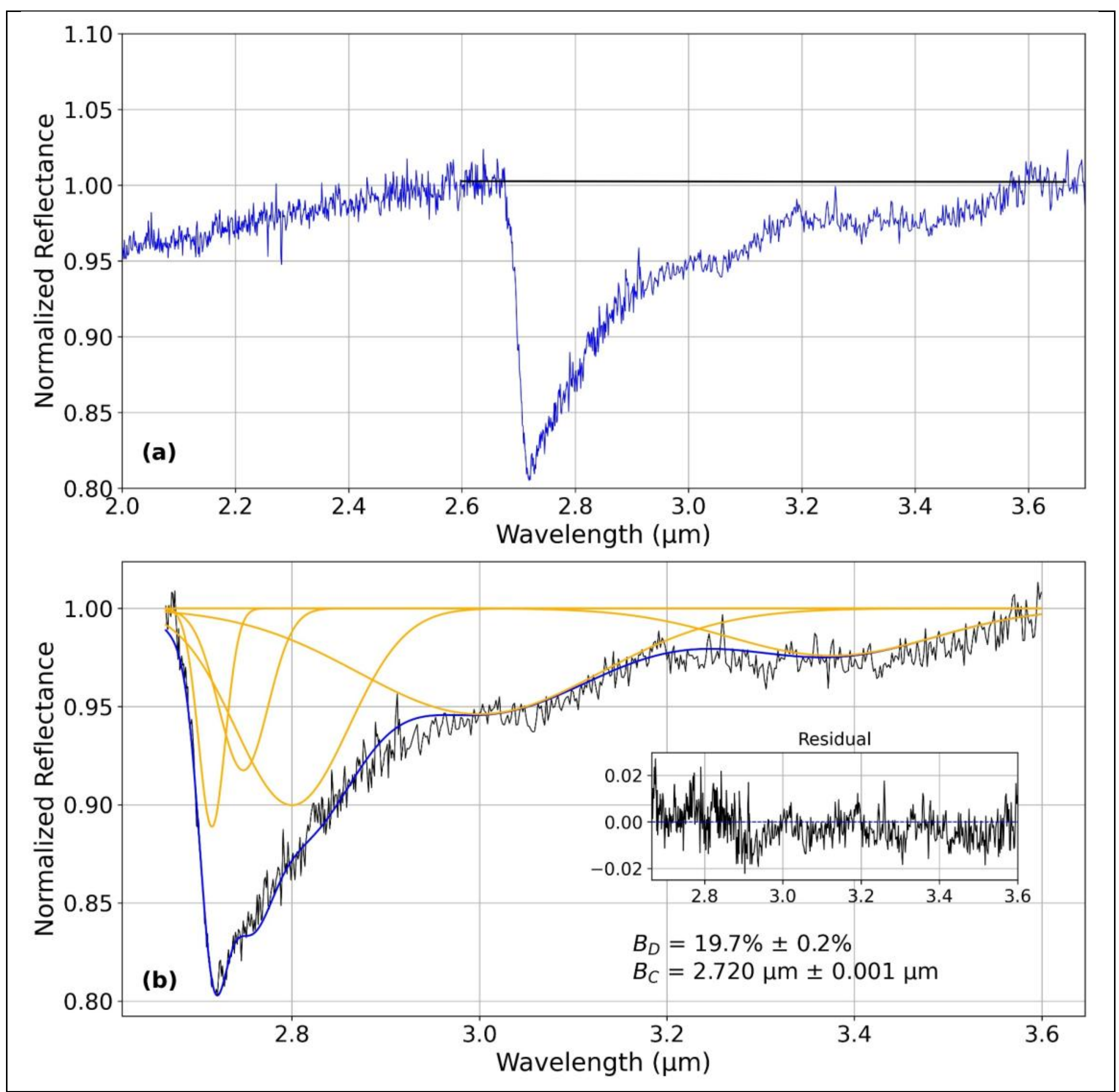

**Figure 5: (a)** Continuum line (black solid line) calculated for the 2.7 µm band for (495) Eulalia's spectrum (in blue). The reflectance is then divided by the continuum line. **(b)** Gaussian fits (in yellow) used to produce a multi-Gaussian-based model of the band (in red). The residual difference of the model and the data as well as the calculated band parameters are shown as well. (For interpretation of the references to color in this figure legend, the reader is referred to the web version of this article).

### 4.2 Band Parameter Analysis of (2441) Hibbs, (6712) Hornstein, and (6769) Brokoff

(2441) Hibbs, (6712) Hornstein, and (6769) Brokoff all exhibit the 2.7-µm band, which was parameterized in terms of $B_C$ and $B_D$ for each object. We invoked 3rd order polynomials (see **Appendix** for more details) to the bottom of the band for each PFA, instead of implementing multi-Gaussian fitting. We note that we do not rule out the possibility of multiple, Gaussian-shaped contributions, but a series of multi-Gaussian-based trials found that best-fit models often did not match their overall band shape(s). The $B_C$ and $B_D$ values and uncertainties are reported in **Table 4**. No minor absorption from 3.0 – 4.0 µm was parameterized via multi-Gaussian or polynomial fitting for any of the three PFAs, due to the progressively lower SNR beyond 3.0 µm.

Visual inspection-based determinations of band parameters of potential absorptions around 3.4 µm are reported in **Table 4** for (2441) Hibbs and (6712) Hornstein (see also **Section 3.2**).

### 4.3 Compositional Interpretations

For all four PEC targets in this work, there is a consistent presence of some spectral features. All four targets' spectra show concavity shortward of 2.6 µm, which is potentially attributable to Fe-bearing minerals (e.g., magnetite or Fe-rich silicates; Takir et al. 2024; Carrozzo et al. 2026). Additionally, each of the targets in this work exhibits a 2.7-µm band, suggesting their surfaces contain phyllosilicates. The shapes and center positions of the bands are consistent with the OH stretch influenced by the presence of the $Mg^{2+}$ cation (**Table 4**).

In (495) Eulalia's spectrum, minor absorptions from 3.0 to 4.0 µm suggest additional surface materials. The shape and center of the minor band around 3.1 µm is potentially consistent with ammoniated clays (*e.g.,* King et al. 1992; see also: Rivkin et al. 2025). The 3.4-µm band is potentially caused by a mixture of organics and carbonates (Dartois et al. 2004; Kaplan et al. 2020; Huang & Kerr. 1960). The center of the 3.4-µm band may specifically indicate an asymmetric stretch of $-CH_2$ bonds in aliphatic organics (Dartois et al. 2004). The presence of the feature centered around 3.9 µm is consistent with many carbonate overtones (Huang & Kerr, 1960), but the center of this potential feature aligns most-closely with that of $CaCO_3$ (Huang & Kerr 1960) The tentative 3.4-µm and 3.9-µm features for (2441) Hibbs and (6712) Hornstein are also centered consistently with organics/carbonates and carbonates, respectively.

In addition to band parameter analysis, numerous equations have been formulated to determine relative abundances of specific elements and compounds in carbonaceous bodies. Additionally, each value's uncertainty can be propagated in the standard manner (i.e., in quadrature). For example, Rivkin et al. 2003 formulated an empirical equation for estimating H/Si:

$H/Si = 7.38 – 7.29 \times (R_{2.9} / R_{2.5})$ … Eq. (1),

where $R_{2.9}$ and $R_{2.5}$ are the reflectance values at 2.9 µm and 2.5 µm, respectively. Using this equation, H/Si ratios for our objects are calculated to be roughly ≤ 0.6 (see **Table 5** for propagated uncertainties), which is consistent with that of (142) Polana, which was both calculated in Arredondo et al. 2025 and re-calculated in this work for consistency. As done in Arredondo et al. 2025, the H/Si ratio can be used to estimate the $H_2O$ wt% by multiplying the H/Si ratio by the atomic weight ratio of $H_2O:SiO_2$ (≈ 18.02 amu/60.08 amu), the weight fraction of $SiO_2$ in carbonaceous chondrites (≈ 30%; Jarosewich 1990), and 0.5 to account for the 2 hydrogen atoms in $H_2O$. Using this methodology, the estimated values for $H_2O$ wt. %. for the targets in this work range from roughly 1.7% to 2.6%, though the uncertainties for the reported values in this work ($\sigma \leq 0.83$, see **Table 5**) encapsulate with the reported value for (142) Polana in Arredondo et al. 2025 and this work (*i.e.,* 1.41%). Further, the H wt. % can be isolated by dividing the $H_2O$ wt. % value by the atomic weight of hydrogen (1.008 amu) and multiplying by 2 (Arredondo et al. 2025). This calculation results in an H wt. % range of roughly 0.2 to 0.3 wt. %, though uncertainties ($0.01 > \sigma < 0.1$) again encapsulate the corresponding H wt. % value for (142) Polana (*i.e.,* 0.16 wt. %; Arredondo et al. 2025 as well as the re-calculation in this work, shown in **Table 5**).

**Table 4**: Band parameters and potential contributor(s) for each absorption discussed in this work.

| Parameter | (495) Eulalia | (2441) Hibbs | (6712) Hornstein | (6769) Brokoff | (142) Polana[a] |
|---|---|---|---|---|---|
| 2.7-µm Band | | | | | |
| Present? | Yes | Yes | Yes | Yes | Yes |
| Center (µm) | 2.720 ± 0.001 | 2.73 ± 0.01 | 2.74 ± 0.01 | 2.73 ± 0.01 | 2.719 ± 0.001[b] |
| Depth (%) | 19.7 ± 0.2 | 14.4 ± 0.3 | 15.3 ± 0.2 | 16.6 ± 0.2 | 19.2 ± 0.2[b] |
| Attribution | Mg-rich phyllosilicates [c] | Mg-rich phyllosilicates | Mg-rich phyllosilicates | Mg-rich phyllosilicates | Mg-rich phyllosilicates |
| 3.1-µm Band | | | | | |
| Present? | Yes | Inconclusive [d] | Inconclusive | Inconclusive | Potentially[d] |
| Center (µm) | 3.09 ± 0.01 | … | … | … | Inconclusive |
| Depth (%) | 1.0 ± 0.1 | … | … | … | < 1 % |
| Attribution | NH-rich clays? [e] | … | … | … | Inconclusive |
| 3.4-µm Band | | | | | |
| Present? | Yes | Potentially [d] | Potentially [d] | Inconclusive | Yes |
| Center (µm) | 3.41 ± 0.01 | ~3.43 | ~3.43 | … | 3.45 ± 0.01[b] |
| Depth (%) | 1.7 ± 0.1 | < 1 | < 1 | … | 2.23 ± 0.04[b] |
| Attribution | Aliph. $CH_2$ and/or $CO_3^{-2}$? [f,g] | Aliph. $CH_2$ and/or $CO_3^{2-}$? | Aliph. $CH_2$ and/or $CO_3^{2-}$? | … | Aliph. $CH_2$ and/or $CO_3^{2-}$? |
| 3.9-µm Band | | | | | |
| Present? | Yes | Potentially [d] | Potentially [d] | Potentially [d] | Yes |
| Center (µm) | 3.91 ± 0.01 | … | … | … | 3.91 ± 0.01[b] |
| Depth (%) | 5.7 ± 0.1 | … | … | … | 2.68 ± 0.03[b] |
| Attribution | Carbonates; $CaCO_3$? [g] | Carbonates? | Carbonates? | Carbonates? | Carbonates; $CaCO_3$? |

[a] Spectrum from Arredondo et al. (2025)

[b] Band parameters for (142) Polana were re-calculated using the methods used for (495) Eulalia, described in **Section 4.1**.

[c] Takir et al. (2013)

[d] These values are tentative and done solely through visual inspection of the region. See Section 3.2 for more details.

[e] King et al. (1992)

[f] Dartois et al. (2004)

[g] Huang & Kerr (1960)

**Table 5**: Calculated abundances for each target in this work

| Name | H/Si | $H_2O$ wt. % (%) | H wt. % |
|---|---|---|---|
| (495) Eulalia | 0.57 ± 0.08 | 2.56 ± 0.37 | 0.29 ± 0.04 |
| (2441) Hibbs | 0.39 ± 0.15 | 1.76 ± 0.70 | 0.20 ± 0.08 |
| (6712) Hornstein | 0.39 ± 0.07 | 1.74 ± 0.33 | 0.19 ± 0.04 |
| (6769) Brokoff | 0.40 ± 0.18 | 1.81 ± 0.82 | 0.20 ± 0.09 |
| (142) Polana[a] | 0.31 ± 0.07 | 1.41 ± 0.32 | 0.16 ± 0.04 |

[a] Values have been calculated for (142) Polana in Arredondo et al. 2025, but we re-calculated them for consistency within this work. We note that our values agree within the reported propagated uncertainties.

## 4.4 Comparison of the 2.7 µm Band Shape

The 2.7-µm bands of the asteroids in this work exhibit key similarities with and differences from carbon-rich meteorites. As shown in **Fig. 6a-c**, the band shapes and centers, though not band depths, for all four PEC targets match well with hydrated carbonaceous chondrites. The Euclidean distance metric especially matches LAP02277 (CM1) with the four JWST targets of this work (**Figure 7a-d**). The spectrum of the Ivuna (CI) meteorite, which is part of the most aqueously-altered meteorite class, is also a relatively close match for (495) Eulalia, (6712) Hornstein, and (6769) Brokoff, whereas (2441) Hibbs shows a relatively lower match to the Ivuna spectrum in comparison. Instead, (2441) Hibbs shows a closer match to the spectral band shape exhibited by Al Rais, a CR2 meteorite.

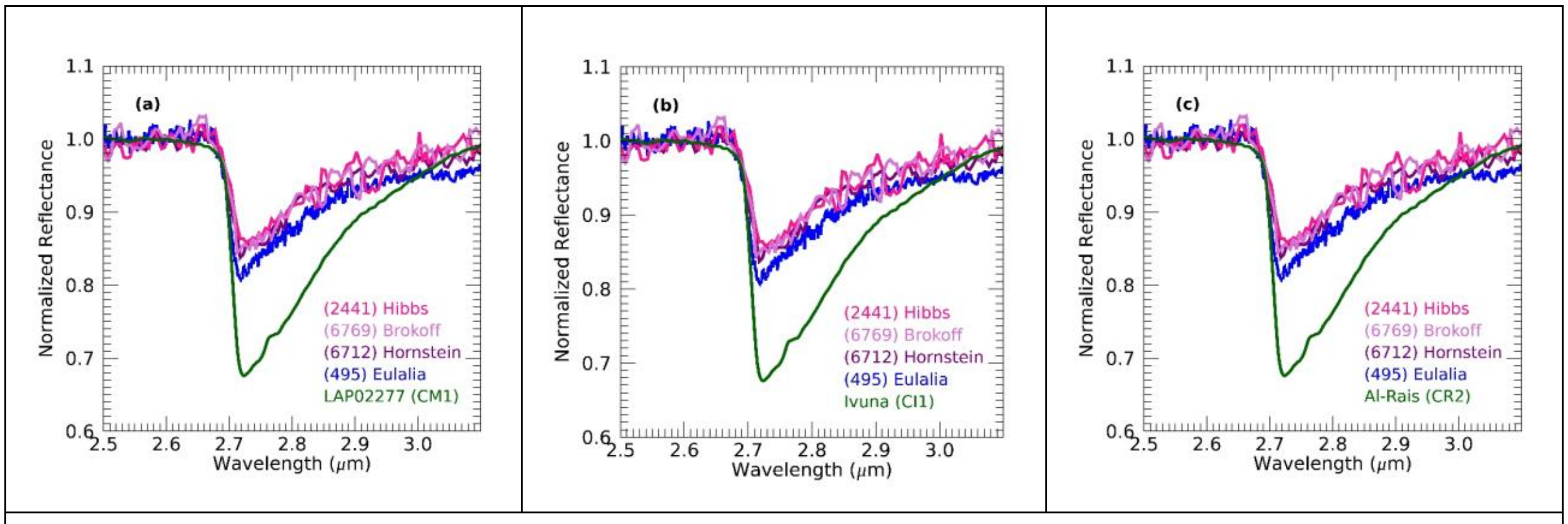

Figure 6: The 2.7-µm bands for the four targets in this work compared with three hydrated carbonaceous chondrite spectra measured by Takir et al. 2013. All spectra are normalized to 1 at 2.6 µm. No error bars are shown for figure clarity. However, error bars are shown for this work's targets in **Fig. 3**, and error bars for the meteorite spectra from Takir et al. 2013 are relatively small due to being measured in a laboratory setting. Shades of pink/purple are the PFAs' spectra, (495) Eulalia's spectrum is shown in blue, and **(a)** a CM1 meteorite, **(b)** a CI1 meteorite, and **(c)** a C2 meteorite are shown in green.

The objects of this work show 2.7-µm band shapes that are comparable to other primitive asteroids. (495) Eulalia, (6712) Hornstein, and (6769) Brokoff show the closest asteroid spectral match to (142) Polana, whereas (2441) Hibbs matches slightly better with the OVIRS spectrum of Bennu (Hamilton et al. 2019; hereafter, Bennu-Remote) than with (142) Polana. Between Bennu and Ryugu, Bennu-Remote consistently yields a smaller Euclidean Distance for all of the targets in this work. We again note that the Euclidean Distance is a shape-based metric, so there are limitations to its use. There is a qualitative similarity between the return-sample spectra of Bennu and Ryugu (hereafter, Bennu-Sample and Ryugu-Sample, respectively). However, Bennu-Sample consistently exhibits a smaller Euclidean metric than Ryugu-Sample. The spectral resolution of Ryugu-Sample is notably lower/noisier than Bennu-Sample, underscoring potential biasing effects from the Euclidean distance's resampling of the JWST spectra with lower SNR.

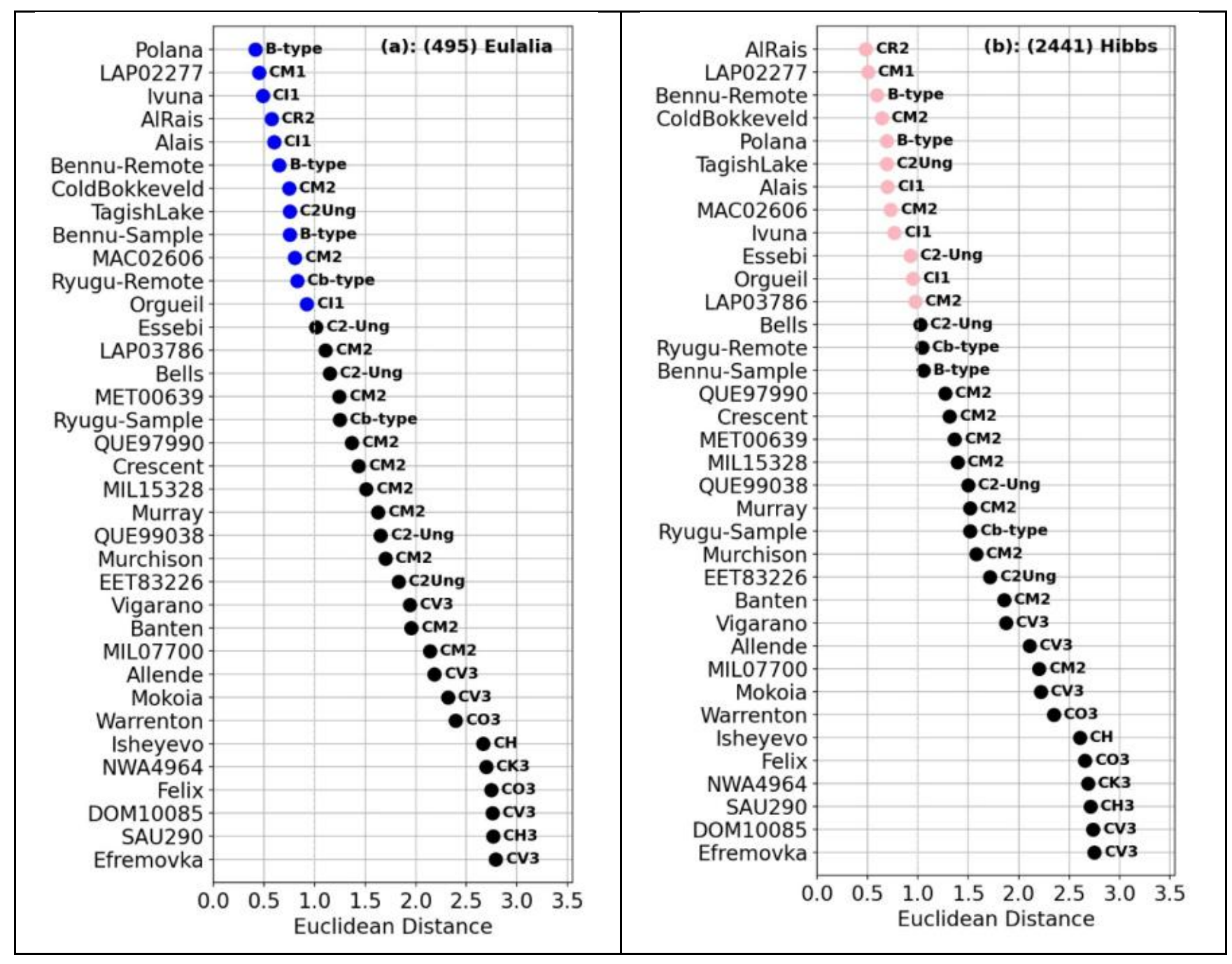

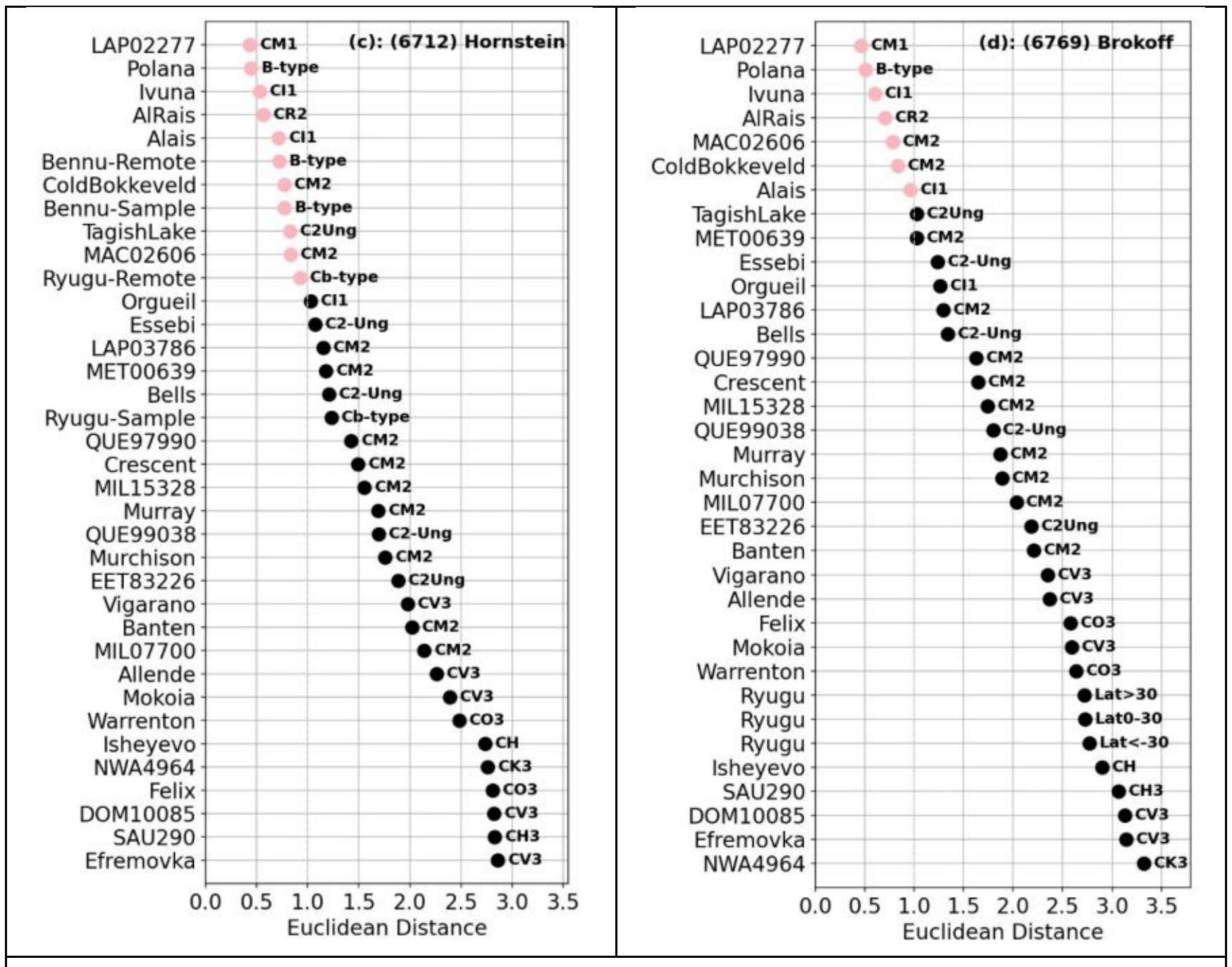

**Figure 7**: Euclidean distances are calculated with respect to a suite of Carbonaceous Chondrites, (142) Polana, and the remotely-sensed and return sample spectra for Bennu and Ryugu. Meteorites/asteroids with Euclidean Distances of < 1 with **(a)** (495) Eulalia (in blue) **(b)** (2441) Hibbs (in pink) **(c)**: (6712) Hornstein (in pink), and **(d)**: (6769) Brokoff (in pink) are highlighted to close matches. (For interpretation of the references to color in this figure legend, the reader is referred to the web version of this article).

## 5. Discussion

### 5.1 Alteration History of the Polana-Eulalia Complex

Based on the 2.7-µm band centers and shapes, spectra of the targets in this work and of (142) Polana from Arredondo et al. (2025) are consistent with CC-type meteorites that are rich in magnesium (Takir et al. 2013; Lawrence & Ehlmann 2025). The 2.7-µm band center of Mg-rich phyllosilicates exhibit band centers at shorter wavelengths (e.g., ~2.70-2.73 µm) than Fe-rich phyllosilicates (e.g., ~2.77-2.90 µm), and Mg-rich phyllosilicates show sharper bands than those produced from their Fe-rich counterparts (Takir et al. 2013; Lawrence & Ehlmann 2025). In all, Mg-rich phyllosilicates correlate with higher degrees of aqueous alteration compared to Fe-rich counterparts. PEC targets consistently show small Euclidean distances from aqueously-altered CCs, like a CI (i.e., Ivuna), CM1 (*i.e.,* LAP02277) and a CR2 (*i.e.,* Al Rais) meteorite (**Fig. 6-7**). CI meteorites are the most aqueously-altered CC-type, followed by CM1 meteorites like LAP02277 (*e.g.,* Takir et al. 2013). That is, CM1s are among the most aqueously-altered members among the broader CM class (Rubin et al. 2007). Refractory element abundances place Al Rais within the relatively less-altered CR2 type, but Al Rais specifically shows grain rims consistent with substantial aqueous alteration as well as relatively high volatile (e.g., water) and serpentine abundances compared to other CR types (Weisburg et al. 1993, Morlok & Libourel 2013). Broad connections to aqueously altered CCs are also consistent with the values of $H_2O$ wt. % reported in **Table 5.** Thus, the spectral similarities between aqueously-altered meteorites and PEC targets provide further evidence that PEC asteroids experienced high degrees of aqueous alteration, though did not reach the end of the aqueous alteration sequence. These results from JWST demonstrate that PEC asteroids originate from ice-rich planetesimals that likely underwent some internal heating (*e.g.,* Dyl et al. 2012; Glavin et al. 2025).

The PEC sample's consistent link to Mg-rich phyllosilicates builds upon visible-wavelength, ground-based spectroscopic observations of PEC asteroids, which do not show 0.7-µm absorptions associated with Fe-bearing phyllosilicates (de Leon et al. 2016). Since the PEC broadly lacks 0.7-µm features and (so far) exhibits 2.7-µm features consistent with Mg-rich phyllosilicates, this work further indicates the PEC's parent bodies underwent high degrees of aqueous alteration. We note that although observed band parameters of PEC asteroids are consistent with high degrees of aqueous alteration in the past, the JWST sample of PEC asteroids does show some variability in their key 2.7-µm band parameters (i.e., band depth and center; see **Table 4**). Minor differences in the 2.7-µm region may underscore a range, albeit restricted, to the high degree of aqueous alteration in the PEC's parent bodies.

Spectral differences between (142) Polana and its smaller family members may reflect key aspects of the family's formation and evolution. With the exception of (6712) Hornstein, the 2.7-µm band center differences between (142) Polana ($B_C \approx 2.719$ µm) and the smaller PFAs ($B_C \approx 2.73$-2.74 µm) are within the uncertainties, whereas the band depth difference are more significant -- $B_D \approx 19\%$ for (142) Polana and $B_D \approx 14$-17% for the PFAs. Differences in band parameters may reflect a size-dependence in the phyllosilicate abundance throughout the family. Slight spectral differences in this wavelength range may underscore a stratification of material within the Polana family parent body, with smaller and larger members representing distinct parts/layers that experienced different levels of melting (e.g., Neumann et al. 2021). Michel et al. (2020) have also shown that collisional families can be heterogeneous in terms of hydration depending on the impact energy within the family's collisional history. Thus, each family member's collisional evolution – which is presumed to be more active for smaller asteroids than larger ones (e.g., Bottke & Morbidelli 2017) – may have played a role in developing a size-based surface evolution trend among Polana family asteroids. Altogether, both the internal structure of the Polana family parent body and the collisional evolution that formed the modern-day Polana family have likely worked together to form the 2.7-µm band shape and center differences exhibited between (142) Polana and the PFAs in this work. The differences in spectral parameters with size extend from previous works that used VISNIR wavelengths measurements. That is, some prior works have associated hydration (as well as grain size) with spectral parameters of specific CC-types and of members with specific sizes throughout numerous asteroid family families (Johnson & Fanale 1973; Shepard et al. 2008; Vernazza et al. 2016; de León et al. 2010; Thomas et al. 2021; Cantillo et al. 2023; McClure et al. 2025).

The large PEC asteroids ((142) Polana and (495) Eulalia) and the smaller PFAs highlight the complex details of the responses to space weathering (SW) on these carbonaceous bodies.

Whereas laboratory simulations of micrometeorite bombardment decreases the 2.7-µm band depth, solar-wind-driven SW has been shown to shift the band to longer wavelengths over prolonged exposure, for both phyllosilicates and CCs (Lantz et al. 2017; Rubino et al. 2020). Although the difference in band depth between the PFAs (shallower, 16-17%) and (142) Polana (deeper, >19%) is consistent with micrometeorite bombardment, aforementioned band parameter differences may add some complexity to the SW arguments between the larger and smaller Polana family asteroids. That is, the surfaces of larger asteroids are expected to be more weathered, and smaller asteroids are in turn predicted to be less weathered, given their smaller surface areas that presumably lead to shorter resurfacing timescales (e.g., Graves et al. 2018). However, (142) Polana (and (495) Eulalia) exhibit relatively short wavelength centers for their 2.7-µm bands compared to one of the smaller PFAs, (6712) Hornstein (**Table 4**). In their SW simulations on different kinds of serpentines, Rubino et al. (2020) found that the SW-induced shift sequence of the 2.7-µm band is generally at shorter wavelengths for chrysotile serpentines (from 2.707 to 2.721 µm) than lizardite serpentines (from 2.708 to 2.736 µm). Thus, the surfaces of the large PEC asteroids (like (142) Polana and (495) Eulalia) and some mid-sized objects (like (2441) Hibbs and (6769) Brokoff) may actually be more weathered, but they may contain more chrysotile on their surfaces than other mid-sized PEC asteroids (like (6712) Hornstein), which in turn harbor more lizardite on their surfaces as indicated by their longer-wavelength 2.7-µm $B_C$ values. O'Hanley & Wicks (1995) found that chrysolite (as well as antigorite) can result from the recrystallization and replacement of lizardite around 523.15 ± 298.15 K, so the potential lack of chrysotile on mid-sized or smaller members could be providing the constraint that some members are not originated from the warmer, deeper regions of the original parent bodies, where lizardite in turn may be more prevalent.

We note that the PFAs in this work may not provide a comprehensive view of a compositional continuum – if present – in the Polana family and the PEC broadly. More JWST-based observations of smaller (D < 9 km) members in both PEC families would be vital to address/explore the possibility of a size-dependence to composition. Additionally, more observations would be able to explore variability in slope and albedo in the context of SW. Given that the families are much older than the space weathering timescales, more observations of PEC members would be critical to ascertaining the extent and history of SW. Moreover, observations of mid-infrared wavelengths, especially the Si-O stretch from 9.5 – 11 µm, would also be vital to more fully characterizing the PEC's alteration history (e.g., Lawrence & Ehlmann 2025) as well as understanding additional effects from grain size and porosity (e.g., Emery et al. 2006; Martin et al. 2022, 2023).

### 5.2 Substantial Similarities & Minor Differences within the Polana-Eulalia Complex

JWST spectra of (495) Eulalia and (142) Polana demonstrate strong similarities, especially in their 2.7-µm band. (142) Polana and (495) Eulalia exhibit similar 2.7-µm band parameters (see **Section 4.1**; **Table 4; Fig. 8**), building upon VISNIR ground-based observations that found those two asteroids are also similar at those wavelengths (de Leon et al. 2016; Pinilla-Alonso et al. 2016). Additionally, two namesake asteroids and the PFAs demonstrate consistency with H/Si, H2O wt. % and H wt. % (**Table 5**), underscoring strong similarities in composition. The consistencies between these two asteroid families in this wavelength range suggest they have nearly identical phyllosilicate mineralogy and abundance, in turn implying similar formation environments for the parent bodies of their respective families.

Minor absorptions in this work's (and previous work's) JWST spectra suggest a possible connection to the bulk differences between the PEC families in NUV-VIS wavelengths (Delbo et al. 2023; McClure et al. 2025; Arredondo et al. 2025). There are non-negligible differences between (142) Polana and (495) Eulalia – and potentially by extension their families. (495) Eulalia exhibits a 3.1-µm band with $B_D \approx 1\%$, whereas (142) Polana's spectrum only exhibits a very shallow ($B_D <$ 1%) feature. The 3.1 µm band in (495) Eulalia's spectrum is tentatively attributed to ammoniated clays (**Table 4**) and is presumably in higher amount on (495) Eulalia's surface than (142) Polana's. The 3.4-µm bands of the (142) Polana and (495) Eulalia are both tentatively attributed to aliphatic organics and carbonates, with (142) Polana's band being only slightly (~0.5%) deeper than (495) Eulalia's (**Table 4**). Conversely, the 3.9-µm band, solely attributed to carbonates in this work, is ≃3% deeper in the (495) Eulalia spectrum than that of the (142) Polana spectrum, potentially indicating higher amounts of carbonates (and NH-bearing compounds) on (495) Eulalia's surface.

Subtle differences in the tentative carbonate absorptions (Fig. 8) are potentially reflective of the formation and/or evolution of the PEC. The 3.9-µm band depth difference between (495) Eulalia and (142) Polana suggests the primary compositional difference between Polana and Eulalia family asteroids is in their carbonate amounts. A difference in carbonates between Polana and Eulalia family asteroids could reflect that their respective parent bodies collected materials from slightly different reservoirs in the outer protoplanetary disk, where they presumably formed (Walsh et al. 2011, 2013). Additionally, distinguishable amounts of carbonates could also indicate slightly different degrees of aqueous alteration within the Polana and Eulalia family parent bodies. Compared to the Polana family parent body, the Eulalia family parent body potentially harbored/produced more Type 2 calcite, which likely form near/at the end of the aqueous alteration sequence (*see* Lee et al. 2014 *and refs. therein*). The Eulalia family parent body could have specifically harbored slightly higher amounts of precursor, carbon-bearing molecules (e.g., CO, $CO_2$, $CH_4$) for carbonates. These precursor materials would have presumably been heated by radiogenic heating in the parent body, leading to carbonate formation (e.g., Froh et al. 2023 and refs. therein). This scenario is more difficult to support with the 3.4-µm band, due to the overlapping absorptions of carbonates and aliphatics in this region. However, (495) Eulalia having a slightly shallower 3.4-µm feature than (142) Polana may imply a slightly higher production of organics within the Polana family parent body. Laboratory-based experiments that simulate space weathering environments have demonstrated that ion-irradiation can reduce the 3.4-µm band caused by organics, independent of the presence of the relatively more-resistant carbonates (*e.g.,* De Sanctis et al. 2024 and refs. therein). Additionally, the relative ages of the Polana and Eulalia families add further complexity to the potential differences in organic matter because the Polana family is older (Walsh et al. 2013) and has thus experienced longer exposure to space weathering than the Eulalia family. In total, more observations, particularly of small members and at longer wavelengths, are crucial to elucidating spectral differences between the PEC families.

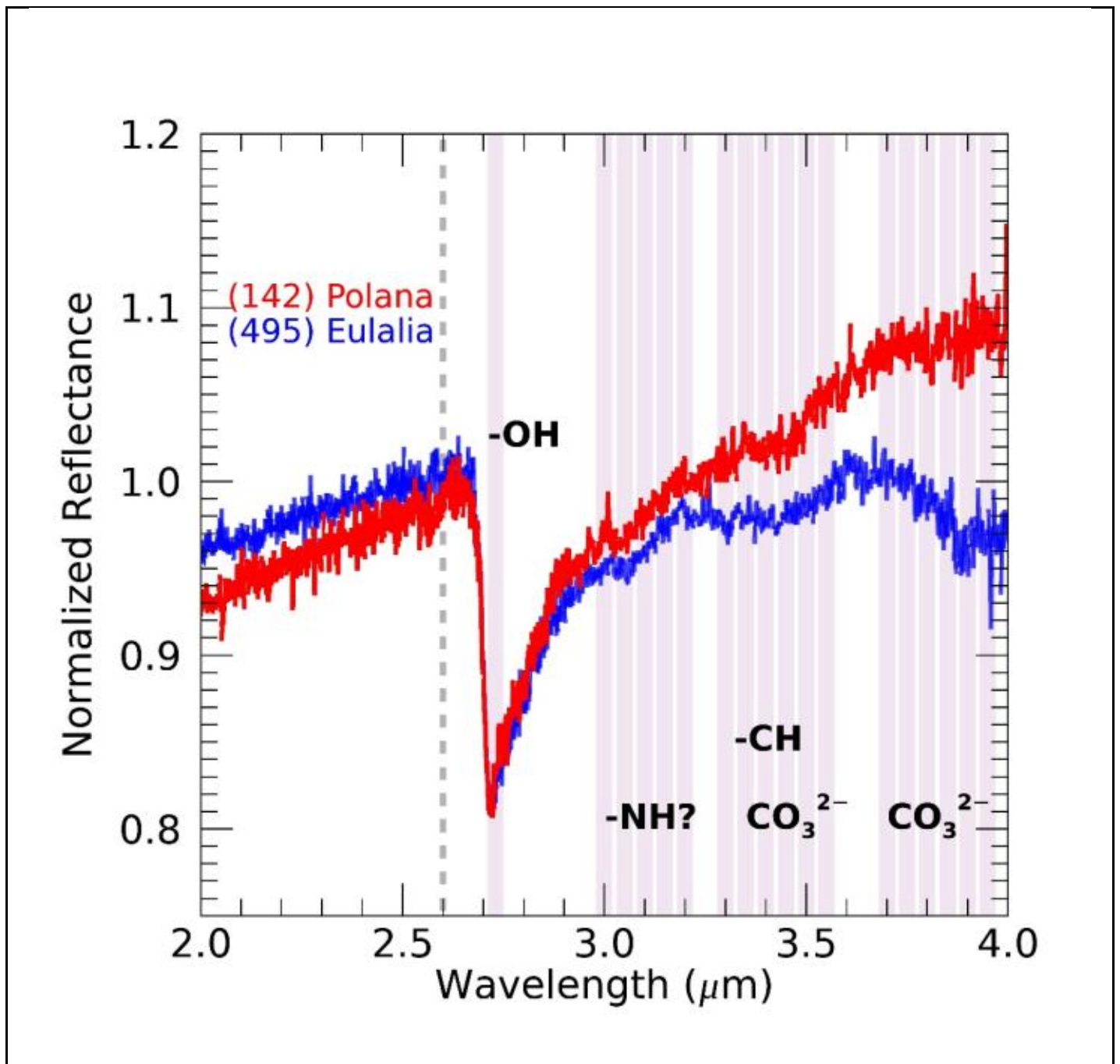

**Figure 8**: Comparison of (142) Polana (red; Arredondo et al. 2025) and (495) Eulalia (this work). Both spectra show features around 2.72 µm, 3.1 µm, 3.4 µm, and 3.9 µm, highlighted by the light purple regions. Both spectra are normalized to 2.6 µm (gray dashed vertical line).

### 5.3 The PEC as the Origin for Bennu and Ryugu

The JWST spectral suite of PEC asteroids – particularly (142) Polana and (495) Eulalia – complement the broader literature pertaining to the determination of the origin family of Bennu and Ryugu. Although dynamical models have mostly favored the origins of Bennu and Ryugu to be the Polana Family (~70%) over the Eulalia family (~30%) (Campins et al. 2010, 2013; Bottke et al. 2015), spectroscopic campaigns of PEC asteroids in VISNIR wavelengths suggested Bennu and Ryugu can only be linked back to the broader origin of the PEC (de Leon et al. 2013; Pinilla-Alonso et al. 2016). The families have been subtly disentangled in terms of NUV- and VIS-wavelength slope, which in turn link Bennu and Ryugu with the Eulalia and Polana families, respectively (Delbo et al. 2023; McClure et al. 2025). Additionally, (142) Polana is currently the only PEC asteroid of which a mid-infrared spectrum has been measured (Arredondo et el. 2025). (142) Polana differs from Bennu's globally-averaged OTES spectrum in the 10 – 15 µm range, perhaps due to a difference between the two bodies' physical structures and/or space weathering histories (Hamilton et al. 2021; Arredondo et el. 2025; see **Sections 5.1-5.2**). Like Bennu and Ryugu, the JWST spectra measured from PEC asteroids include key features associated with carbonaceous materials and hydrated phyllosilicates (Fig. 3; Fig 7; Table 4; Pilogret et al. 2022; Lauretta et al. 2024; Arredondo et al. 2025). With the current JWST repository of PEC asteroids, we are unable to claim a specific family of origin for either Bennu or Ryugu, but our observations still support a broad origin in the PEC. More observations, particularly of smaller members, will be crucial to determining specific family membership.

The shapes and centers 2.7-µm bands from this work and from Arredondo et al. 2025 (*i.e.,* (142) Polana) provide broad support that the PEC is the origin population of Bennu and Ryugu (**Fig. 9**; *e.g.,* Campins et al. 2013; Bottke et al. 2015; de Leon et al. 2016; Pinilla-Alonso et al. 2016; Delbo et al. 2023; McClure et al. 2025; Arredondo et al. 2025). All four targets exhibit small Euclidean distances (**Fig. 7**) with Bennu-Remote, suggesting that the overall band shape for the targets of this work are most consistent with the globally-averaged surface observation of Bennu (though we note that the other remote and return-sample spectra are relatively close matches as well). The PFAs' 2.7-µm band centers ($B_c$ ≈ 2.73-2.74 µm) reinforce the relatively close match between Bennu-Remote ($B_c$ = 2.74 µm ± 0.01 µm; Hamilton et al. 2019). Although similar to Bennu-Remote in terms of the Euclidean distance metric, the larger PEC asteroids (*i.e.,* (142) Polana and 495 Eulalia) have 2.7 µm band centers ($B_c$ ≈ 2.72 µm) that are closely aligned with that of the Bennu-Sample/Ryugu-Sample ($B_c$ ≈ 2.72 µm; Pilogret et al. 2022; Lauretta et al. 2024), which include subsurface materials that were captured during the sample collections on these NEA (e.g., Le Pivert-Jolivet et al. 2023).

The PEC, Bennu, and Ryugu show a range of degrees to the aqueous alteration process, highlighting yet another area of commonality. Remote observations of Bennu and Ryugu suggest an origin in an aqueously-altered parent body (*e.g.,* Hamilton et al. 2019; Tatsumi et al. 2021), and the Bennu and Ryugu returned samples further detail how – and to what extent – aqueous alteration occurred within the Bennu and Ryugu parent body asteroid(s) (e.g., Nakamura et al. 2023; Brunetto et al. 2023, Matsumoto et al. 2024; Lauretta et al. 2024; Barnes et al. 2025; Pilogret et al. 2025; Le Pivert-Jolivet et al. 2025b). In particular, Pilogret et al. (2025) show that Bennu and Ryugu samples are indicative of extensive aqueous alteration in the past, but the bulk-averaged spectrum of Ryugu's samples shows a deeper 2.7-µm band than those of Bennu's samples. Pilogret et al. (2025) suggest that band depth difference between Bennu and Ryugu could be due in part to differences in the alteration sequence. Differences in the extent of aqueous alteration is also supported by the range of 2.7-µm band shape, depths, and centers features exhibited by the PEC asteroids (**Table 4; Fig. 9**). Additionally, based on the Euclidean distance metric (**Fig 7**), the spectra of PEC asteroids are consistent with a range of – not a specific class of – CC types that underwent high degrees of aqueous alteration (*i.e.,* CMs and CIs), so spectral variation resulting from differences in the alteration sequence is perhaps expected throughout the PEC members, like Bennu and Ryugu.

An extension of the discussion from **Section 5.1** to include Bennu and Ryugu may allow for further evidence for the kinds of phyllosilicates present on PEC asteroid surfaces/subsurfaces. The range of $B_C$ values for the 2.7-µm bands measured from Bennu-Remote and Ryugu-Remote spans from 2.72– 2.74 µm (Kitazato et al. 2019; Hamilton et al. 2019), which is a range that is consistent with that of the smaller PFAs and that is potentially linked to surfaces with greater amounts of lizardite than chrysotile. The $B_C$ range between Bennu-Remote and Ryugu-Remote is wider and over longer wavelengths than the $B_C$ range between their returned samples (~2.72 µm; Pilogret et al. 2022; Lauretta et al. 2024), which likely probe the less-weathered subsurfaces (e.g.,

Le Pivert-Jolivet et al. 2023). Furthermore, neither Bennu-Return nor Ryugu-Return have been shown to exhibit clear signs of chrysotile, with lizardite being either detected or inferred for both samples (Roskosz et al. 2024 Lauretta et al. 2024; Ishimaru et al. 2024). Thus, $B_C$ differences in the JWST observations of big and small PEC members (see **Section 5.1**), lab-based SW experiments of Mg-rich phyllosilicates and CCs (Lantz et al. 2017; Rubino et al. 2020), and detailed analyses of both NEAs' returned-samples may be converging to the conclusion that the serpentinization of Mg-rich phyllosilicates may be occurring to different extents throughout a range of sizes within PEC, as indicated by the response to SW. That is, smaller and larger members may show more evidence of higher amounts of lizardite and chrysotile, respectively (see **Section 5.1** for more details). We note that the PFAs in this work are considerably larger than both Bennu and Ryugu, underscoring the importance of acquiring more observations of smaller members from both PEC families.

Minor absorptions in the spectrum of (495) Eulalia, (142) Polana, and (tentatively) two PFAs serve as additional spectral links between the PEC and Bennu/Ryugu, particularly in the era of returned-sample analyses. Bennu-Remote (**Fig. 9**; at 3.4 µm; Hamilton et al. 2019; Kaplan et al. 2020), Bennu-Sample (at 3.1, 3.4, and 3.9 µm; Hamilton et al. 2026; Lauretta et al. 2024), and Ryugu-Sample spectra (at 3.1, 3.4, and 3.9 µm; Pilogret et al. 2022) each exhibit features from 3.0 – 4.0 µm. Similar to the features exhibited by (142) Polana and (495) Eulalia (and potentially the smaller PFAs), the features in the Bennu and Ryugu spectra have largely been linked to NH-bearing minerals (for the 3.1-µm band; Hamilton et al. 2026). In particular, Hamilton et al. 2026 found that some of Bennu's samples exhibit a feature centered around 3.06 µm, which they tentatively associate with ammonium-bearing phosphates. Comparative analysis of the Bennu and Ryugu return samples has shown that both exhibit similar amounts of NH-bearing phyllosilicates (Jiang et al. 2026). At longer wavelengths in Bennu-Sample/-Remote and Ryugu-Sample, a 3.4-µm band has been associated with C-H stretches in organics, and both the 3.4-µm and/or 3.9-µm bands have been linked with surface carbonates (Hamilton et al. 2026; Angrisani et al. 2026; Pilogret et al. 2022; Lauretta et al. 2024 Mahlke et al. 2026; see **Table 4**). Of particular note, the 3.4-µm features present among the returned samples for Bennu and Ryugu are specifically consistent with dolomite ($CaMg(CO_3)_2$) and breunnerite ($(Mg,Fe)CO_3$) (Mahlke et al. 2026), which in part highlights the disconnect with the potential detection of calcite from Bennu-Remote spectrum. (**Table 4**; Kaplan et al. 2020). Although the relative depths of the 3.4-µm bands are similar between (142) Polana and (495) Eulalia, the exclusively-carbonate (seemingly calcite-like), 3.9-µm band is deeper for (495) Eulalia, which may underscore differences that permeate between the families. In total, the strong similarities between the Bennu and Ryugu return samples' spectra may indicate that observed spectral differences (from 3.0 to 4.0 µm) between Polana and Eulalia family asteroids could help identify a specific family of origin for both NEAs. Future, dedicated work should be dedicated to comparing all PEC, Bennu, and Ryugu spectra more comprehensively.

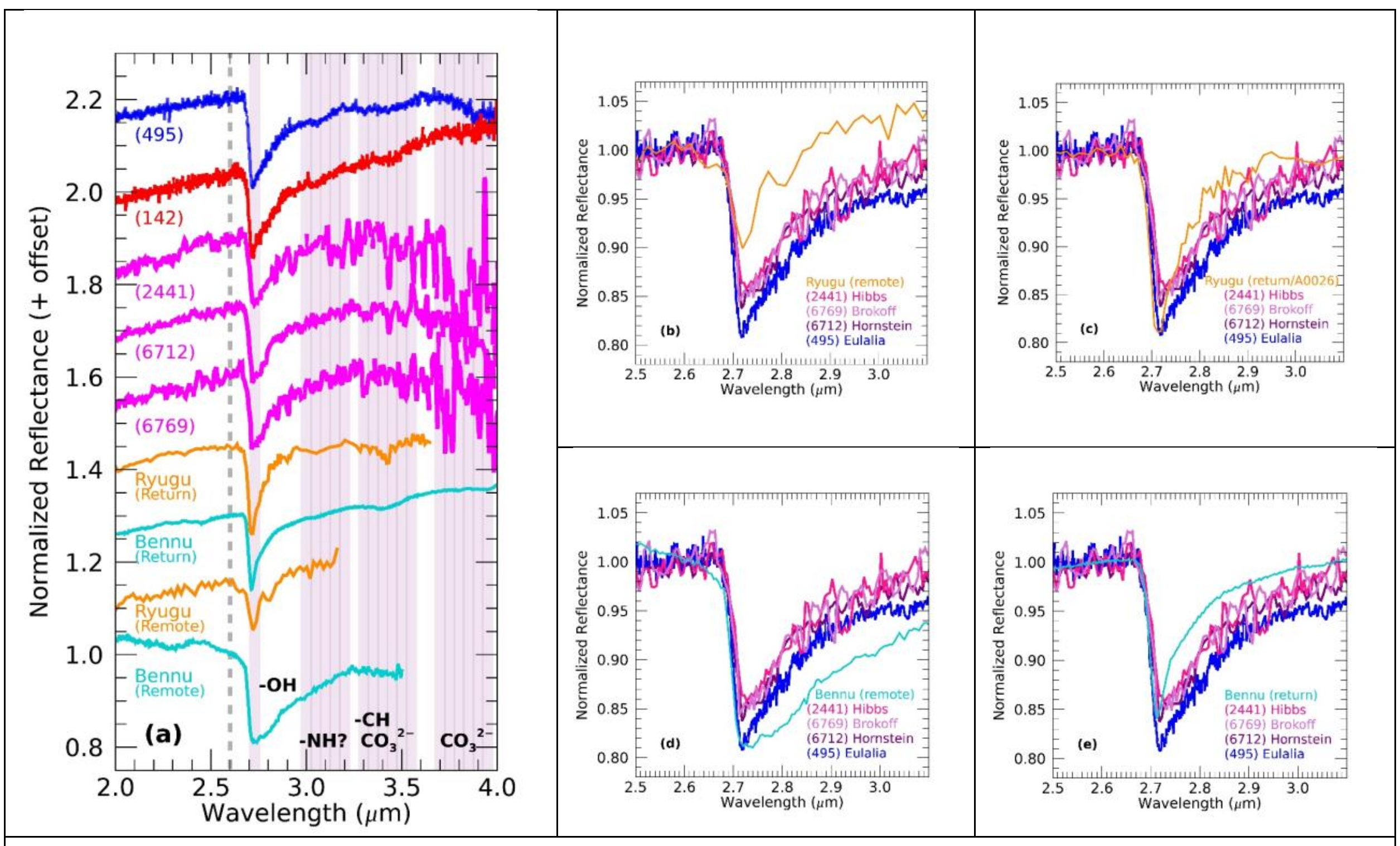

**Figure 9: (a)** Offset comparison between the JWST spectral repository of PEC asteroids ((495) Eulalia = blue; (142) Polana = red (Arredondo et al. 2025); Polana family objects = magenta), Bennu (turquoise; Hamilton et al. 2019; Lauretta et al. 2024), and Ryugu (orange; Kitazato et al. 2019; Pilogret et al. 2022; *see also:* Yada et al. 2022; Nakamura et al. 2023). Both the remote, spacecraft spectra and the returned-sample spectra are presented for Bennu and Ryugu. All spectra are normalized to 1 at 2.6 µm (by the vertical, dashed line (gray), with offset increments of 0.15 in reflectance for clarity. Each spectrum shows a feature centered around 2.72 µm, as referenced by the light purple regions. (142) Polana, (495) Eulalia, the retuned-samples from both NEAs, and the remote spectrum of Bennu also exhibit minor absorptions longward of 2.72 µm, also referenced by the light purple regions. **(b)** Overlapping comparison between SAMBA-3 PEC targets and Ryugu's NIRS3 spectrum (Kitazato et al. 2019), **(c)** Ryugu's A0026 sample spectrum (Pilogret et al. 2022; *see also* Yada et al. 2022; Nakamura et al. 2023), **(d)** Bennu's globally-averaged spectrum (Hamilton et al. 2019), and **(e)** Bennu's averaged sample spectrum (Lauretta et al. 2024). Normalization and color-coding remain the same throughout **(a)**-**(e)**. (For interpretation of the references to color in this figure legend, the reader is referred to the web version of this article).

## 5.4 The PEC's Outer Solar System Origins

The JWST spectra in this work provides key insights into the Polana and Eulalia Family parent bodies. As mentioned in **Section 5.1**, the planetesimals, from which the Polana and Eulalia

Families originated, presumably formed in the outer Solar System (e.g., Walsh et al. 2013). This formation is supported by their connection to aqueously-altered CC-type and in turn their compositions (*e.g.,* Rivkin et al. 2022; Takir et al. 2023). Tracers of an Outer Solar System origin can be reinforced by comparisons between PEC and other bodies within and beyond the Main Belt. The PEC's expression of the 2.7-μm band indicates a connection to the outer Solar System. Additional support is provided by the minor absorptions from 3.0 to 4.0 μm, present in (142) Polana and (495) Eulalia's spectra.

- The 2.7-μm Feature: All five PEC asteroid spectra obtained by NIRSpec exhibit 2.7-μm absorptions that are associated with Mg-rich phyllosilicates on their surfaces and characterized by a "sharp" minimum. (1) Ceres, (2) Pallas, and (10) Hygiea are primitive carbonaceous asteroids farther in the Main Belt (middle Main Belt for Ceres and Pallas; outer main belt for Hygiea. The JWST spectra of these bodies exhibit "sharp" minima in their 2.7 μm bands (Rivkin et al. 2025). The relatively high sharpness of (6712) Hornstein spectrum's 2.7 μm band is especially similar to that which is exhibited by (2) Pallas (**Fig. 9;** Rivkin et al. 2025), whereas the rest of the PEC resembles the less-sharp 2.7-μm bands of (1) Ceres and (10) Hygiea (**Fig. 10**).

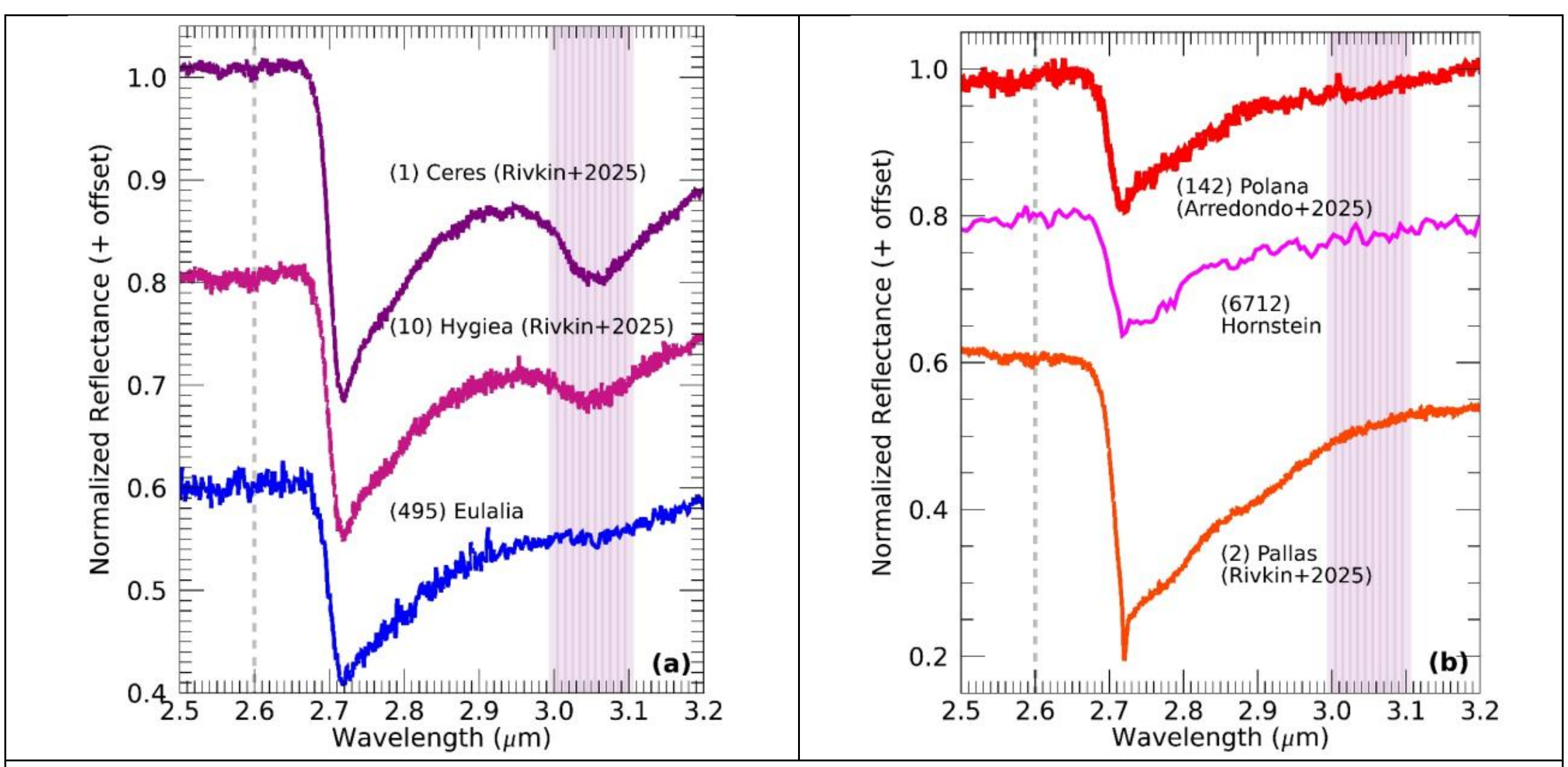

**Figure 10:** Spectral comparison of the 2.6 – 3.2 μm region among main belt asteroids **(a) (**1) Ceres (purple), (10) Hygiea (medium dark violet), and 495 Eulalia (blue), which show "sharp"-type 2.7-μm bands and various depths in their absorptions around 3.0-3.2 μm. **(b)** (142) Polana (red), (6712) Hornstein (magenta), and (2) Pallas (orange red) also show sharp-type 2.7-μm bands, but lack appreciable absorptions in the 3.0—3.2 μm range (Rivkin et al. 2025; Arredondo et al. 2025). Given their overall similarities in the 3.0 – 3.2 μm range, we do not show the other 2 PFAs in this work for the sake of space. Spectra below (1) Ceres and (142) Polana are offset in reflectance increments of 0.2, and a vertical, dashed, gray line highights the normalization wavelength of 2.6 μm (For interpretation of the references to color in this figure legend, the reader is referred to the web version of this article).

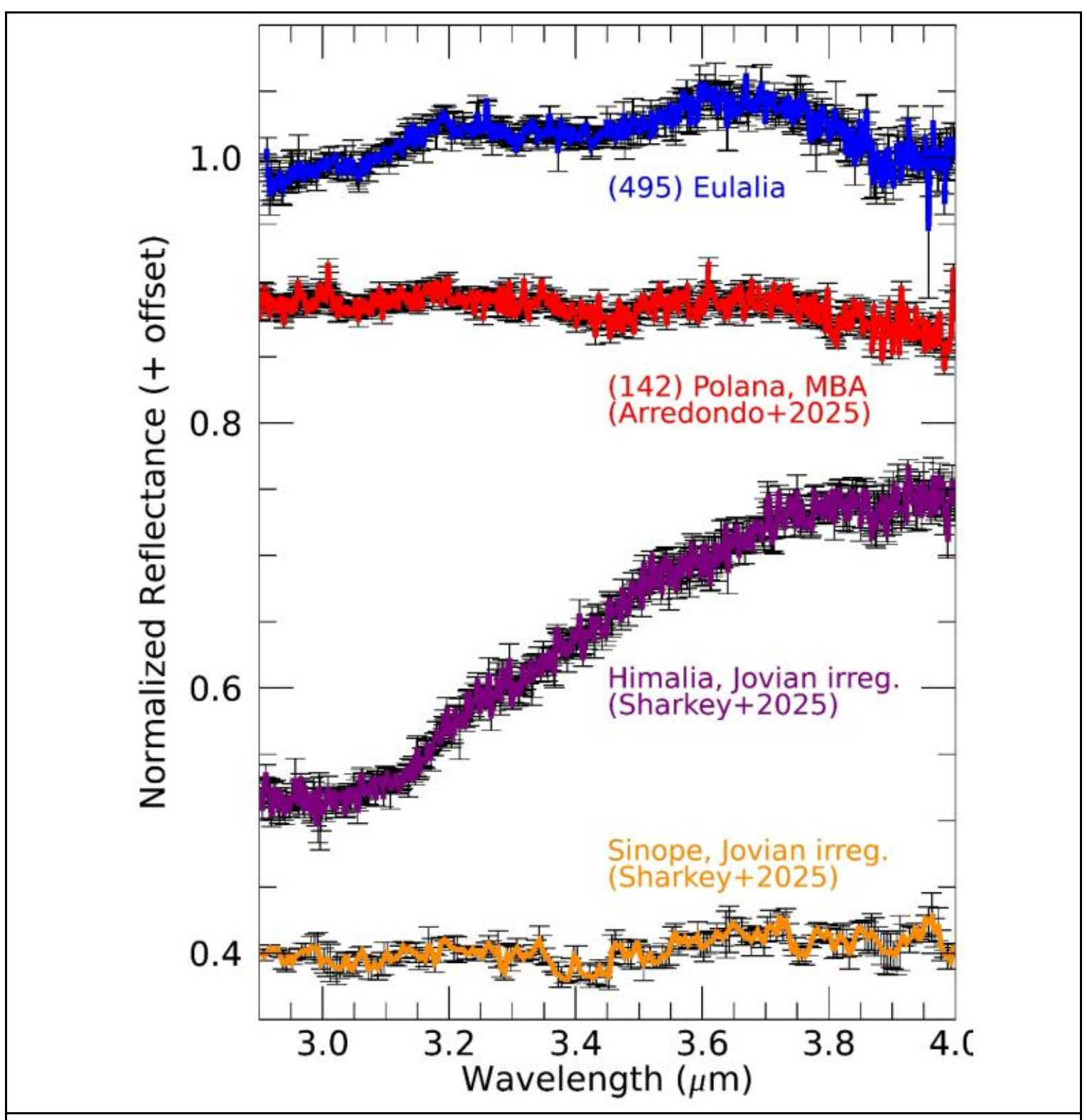

**Figure 11**: The spectrum of (495) Eulalia (in blue) exhibits minor absorptions from 3.0 – 4.0 µm that are comparable to small bodies within and beyond the Main Belt. Examples include (142) Polana (red; Arredondo et al. 2025), Jovian irregular satellite Himalia (purple; Sharkey et al. 2025), and organic-rich Jovian irregular satellite Sinope (dark orange; Pinilla-Alonso et al. 2025). Reflectance values are normalized to 4.0 µm and offset by increments of 0.12 µm for (142) Polana (-0.12), Himalia (-0.48), and Sinope (-0.60). (For interpretation of the references to color in this figure legend, the reader is referred to the web version of this article).

- The 3.1-µm Feature: This feature has been detected and often attributed to ammoniated minerals/unaltered silicates on a variety of small Solar System bodies that reside at similar or farther distances from the Sun than asteroids in the PEC. Although (495) Eulalia exhibits a band in this region, (142) Polana exhibits only a potential 3.1 µm feature (see **Table 4, Fig 10, Appendix**), with it being weak enough to have not been explored in the initial analysis of its spectrum (Arredondo et al. 2025). The 3.1 µm band center in the (495) Eulalia spectrum ($B_C$ = 3.086 µm ± 0.004 µm) is most comparable to the presence of ammoniated illite, which exhibits feature around 3.06 µm (Ferrari et al. 2019). When ammoniated and varied in temperature/pressure, meteorite Cold Bokkeveld, which shows a relatively small Euclidean distance with each PEC asteroid spectra in this work (see **Figure 7**), exhibits a band center at longer-wavelengths ($B_C$ = 3.09; Ehlmann et al. 2018) that is in turn comparable to that which is measured for (495) Eulalia (**Table 4**). Although not present on (2) Pallas, (1) Ceres and (10) Hygiea exhibit both a feature in this region, attributed to NH-bearing molecules on their surfaces (King et al. 1992; De Sanctis et al. 2013; Kurokawa et al. 2020; Rivkin et al. 2025). The JWST spectra of (1) Ceres and (10) Hygiea also exhibit features around 3.3-µm, which are attributed to ammoniated salts. Beyond the asteroid belt, Jupiter's irregular satellites (Sharkey et al. 2025) have shown features in this region. In particular, the irregular Jovian moon Himalia exhibits a deeper, wider expression of the band compared with (495) Eulalia (**Fig. 11**). We note that Themisto's band center is most aligned with that of (495) Eulalia (Sharkey et al. 2025). Although NH-bearing molecules have not been directly detected on medium-sized (Pinilla-Alonso et al. 2025; McClure et al. 2026) or dwarf planet (Emery et al. 2024) trans-Neptunian Objects (TNOs) in this wavelength range, the presence of NH-bearing molecules is predicted to be on TNO surfaces (e.g., Barucci et al. 2008; Dalle Ore et al. 2009), with studies (*e.g.,* Cryan et al. 2025) often being dedicated to longer-wavelength bands since an $H_2O$ overtone makes assessments of this wavelength range challenging. Future JWST-based spectral modeling may help to discern NH-based connections between primitive asteroids in the IMB and TNOs.

- The 3.4-µm Feature: The presence of the 3.4-µm band in the (142) Polana and (495) Eulalia spectra not only indicates that the PEC contains organics and carbonates, but it also highlights key processes in the PEC parent bodies (*see* **Sections 5.1** & **5.2**). The center and depth of (495) Eulalia's 3.4-µm band in particular is shorter-wavelength and shallower than that of (142) Polana's (see **Table 4**; **Fig. 11**; Arredondo et al. 2025). The slight band parameter difference in this region may in part highlight an area of compositional difference between these two similar families, particularly for carbonates abundance (see **Section 5.2**).

Small bodies near and beyond the PEC also exhibit this band. (84) Klio also has a possible 3.4-µm feature (Le Pivert-Jolivet et al. 2025a) consistent with a combination of organics and carbonates. Though they exhibit a feature in this range, (1) Ceres and (10) Hygeia exhibit

3.4-µm bands that are centered at short-enough wavelengths to be considered associated with ammoniated minerals (Kurokawa et al. 2020; Rivkin et al. 2025). Jupiter Trojans Orus, Leucus, Patroclus, and Eurybates show features around 3.4-µm associated with organics and with band depths similar to (142) Polana and (495) Eulalia (Table 4; Brown et al. 2025; Wong et al. 2025). Further, the 3.4-µm band centers of Leucus ($B_C$ = 3.440 µm ± 0.015 µm; Wong et al. 2025) and Patroclus ($B_C$ = 3.409 µm ± 0.015 µm; Wong et al. 2025) are closest to those of (142) Polana ($B_C$ = 3.445 µm ± 0.004 µm) and (495) Eulalia ($B_C$ = 3.41 µm ± 0.01 µm), respectively. Jovian irregular moons also display a range of band shapes in this region, with each being associated with the presence of surface organics. In particular, Sinope's 3.4-µm band associated with aliphatic organics is the closest match to that of (495) Eulalia's in terms of center and depth (**Fig. 11**; Sharkey et al. 2025). Comet 67P Churyumov–Gerasimenko (Raponi et al. 2020), centaurs (specifically, 2002 KY14; see Licandro et al. 2024), small-sized (organic-rich) TNOs (such as 1999 $OX_9$ in Pinilla-Alonso et al. 2025; McClure et al. 2026), and medium-sized TNOs (Emery et al. 2024) also exhibit a range of generally deeper bands in this region that have been broadly linked to simple to complex organic molecules.

- The 3.9-µm Feature: The namesake asteroids for the PEC families both exhibit a feature around 3.9 µm that are attributed to the surface presence of carbonates. As mentioned in **Table 4** and **Section 5.2**, (495) Eulalia exhibits a deeper band than (142) Polana, reinforcing the notion that carbonate abundance may be the main compositional difference between these two very similar asteroid populations. Surface carbonates have been inferred through this wavelength region for (1) Ceres (Kurokawa et al. 2020), (10) Hygiea (Rivkin et al. 2025), (84) Klio (Le Pivert-Jolivet et al. 2025a) and a few of Jupiter's irregular moons (*e.g.,* Himalia in **Fig. 11**; Sharkey et al. 2025). Carbonates have also been associated with a slightly longer-wavelength (*i.e.,* ~4.02 µm) feature for the Uranian Moon Ariel (Cartwright et al. 2025).

Observations of other small bodies, especially in the era of JWST, allow for some constraints to the formation region(s) of the PEC parent bodies. Altogether, the major and minor absorptions present in the JWST-based repository of PEC asteroid spectra indicate that the PEC likely formed from materials present in the early, outer Solar System. The PEC spectra's of the 2.7-µm band centers/shapes are similar to other primitive Main Belt Asteroids (*e.g.,* (1) Ceres, (4) Pallas, and (10) Hygiea; Rivkin et al. 2025) that are linked to outer Solar System formation and subsequent implantation into the Main Belt through interactions with migrating Gas Giants (Walsh et al. 2013). In particular, (1) Ceres and (10) Hygiea exhibit additional features around 3.3-µm, which are associated with NH-bearing minerals (Rivkin et al. 2025). (142) Polana and (495) Eulalia exhibit shallow, tentative features around 3.1 µm, which this work associates with NH-bearing minerals. However, (142) Polana and (495) Eulalia do not show longer-wavelength absorptions like the 3.3-µm feature and instead exhibit features consistent with surfaces containing organics and, especially, carbonates. Thus, the PEC parent bodies, particularly the Eulalia family parent body, may have formed slightly interior to the ammonia snow line (*see* Rivkin et al. 2022), but they perhaps formed farther than the parent body of (2) Pallas, which exhibits no spectral indication of NH-bearing minerals (**Fig. 10**). Additionally, this formation scenario would in turn allow for the acquisition of precursor materials for organics and carbonates, reinforcing aforementioned interpretations and connecting the PEC to bodies currently within and beyond the Main Belt.

## 6. Conclusions

This work provided new JWST observations of 4 members of the Polana-Eulalia Complex, including the namesake asteroid of the Eulalia Family and three members of the Polana Family (Walsh et al. 2013). These observations were part of a lager campaign, called SAMBA-3, that seeks to observe primitive, inner main belt asteroids with NIRSpec. This work's sample also complemented an observation of (142) Polana from Cycle 2 (Arredondo et al. 2025), an observation from the broader SAMBA-3 campaign (i.e., Le Pivert-Jolivet et al. 2025a) and observations from sample-return, spacecraft missions to NEAs, Bennu and Ryugu (Hamilton et al. 2019; Kitazato et al. 2019; Pilogret et al. 2022; Lauretta et al. 2024).

Each PEC target observed by JWST exhibits a feature centered around 2.72-2.74 µm. This feature was determined to be consistent with surfaces containing Mg-rich phyllosilicates, based on the relatively short-wavelength band centers and "sharp" band shapes (see Takir & Emery 2012; Takir et al. 2013; Rivkin et al. 2022). Some minor absorptions are also detected for (495) Eulalia (and more tentatively for 2441 Hibbs and 6712 Hornstein). These absorptions are consistent with the presence of NH-bearing minerals (~3.1 µm), organics (~3.4 µm), and carbonates (~3.4 & 3.9 µm).

Analysis of the features in the JWST spectra, particularly when placed in context with aqueously-altered carbonaceous chondrites, provide insights into the alteration histories of these bodies. The targets of this work broadly support the idea of PEC parent bodies that were highly aqueously-altered, perhaps due to internal heating in the original planetesimals of the PEC. However, we note that such alteration may not have occurred to the fullest extent possible. Additionally, differences between (142) Polana (Arredondo et al. 2025) and the smaller PFAs in terms of the 2.7-µm band depth and center may underscore a size-based spectral variability for the band. Band differences could be attributed to variability within the parent body and/or differences in the response to space weathering. We note that future JWST observations could further elucidate potential size-dependent variability.

The PEC exhibits broad spectral homogeneity, but we note key differences between the namesake asteroids for the families of the PEC. Namely, (142) Polana and (495) Eulalia exhibit similar 2.7-µm bands in terms of shape, center, and depth. However, the relative depths of their minor absorptions from 3.0 to 4.0 µm might underscore some compositional differences between these namesake asteroids. (142) Polana and (495) Eulalia (and by extension their families) potentially show differences in the relative amounts of carbonates and NH-bearing minerals on their surfaces.

With the current JWST repository of PEC asteroids, we are unable to determine whether Bennu and Ryugu came specifically from either the Polana family or Eulalia family. Our results support a broad PEC-based origin for these NEAs based on the PEC's expression of the 2.7-µm feature, the minor absorptions present in the (142) Polana and (495) Eulalia spectra, and the Euclidean distances between this work's targets and the Bennu/Ryugu spectral repository. We note that more JWST spectra, particularly from smaller PEC members of both families, may be required to more confidently link Bennu and Ryugu to a specific PEC family.

Contextualization of our observations with previous JWST spectra of other bodies allow for plausible constraints on the formation region(s) of the PEC's parent bodies. The spectra of (142) Polana and (495) Eulalia indicate their surfaces have relatively low abundances of NH-bearing minerals compared to (1) Ceres and (10) Hygiea, but the large PEC asteroids are perhaps not as bare in NH-bearing molecules as (2) Pallas. That is, the 3.1-µm feature exhibited by both (142) Polana and (495) Eulalia might be indicating that the Eulalia family parent body formed near/beyond the ammonia snowline, whereas the Polana family parent body formed interior to it. Thus, we tentatively conclude that the PEC's parent bodies formed in a region of the outer Solar System intermediate to that of (2) Pallas and (1) Ceres/(10) Hygiea. This formation scenario would in turn be consistent with additional features associated with organics and carbonates, which are potentially present on the surfaces of (142) Polana and (495) Eulalia.

## Acknowledgments

This work is based [in part] on observations made with the NASA/ESA/CSA James Webb Space Telescope. The JWST data presented in this article were obtained from the Mikulski Archive for Space Telescopes (MAST) at the Space Telescope Science Institute. The specific observations analyzed can be accessed via doi: 10.17909/nsyz-fs95. The Space Telescope Science Institute is operated by the Association of Universities for Research in Astronomy, Inc., under NASA contract NAS 5-03127 for JWST. These observations are associated with program #6384. Support for program 6384 was provided by NASA through a grant from the Space Telescope Science Institute. L. M. would like to acknowledge support from the Future Investigators in NASA Earth and Space Science and Technology (FINESST) research grant (solicitation: ROSES-2023 NNH22ZDA001N-FINESST; grant number: 80NSSC23K1587) as well as the Lawson Scholarship through the Achievement Rewards for College Scientists (ARCS) Foundation. N.P.-A. acknowledges funding through the ATRAE program of the Ministry of Science, Innovation, and Universities (MICIU) and the State Agency for Research (AEI) in Spain. N.P.-A. acknowledges support from the Agencia Estatal de Investigacion del Ministerio de Ciencia e Innovacion under grant ``Exploring the Solar System with the James Webb Space Telescope'' (ESSTE-JWST), with reference PID2024-160618NB-C21.

## Appendix

In this section, we present the polynomial fits (**Figure A-B**) and band parameter analysis definitions (**Table A**) for the the 2.7-µm band of the PFAs as well as for the minor absorptions present in the (142) Polana and (495) Eulalia spectrum. Values for band center and depth for each absorption are presented in **Table 4** in the main text. Explanation of the polynomial fitting is provided in **Section 4.1** (for 495 Eulalia) and **Section 4.1** (for PFAs).

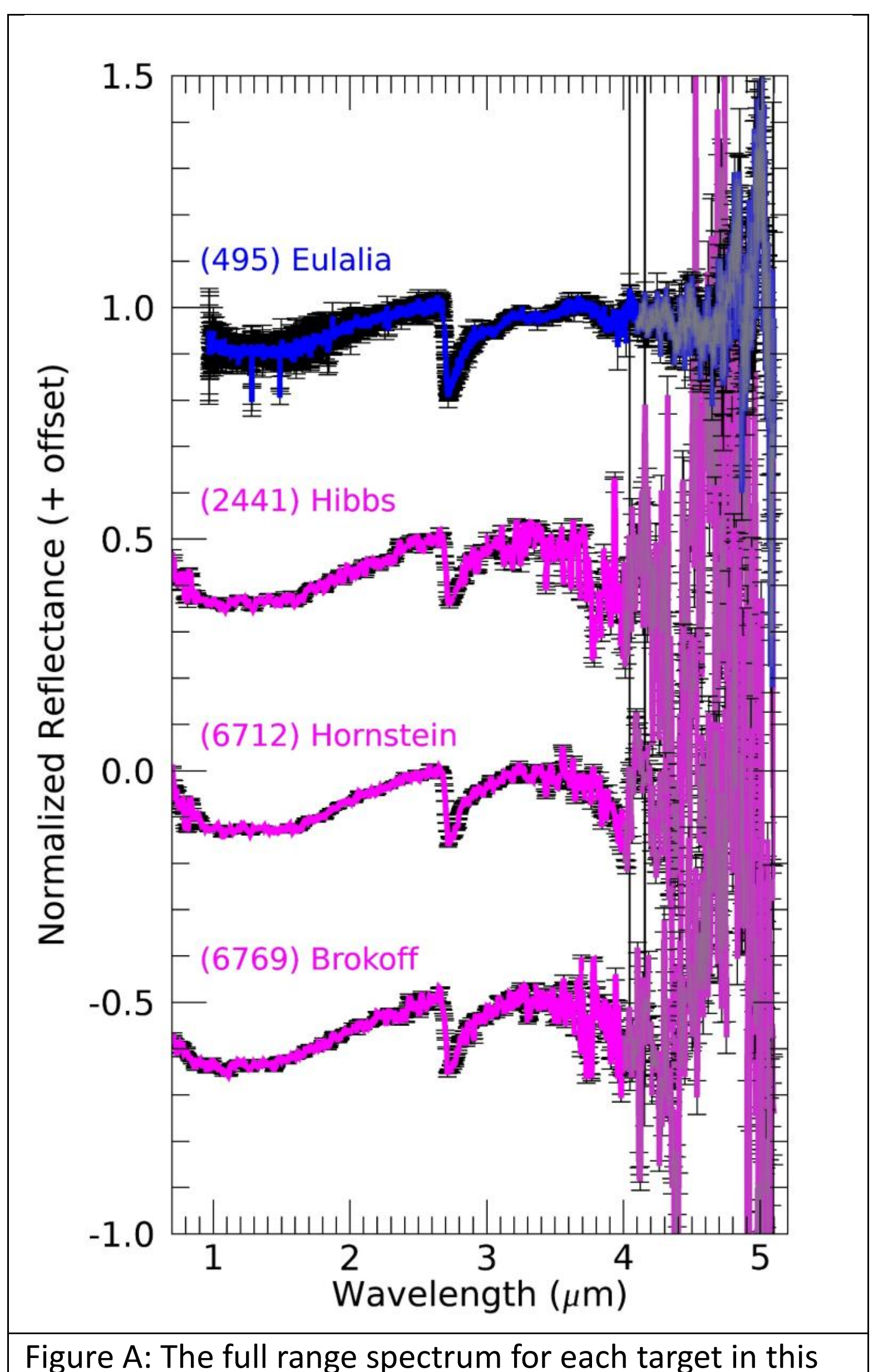

Figure A: The full range spectrum for each target in this work. Longward of 4.1 µm for (495) Eulalia (and longward of 4.0 µm for the PFAs), data were largely unusable due to low SNR in this wavelength range. Eulalia is in blue, whereas the PFAs are shown in magenta. Grayed out regions indicate ranges where data was not used in this work. (For interpretation of the references to color in this figure legend, the reader is referred to the web version of this article).

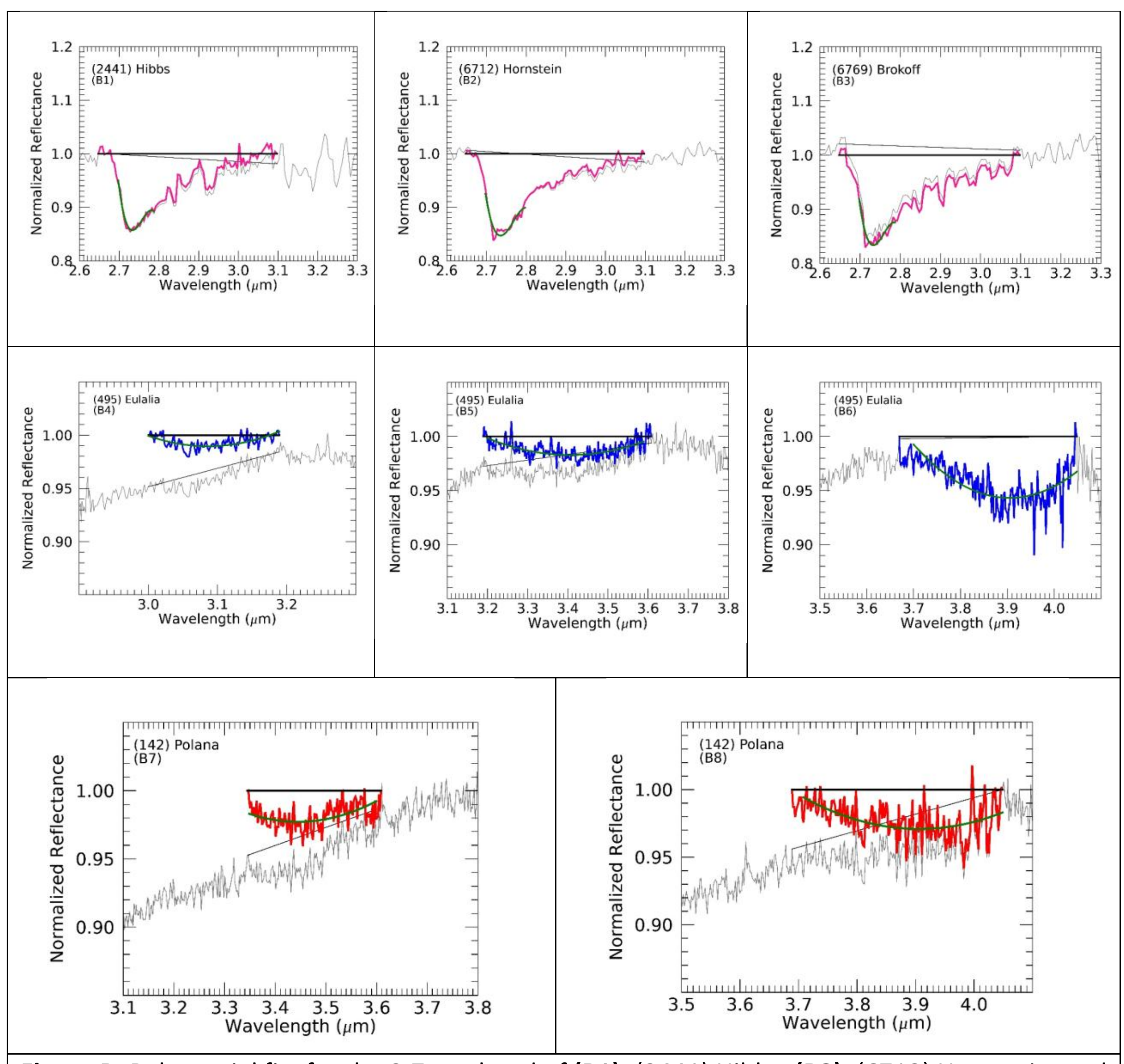

**Figure B:** Polynomial fits for the 2.7-µm band of **(B1)**: (2441) Hibbs, **(B2)**: (6712) Hornstein, and **(B3)**: (6769) Brokoff, with each spectrum normalized to 1 at 2.67 µm. **(B4)**: The (495) Eulalia spectrum was also fit with polynomials to parameterize minor absorptions around 3.1 µm (normalized at 3.65 µm), **(B5)**: 3.4 µm (normalized at 3.69 µm), and **(B6)**: 3.9 µm (normalized at 4.05 µm). **(B7)** The (142) Polana spectrum was similarly fit with a 2$^{nd}$ polynomial for the 3.4-

µm and **(B8)** 3.9-µm absorptions (spectral data from Arredondo et al. 2025). The solid black lines represent the continuum for each band, prior (thinner line) and after (thicker line) division/isolation of the band. Polynomial fits are presented in green, overlaying either pink (for PFAs), blue (for 495 Eulalia's bands), or red (for (142) Polana's bands) spectra. (For interpretation of the references to color in this figure legend, the reader is referred to the web version of this article).

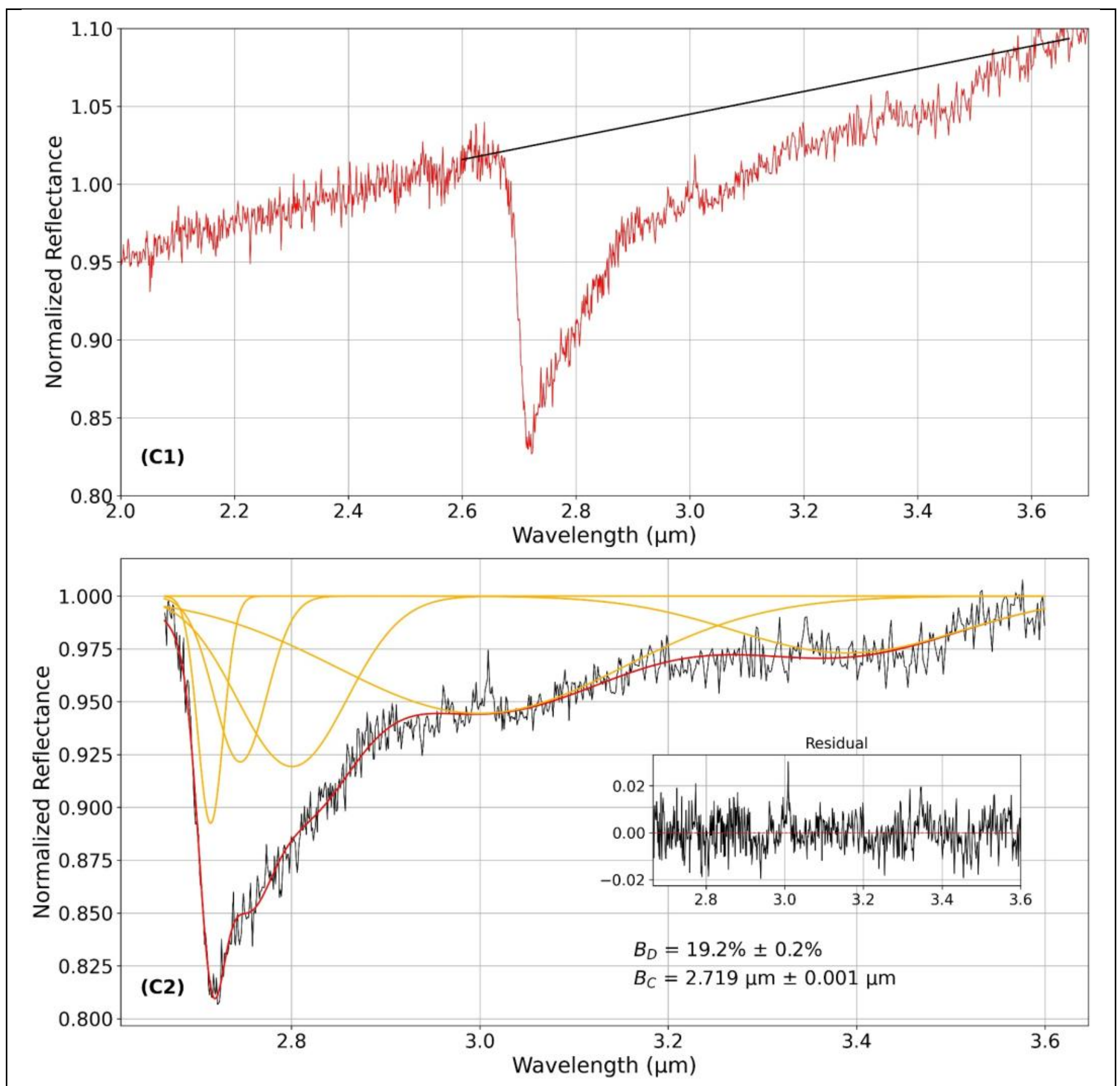

**Figure C: (C1)** Continuum line (black solid line) calculated for the 2.7 µm band for (142) Polana's spectrum (in red). The reflectance is then divided by the continuum line. **(C2)** Gaussian fits (in yellow) used to produce a multi-Gaussian-based model of the band (in red). The residual difference of the model and the data as well as the calculated band parameters are shown as well. (For interpretation of the references to color in this figure legend, the reader is referred to the web version of this article).

**Table A**: Band Parameter Analysis Definitions.

| Object | Continuum Range (µm) | Polynomial (P) or Guassian (G) Fit Range (µm) | Normalizing Wavelength (µm) |
|---|---|---|---|
| 2.7-µm Band | | | |
| (142) Polana | 2.60 – 3.67 | 2.665 – 3.60 (G) | 2.67 |
| (495) Eulalia | 2.60 – 3.67 | 2.665– 3.60 (G) | 2.67 |
| (2441) Hibbs | 2.65 – 3.1 | 2.70 – 2.79 (P) | 2.67 |
| (6712) Hornstein | 2.65 – 3.1 | 2.70 – 2.80 (P) | 2.67 |
| (6769) Brokoff | 2.65 – 3.1 | 2.70 – 2.79 (P) | 2.67 |
| 3.1-µm Band | | | |
| (495) Eulalia | 3.00 – 3.19 | 3.00 – 3.19 (P) | 3.65 |
| 3.4-µm Band | | | |
| (142) Polana | 3.19 – 3.61 | 3.20 – 3.59 (P) | 3.69 |
| (495) Eulalia | 3.347 – 3.61 | 3.35 – 3.60 (P) | 3.69 |
| 3.9-µm Band | | | |
| (142) Polana | 3.67 – 4.05 | 3.70 – 4.05 (P) | 4.05 |
| (495) Eulalia | 3.69 – 4.05 | 3.71 – 4.05 (P) | 4.05 |

Figures 1-11

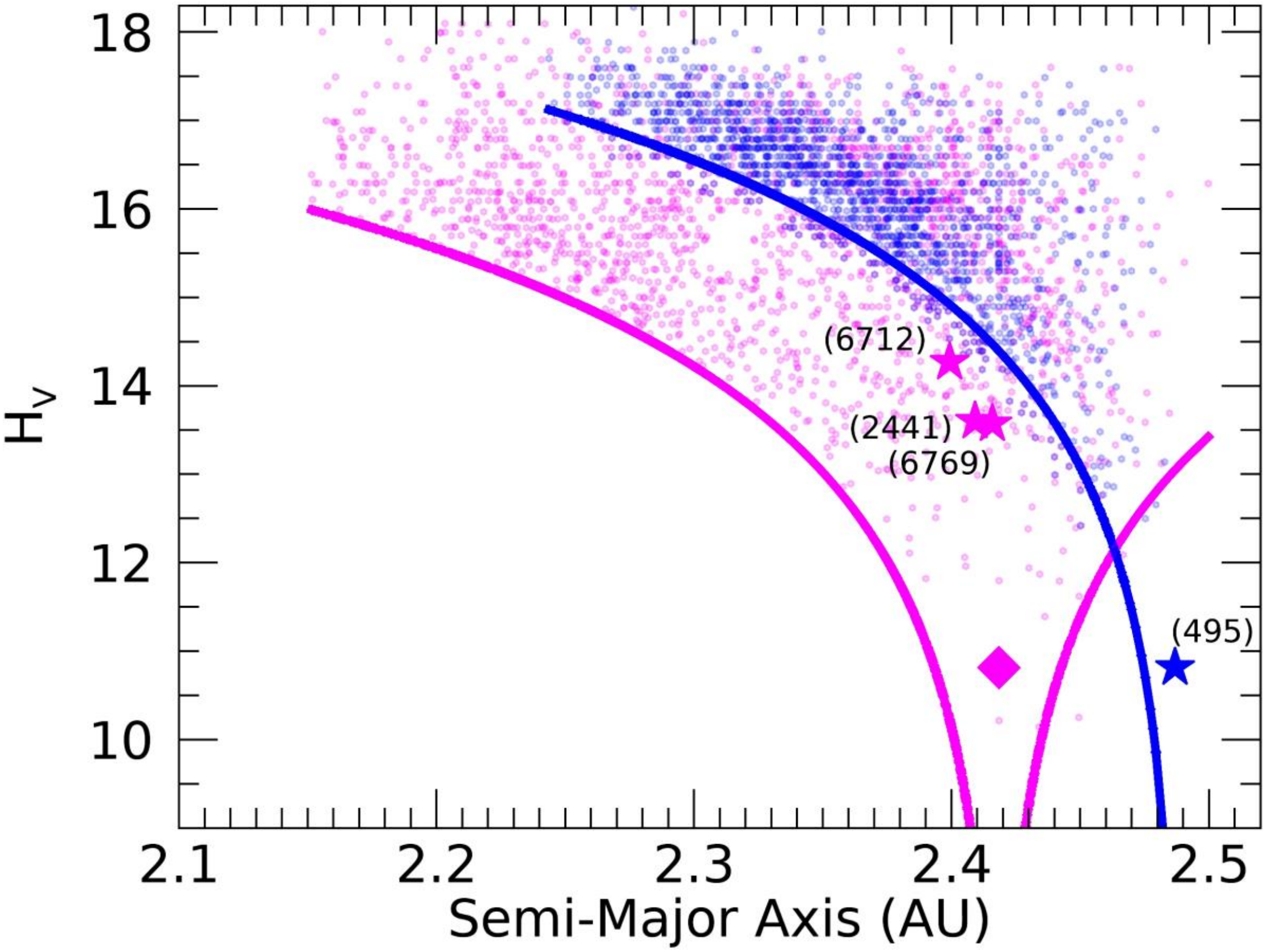
18
16
14
12
10
$H_V$
2.1
2.2
2.3
2.4
2.5
Semi-Major Axis (AU)
(6712)
(2441)
(6769)
(495)

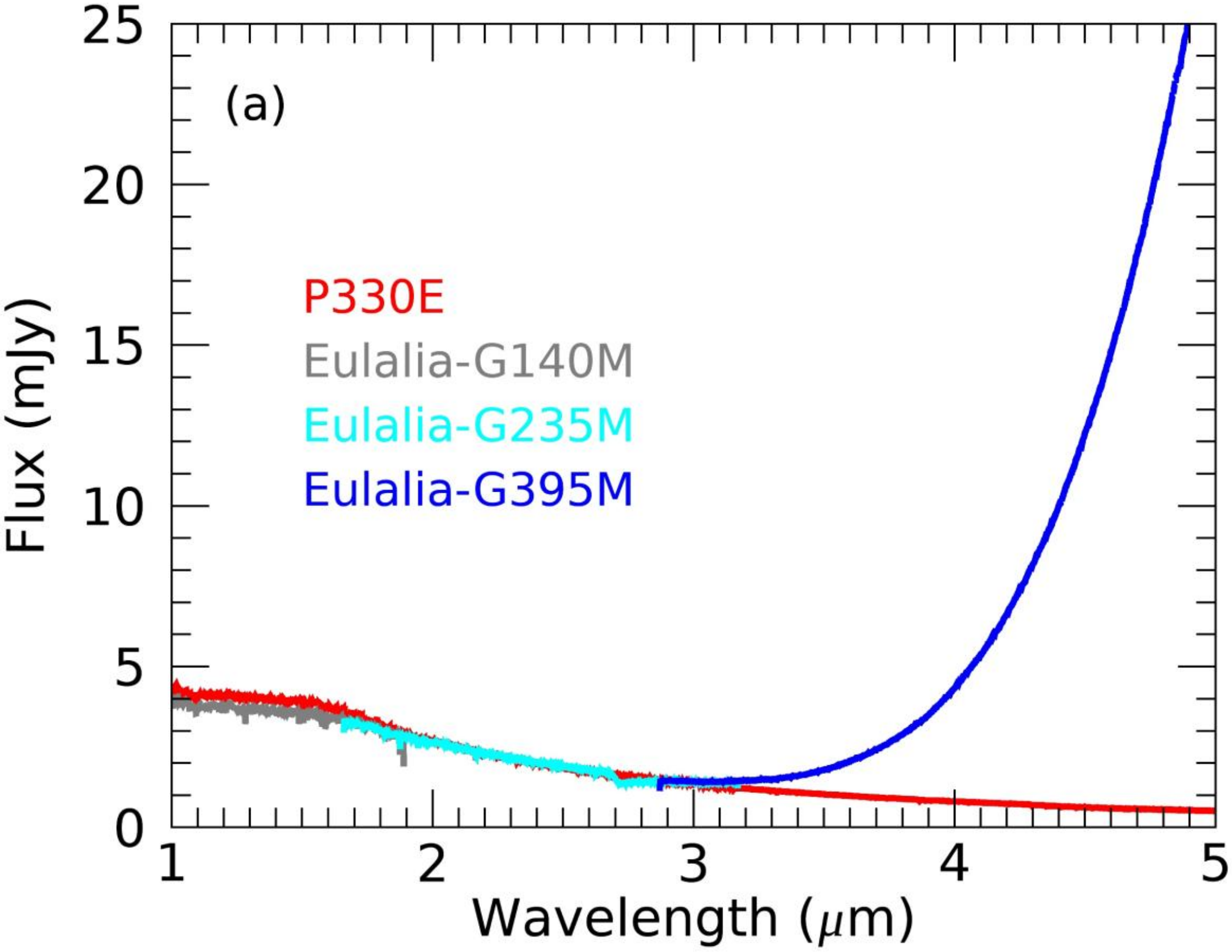
(a)
P330E
Eulalia-G140M
Eulalia-G235M
Eulalia-G395M
Flux (mJy)
Wavelength (μm)
0
5
10
15
20
25
1
2
3
4
5

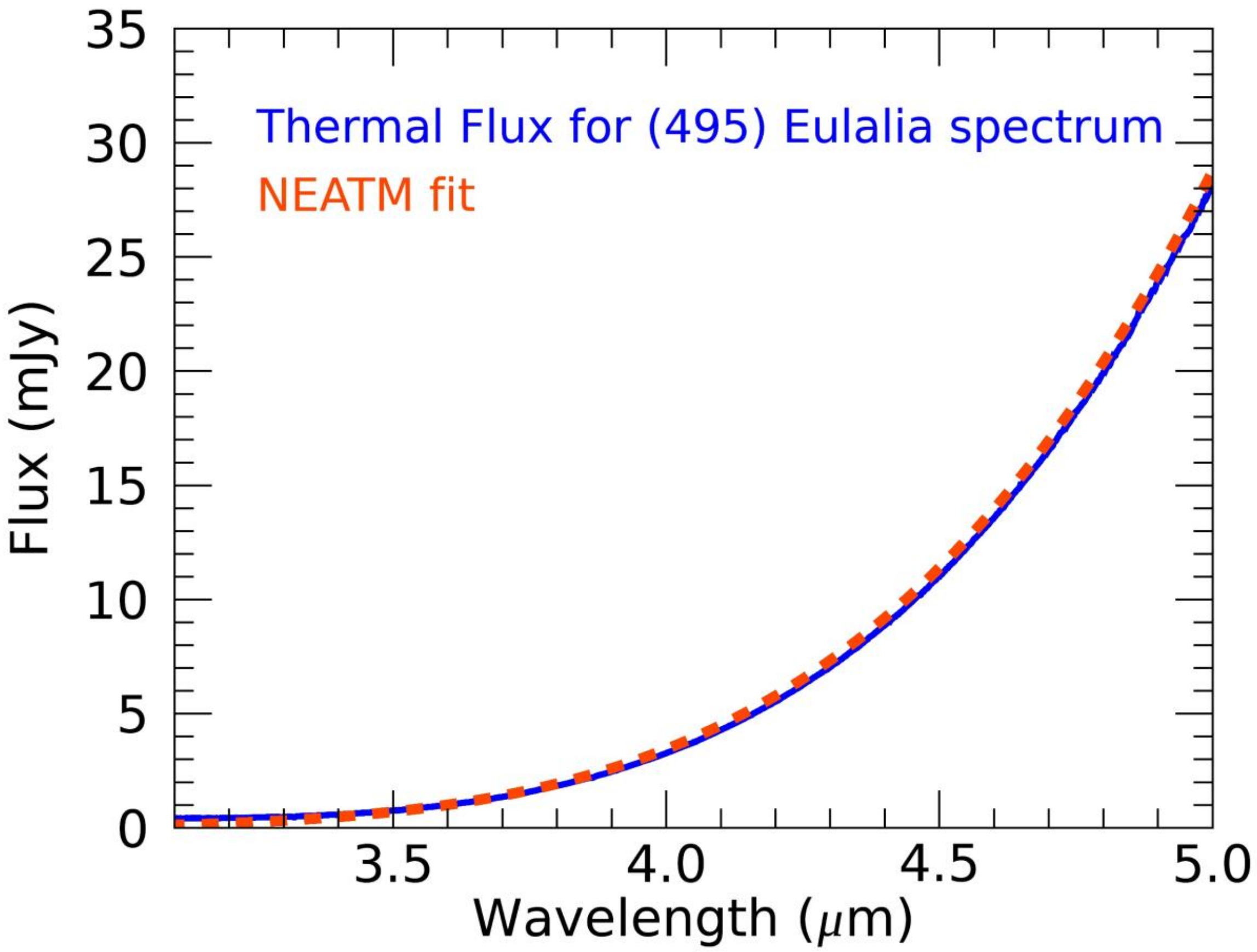
Thermal Flux for (495) Eulalia spectrum
NEATM fit
Flux (mJy)
Wavelength (μm)
0
5
10
15
20
25
30
35
3.5
4.0
4.5
5.0

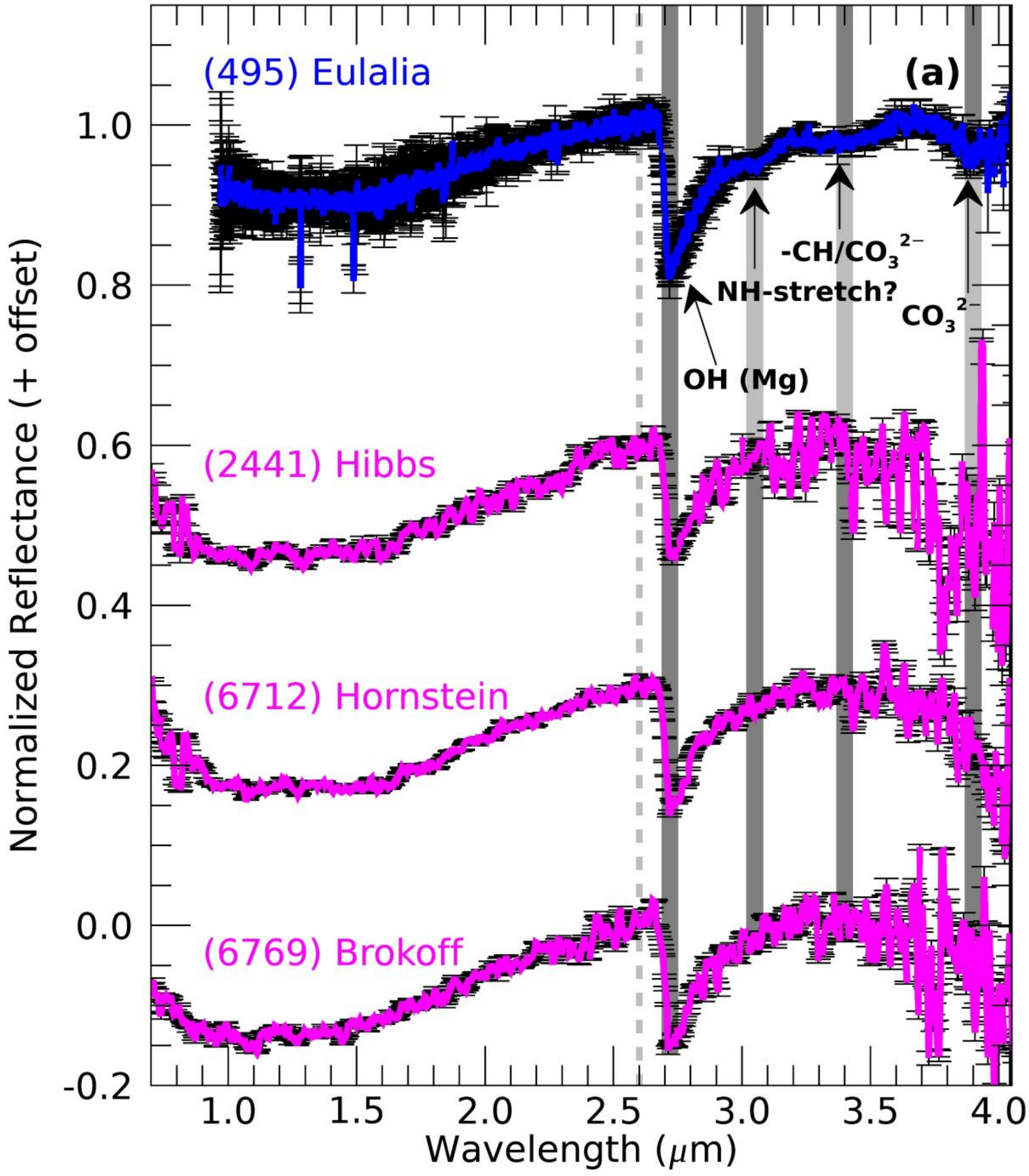
(495) Eulalia
(a)
-CH/$CO_3^{2-}$
NH-stretch?
$CO_3^{2-}$
OH (Mg)
(2441) Hibbs
(6712) Hornstein
(6769) Brokoff
Normalized Reflectance (+ offset)
Wavelength (μm)
1.0
0.8
0.6
0.4
0.2
0.0
-0.2
1.0
1.5
2.0
2.5
3.0
3.5
4.0

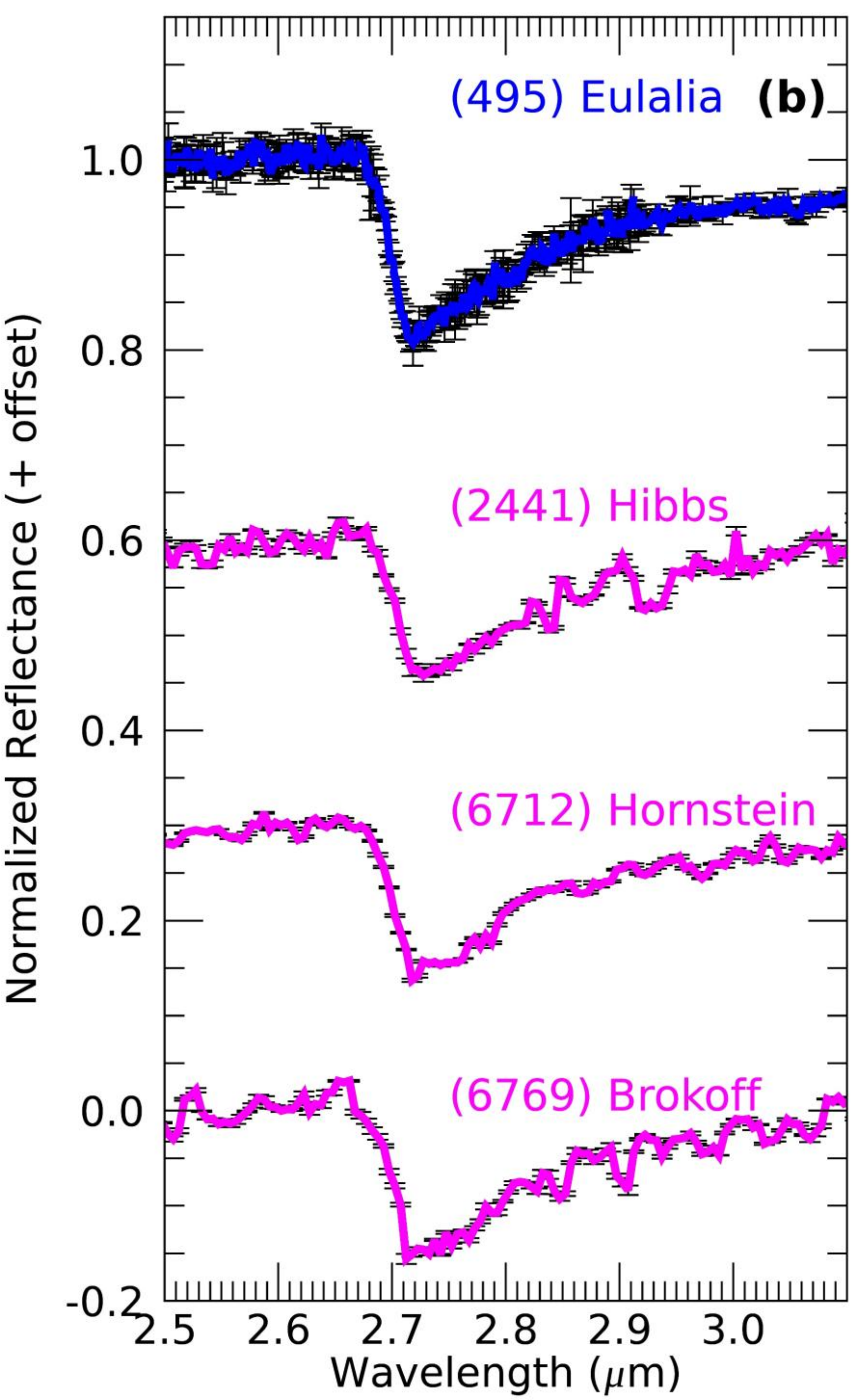
(495) Eulalia
(b)
(2441) Hibbs
(6712) Hornstein
(6769) Brokoff
Normalized Reflectance (+ offset)
1.0
0.8
0.6
0.4
0.2
0.0
-0.2
2.5
2.6
2.7
2.8
2.9
3.0
Wavelength (μm)

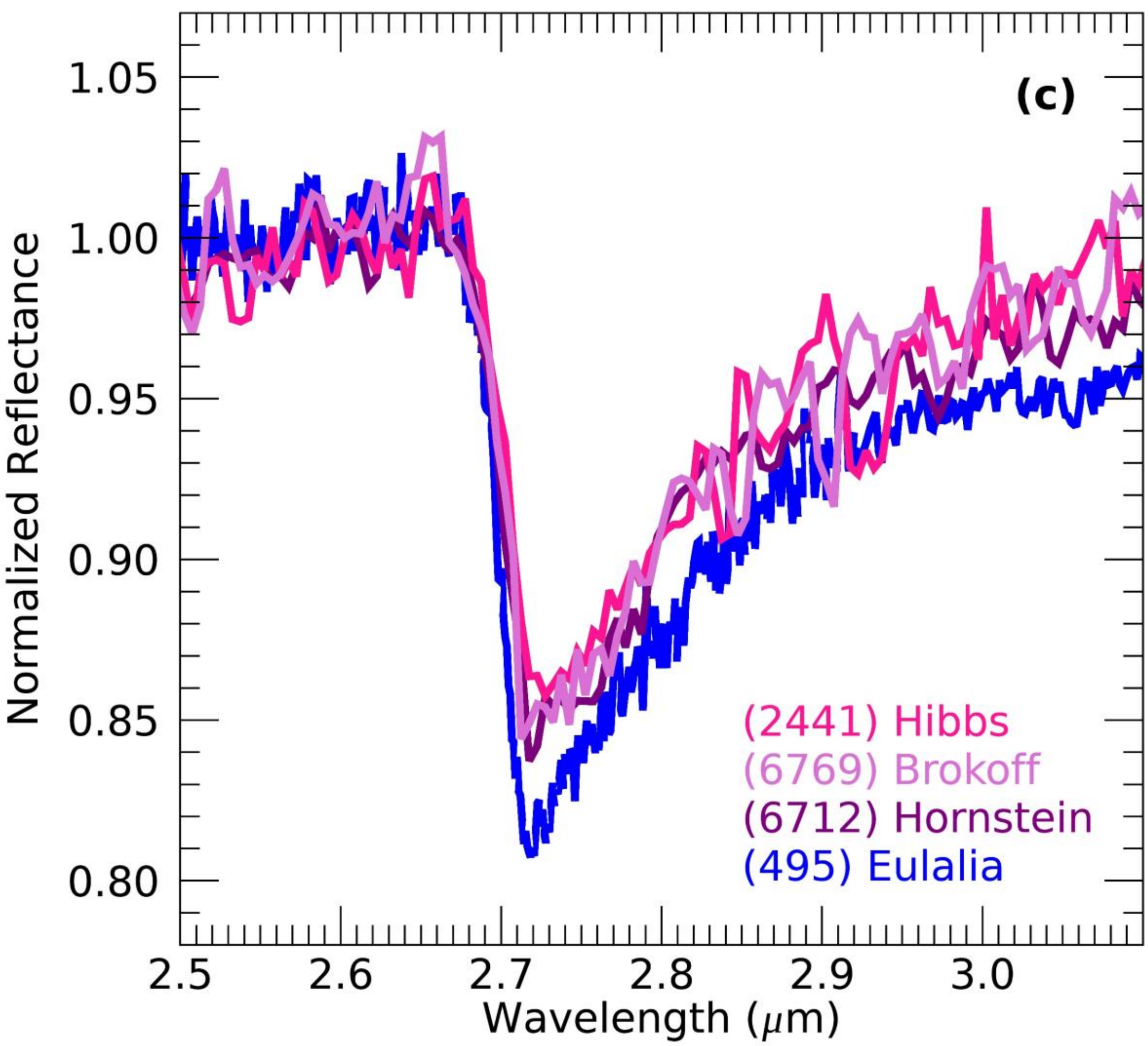

(c)
Normalized Reflectance
1.05
1.00
0.95
0.90
0.85
0.80
2.5
2.6
2.7
2.8
2.9
3.0
Wavelength ($\mu$m)
(2441) Hibbs
(6769) Brokoff
(6712) Hornstein
(495) Eulalia

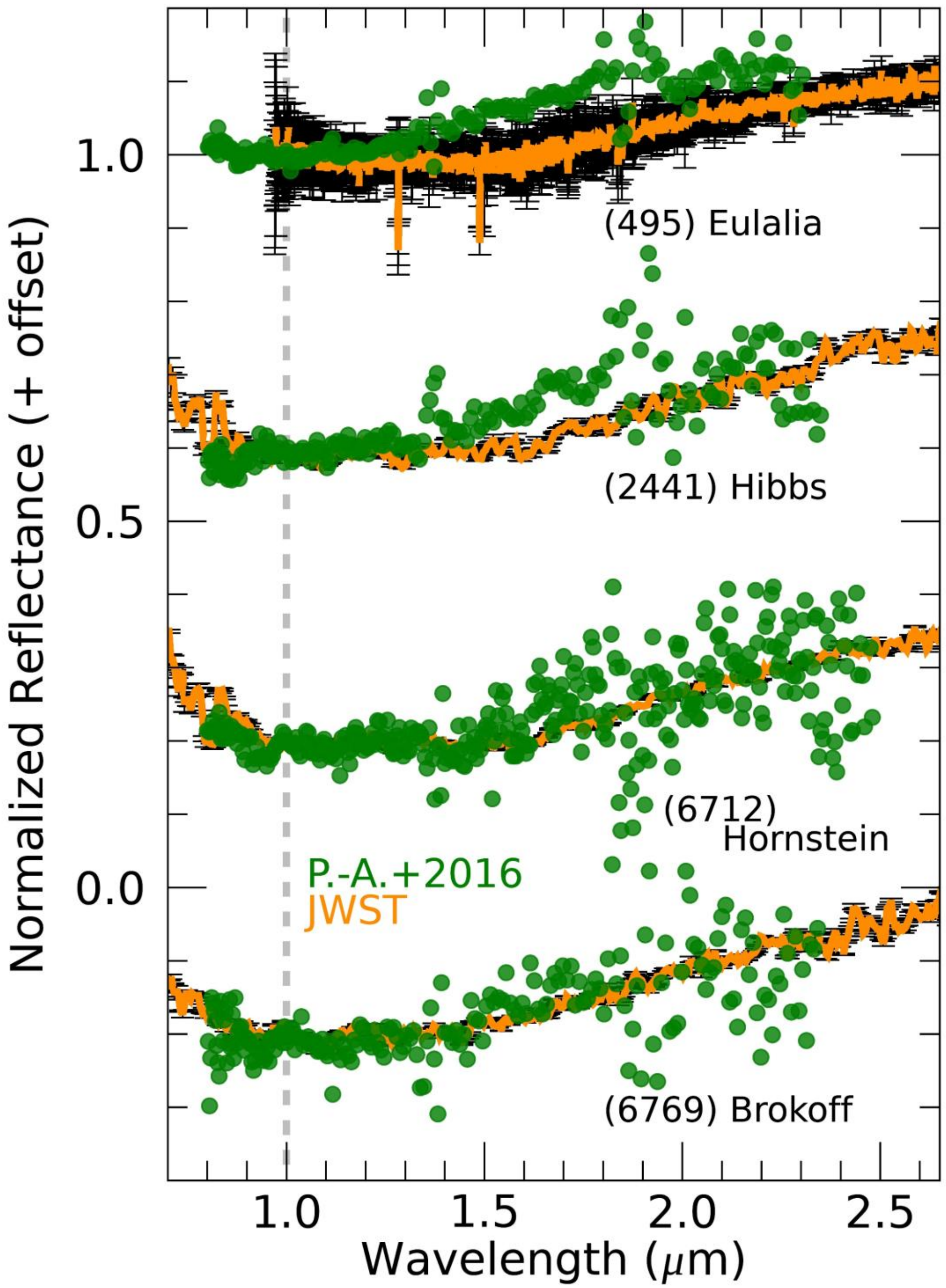
Normalized Reflectance (+ offset)
Wavelength (μm)
(495) Eulalia
(2441) Hibbs
(6712) Hornstein
(6769) Brokoff
P.-A.+2016
JWST
1.0
0.5
0.0
1.0
1.5
2.0
2.5

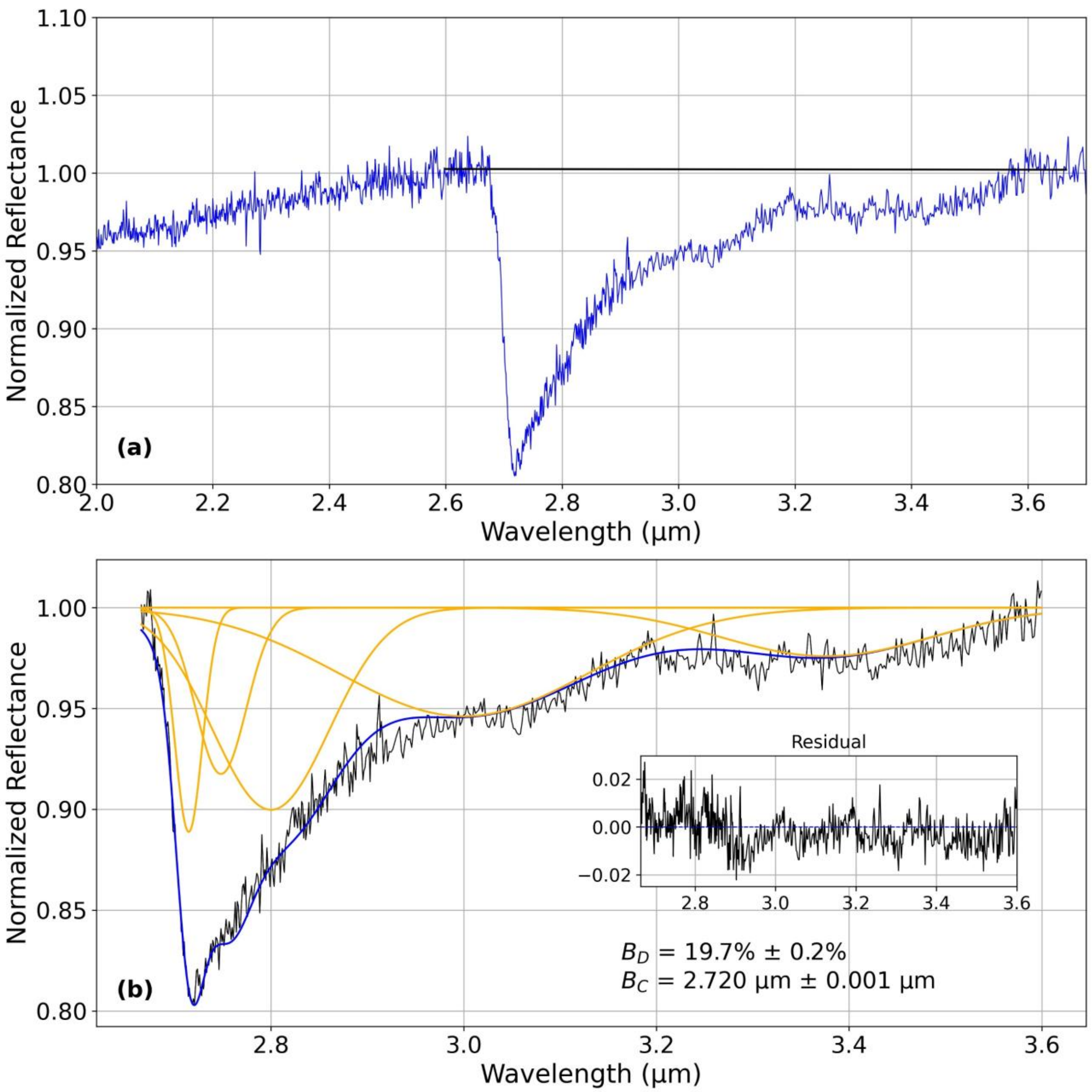
Normalized Reflectance
Wavelength (µm)
(a)
(b)
Residual
$B_D$ = 19.7% ± 0.2%
$B_C$ = 2.720 µm ± 0.001 µm

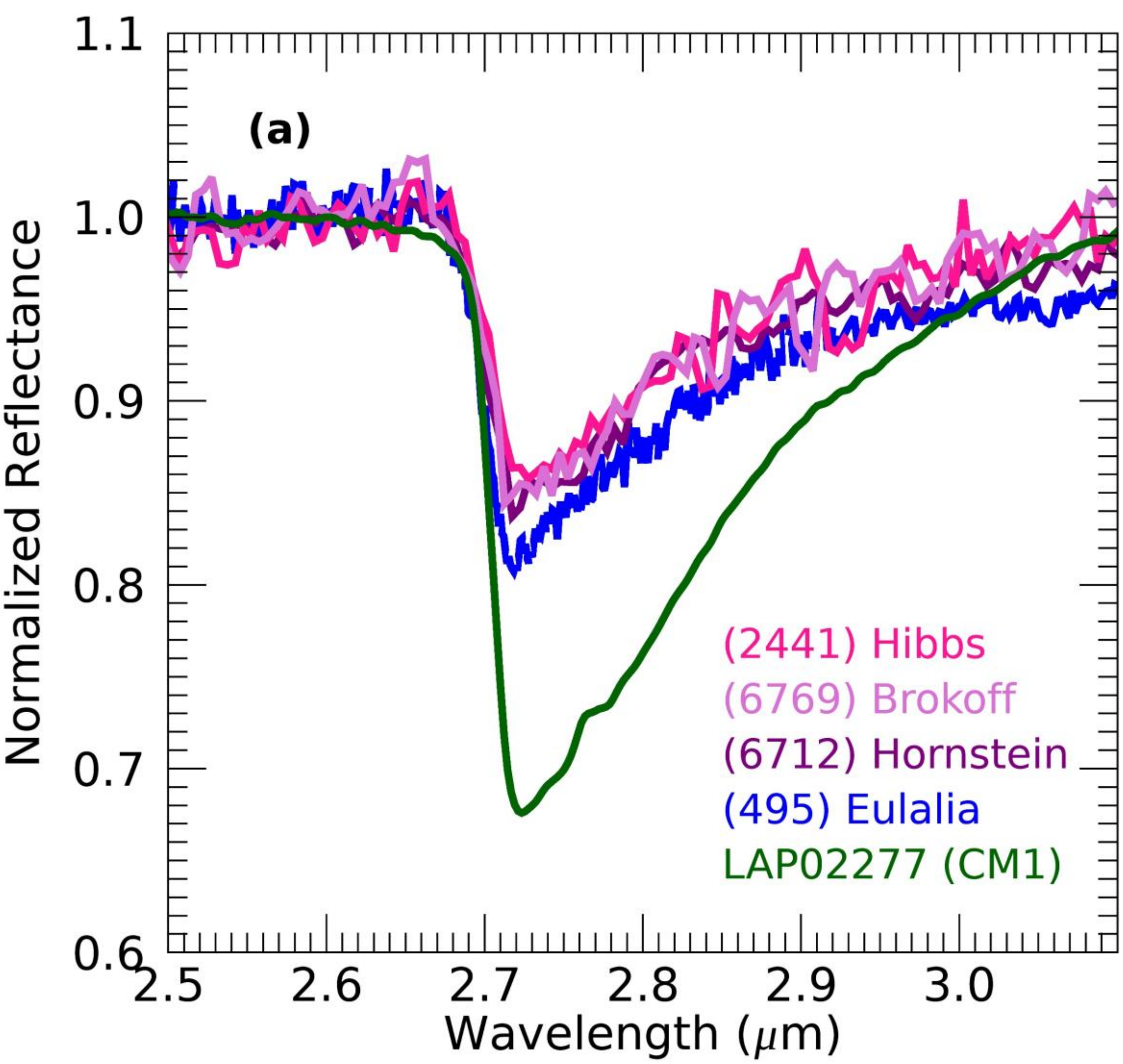
(a)
Normalized Reflectance
Wavelength (μm)
1.1
1.0
0.9
0.8
0.7
0.6
2.5
2.6
2.7
2.8
2.9
3.0
(2441) Hibbs
(6769) Brokoff
(6712) Hornstein
(495) Eulalia
LAP02277 (CM1)

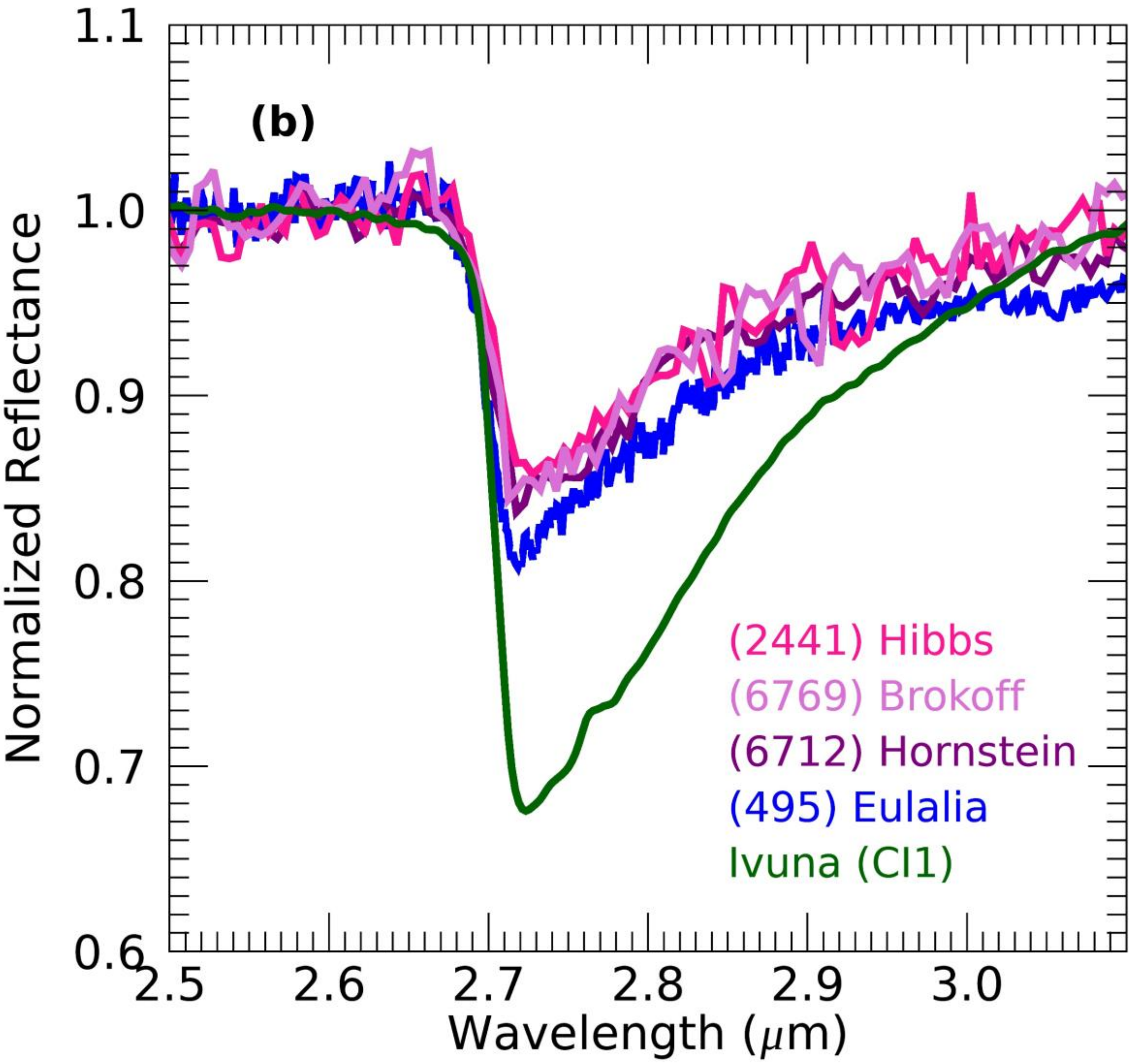
(b)
Normalized Reflectance
Wavelength (μm)
1.1
1.0
0.9
0.8
0.7
0.6
2.5
2.6
2.7
2.8
2.9
3.0
(2441) Hibbs
(6769) Brokoff
(6712) Hornstein
(495) Eulalia
Ivuna (CI1)

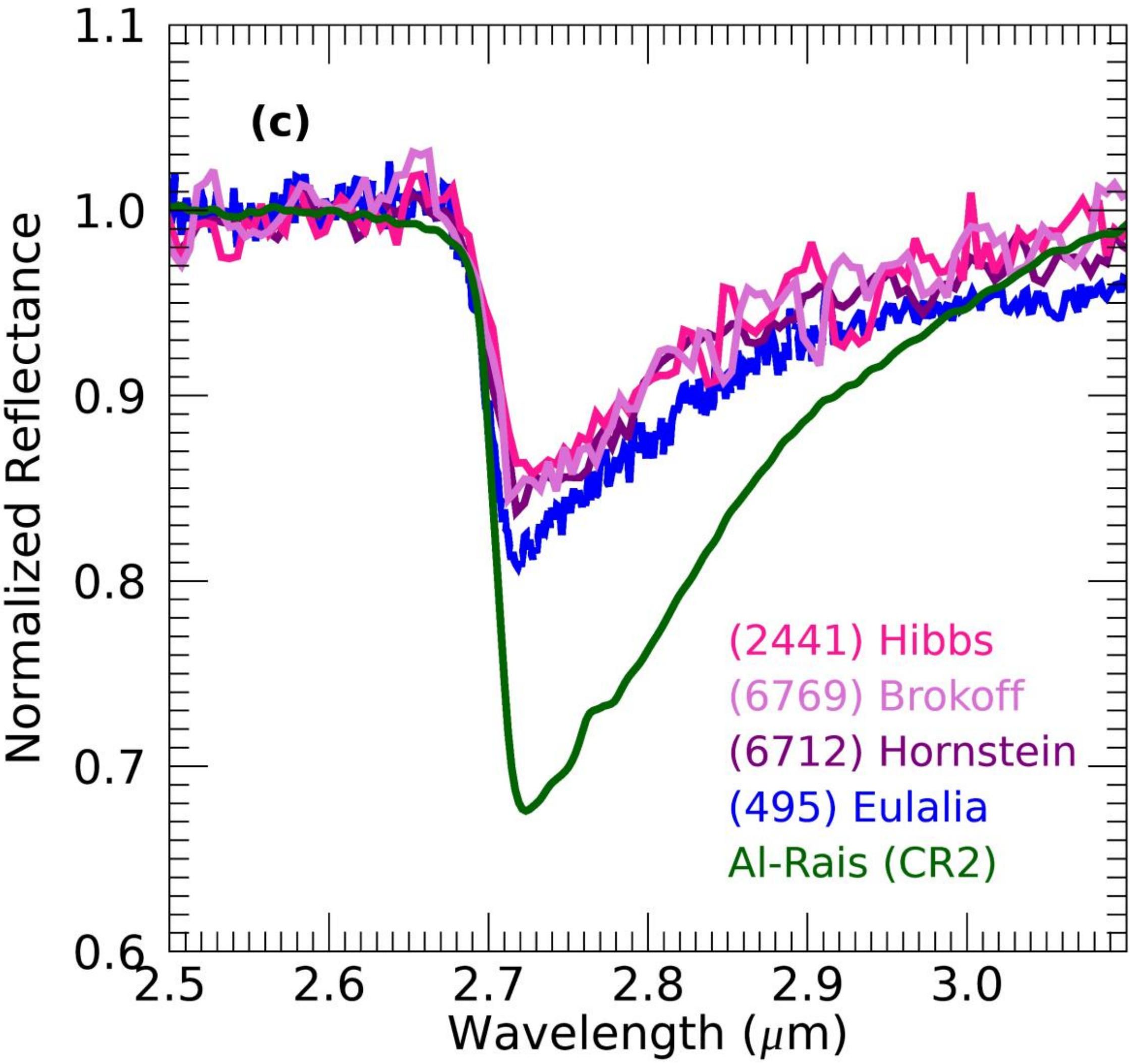

(c)
1.1
1.0
0.9
0.8
0.7
0.6
Normalized Reflectance
2.5
2.6
2.7
2.8
2.9
3.0
Wavelength (μm)
(2441) Hibbs
(6769) Brokoff
(6712) Hornstein
(495) Eulalia
Al-Rais (CR2)

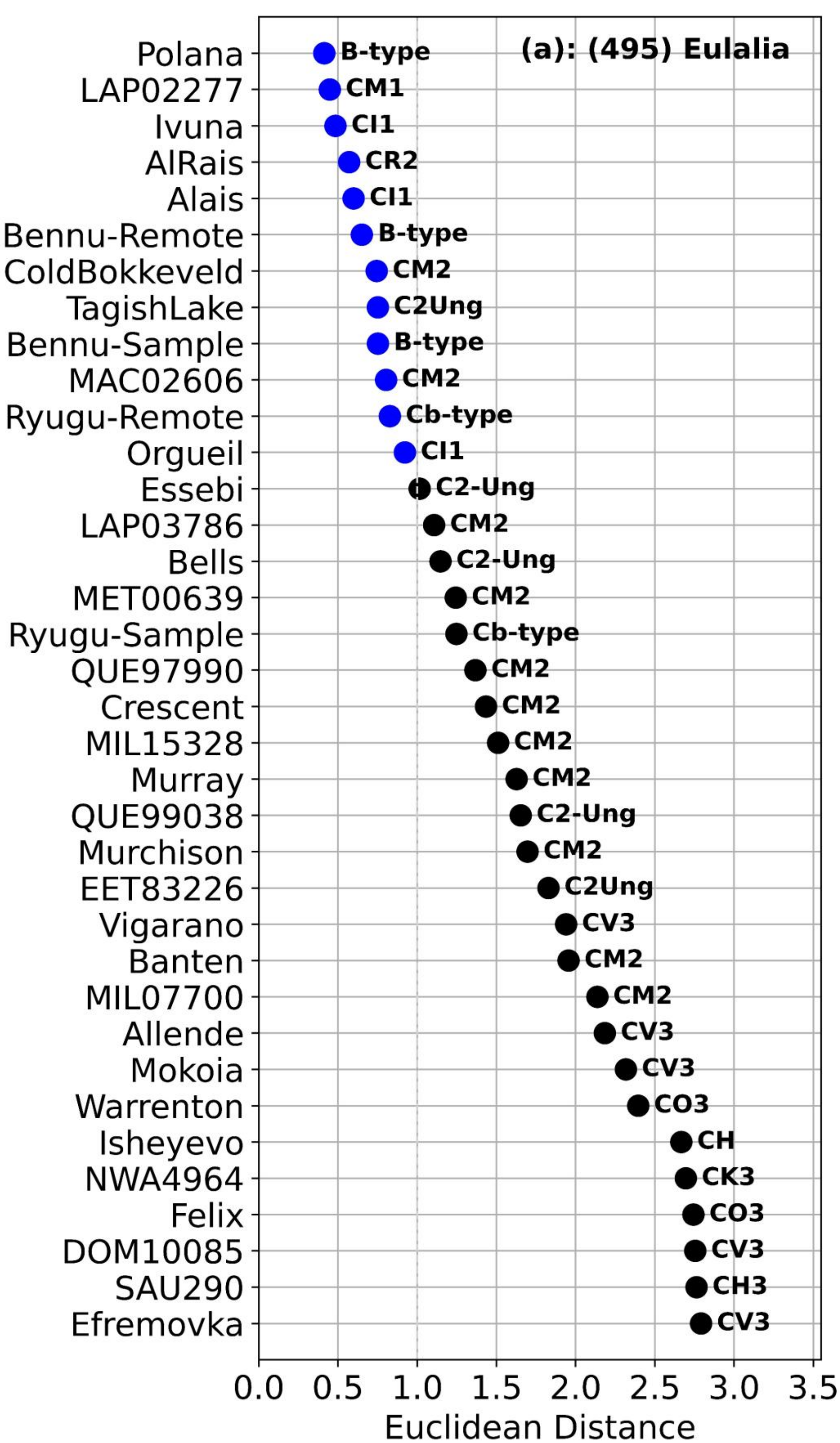
(a): (495) Eulalia
Polana
B-type
LAP02277
CM1
Ivuna
CI1
AlRais
CR2
Alais
CI1
Bennu-Remote
B-type
ColdBokkeveld
CM2
TagishLake
C2Ung
Bennu-Sample
B-type
MAC02606
CM2
Ryugu-Remote
Cb-type
Orgueil
CI1
Essebi
C2-Ung
LAP03786
CM2
Bells
C2-Ung
MET00639
CM2
Ryugu-Sample
Cb-type
QUE97990
CM2
Crescent
CM2
MIL15328
CM2
Murray
CM2
QUE99038
C2-Ung
Murchison
CM2
EET83226
C2Ung
Vigarano
CV3
Banten
CM2
MIL07700
CM2
Allende
CV3
Mokoia
CV3
Warrenton
CO3
Isheyevo
CH
NWA4964
CK3
Felix
CO3
DOM10085
CV3
SAU290
CH3
Efremovka
CV3
0.0
0.5
1.0
1.5
2.0
2.5
3.0
3.5
Euclidean Distance

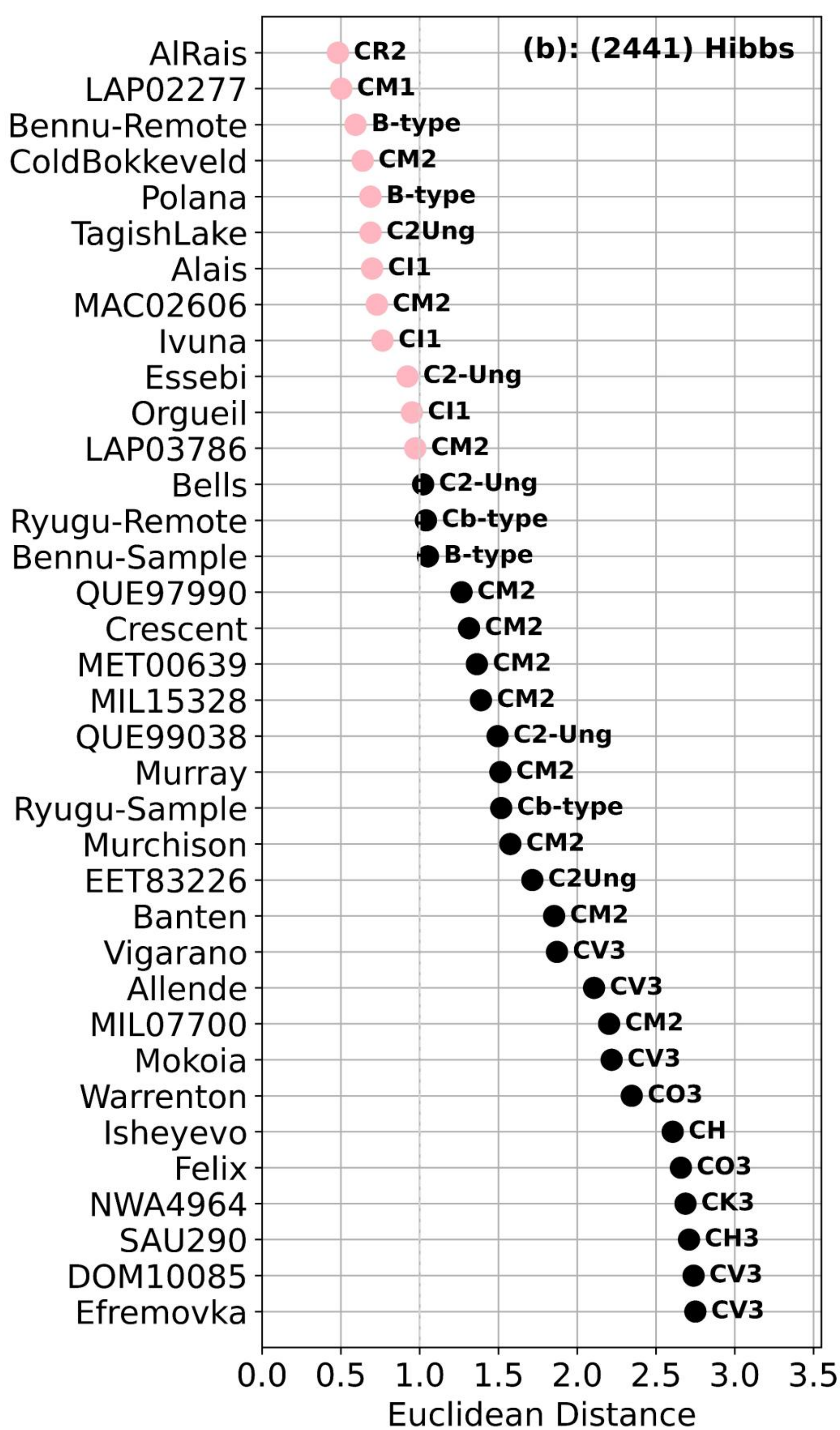

(b): (2441) Hibbs
AlRais
CR2
LAP02277
CM1
Bennu-Remote
B-type
ColdBokkeveld
CM2
Polana
B-type
TagishLake
C2Ung
Alais
CI1
MAC02606
CM2
Ivuna
CI1
Essebi
C2-Ung
Orgueil
CI1
LAP03786
CM2
Bells
C2-Ung
Ryugu-Remote
Cb-type
Bennu-Sample
B-type
QUE97990
CM2
Crescent
CM2
MET00639
CM2
MIL15328
CM2
QUE99038
C2-Ung
Murray
CM2
Ryugu-Sample
Cb-type
Murchison
CM2
EET83226
C2Ung
Banten
CM2
Vigarano
CV3
Allende
CV3
MIL07700
CM2
Mokoia
CV3
Warrenton
CO3
Isheyevo
CH
Felix
CO3
NWA4964
CK3
SAU290
CH3
DOM10085
CV3
Efremovka
CV3
0.0
0.5
1.0
1.5
2.0
2.5
3.0
3.5
Euclidean Distance

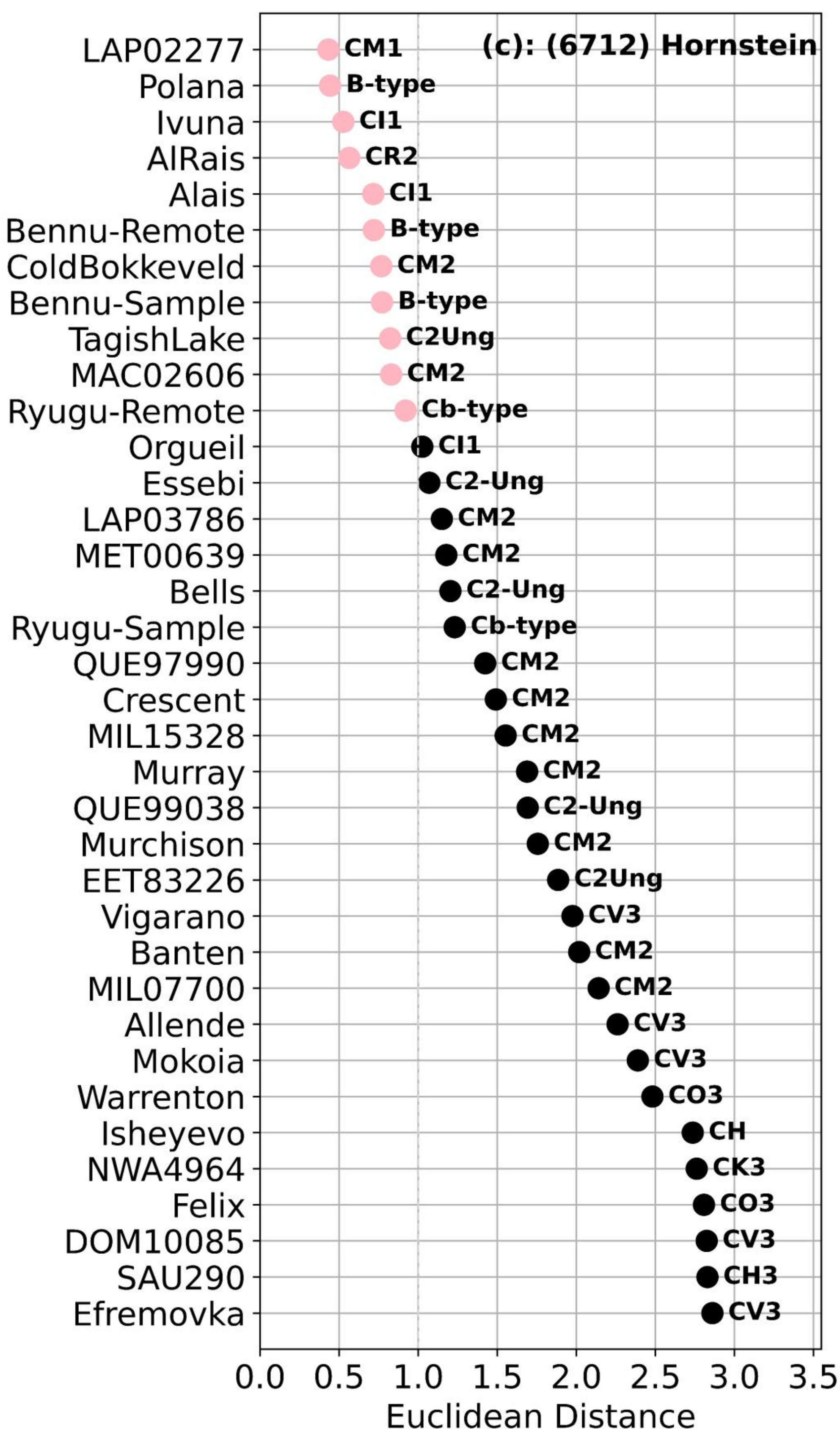
(c): (6712) Hornstein
LAP02277 CM1
Polana B-type
Ivuna CI1
AlRais CR2
Alais CI1
Bennu-Remote B-type
ColdBokkeveld CM2
Bennu-Sample B-type
TagishLake C2Ung
MAC02606 CM2
Ryugu-Remote Cb-type
Orgueil CI1
Essebi C2-Ung
LAP03786 CM2
MET00639 CM2
Bells C2-Ung
Ryugu-Sample Cb-type
QUE97990 CM2
Crescent CM2
MIL15328 CM2
Murray CM2
QUE99038 C2-Ung
Murchison CM2
EET83226 C2Ung
Vigarano CV3
Banten CM2
MIL07700 CM2
Allende CV3
Mokoia CV3
Warrenton CO3
Isheyevo CH
NWA4964 CK3
Felix CO3
DOM10085 CV3
SAU290 CH3
Efremovka CV3
0.0 0.5 1.0 1.5 2.0 2.5 3.0 3.5
Euclidean Distance

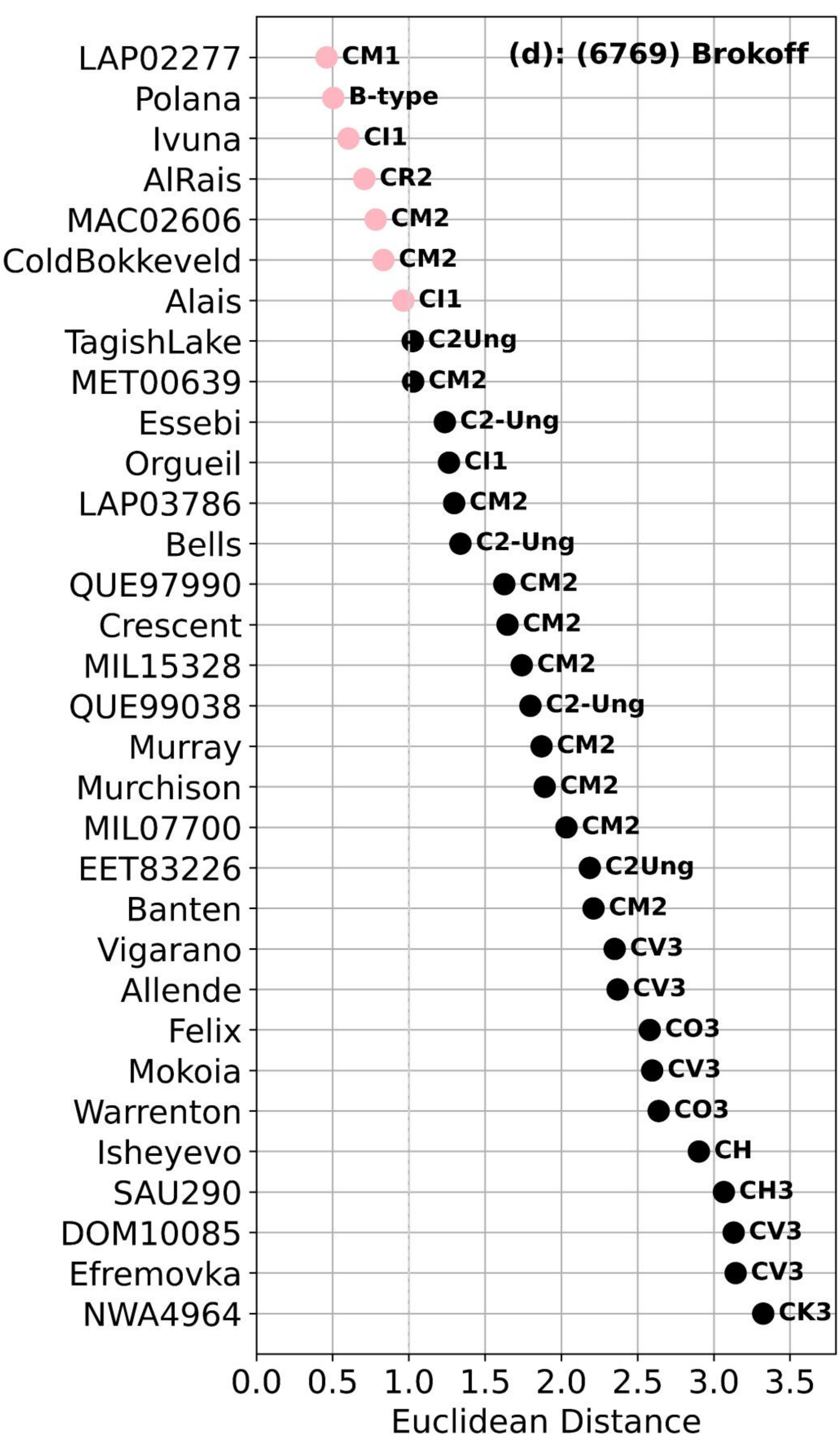
(d): (6769) Brokoff
LAP02277
CM1
Polana
B-type
Ivuna
CI1
AlRais
CR2
MAC02606
CM2
ColdBokkeveld
CM2
Alais
CI1
TagishLake
C2Ung
MET00639
CM2
Essebi
C2-Ung
Orgueil
CI1
LAP03786
CM2
Bells
C2-Ung
QUE97990
CM2
Crescent
CM2
MIL15328
CM2
QUE99038
C2-Ung
Murray
CM2
Murchison
CM2
MIL07700
CM2
EET83226
C2Ung
Banten
CM2
Vigarano
CV3
Allende
CV3
Felix
CO3
Mokoia
CV3
Warrenton
CO3
Isheyevo
CH
SAU290
CH3
DOM10085
CV3
Efremovka
CV3
NWA4964
CK3
0.0
0.5
1.0
1.5
2.0
2.5
3.0
3.5
Euclidean Distance

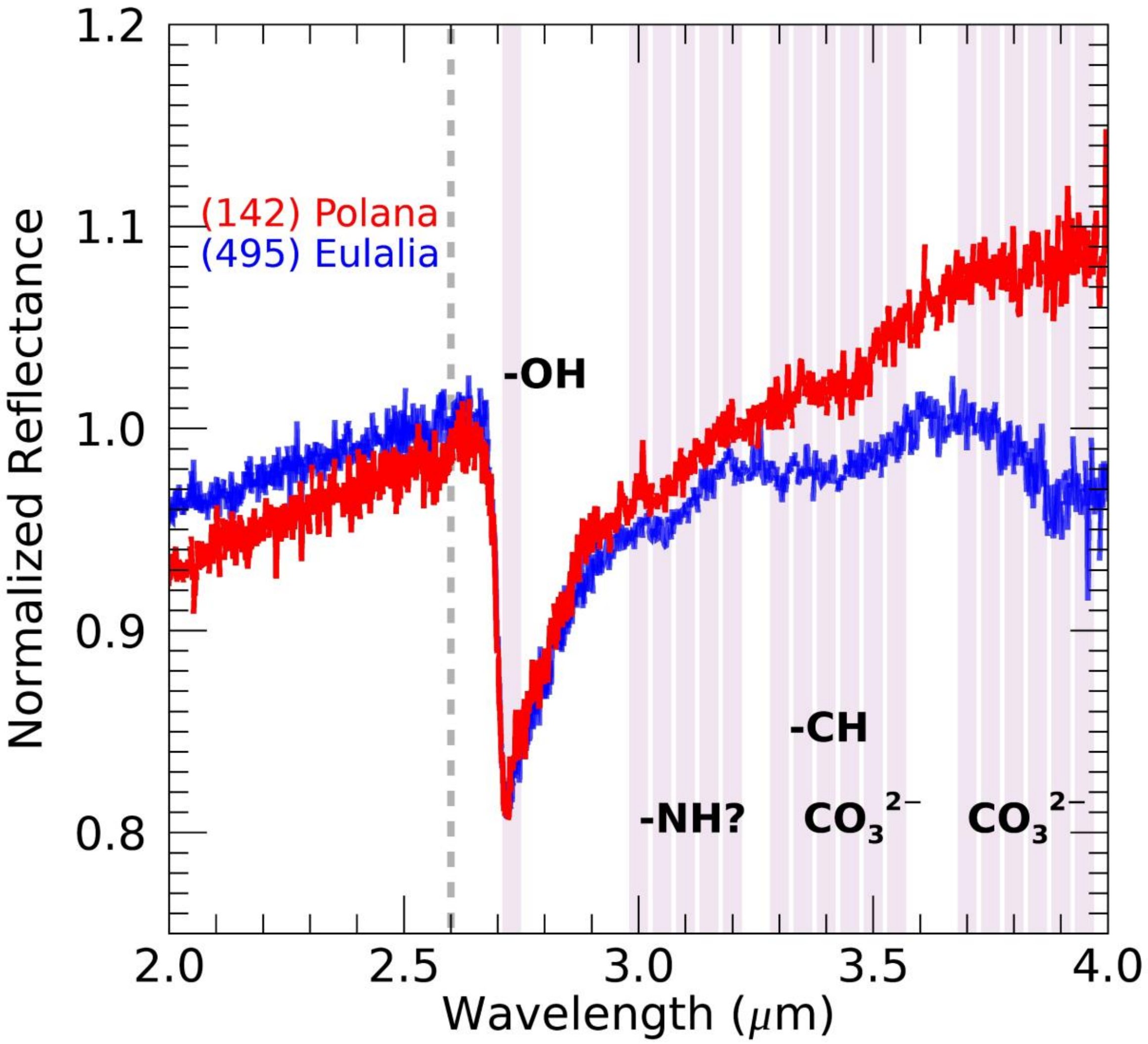

1.2
1.1
1.0
0.9
0.8
Normalized Reflectance
(142) Polana
(495) Eulalia
-OH
-CH
-NH?
$CO_3^{2-}$
$CO_3^{2-}$
2.0
2.5
3.0
3.5
4.0
Wavelength (μm)

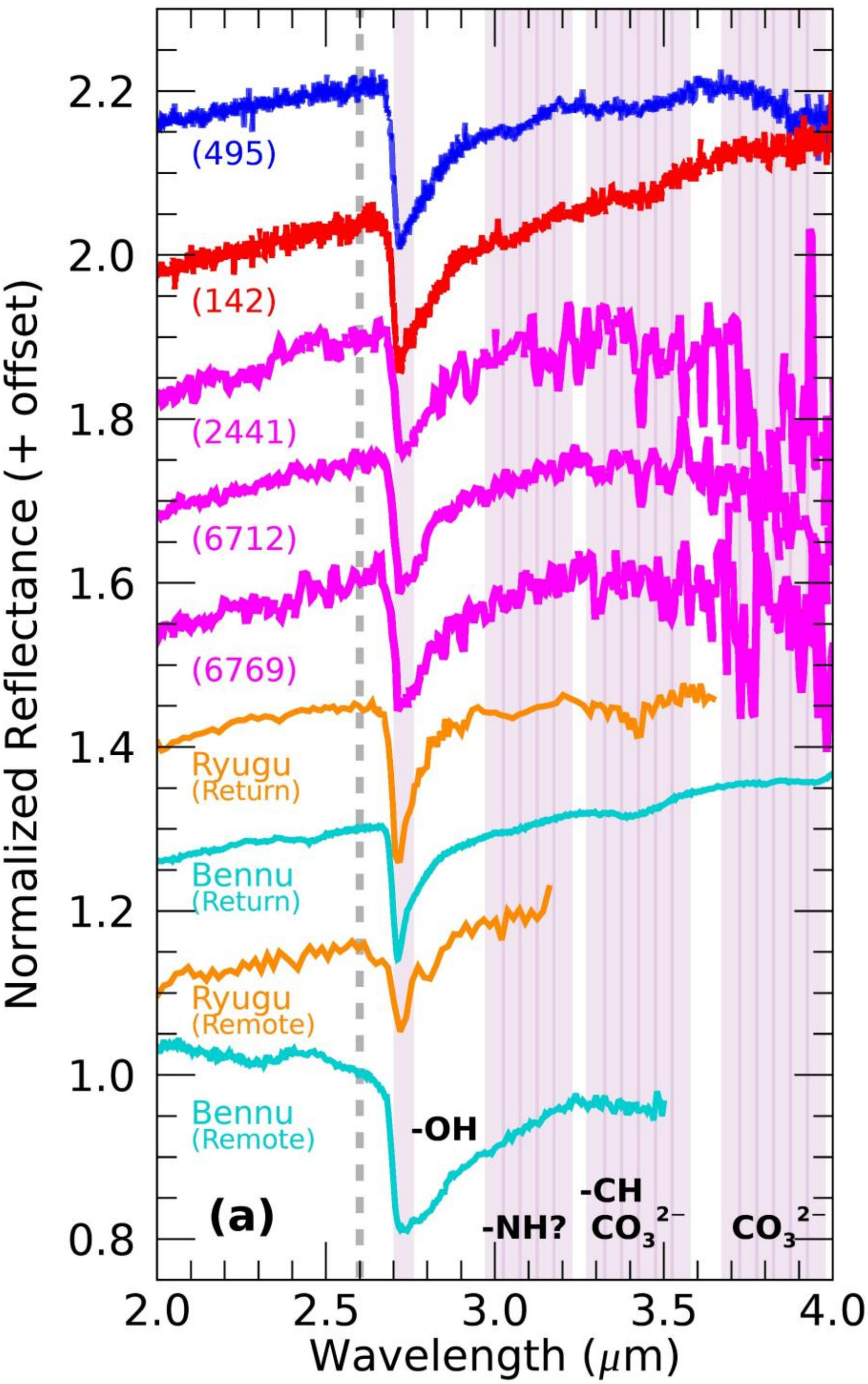
2.2
2.0
1.8
1.6
1.4
1.2
1.0
0.8
Normalized Reflectance (+ offset)
2.0
2.5
3.0
3.5
4.0
Wavelength (μm)
(495)
(142)
(2441)
(6712)
(6769)
Ryugu
(Return)
Bennu
(Return)
Ryugu
(Remote)
Bennu
(Remote)
-OH
-NH?
-CH
$CO_3^{2-}$
$CO_3^{2-}$
(a)

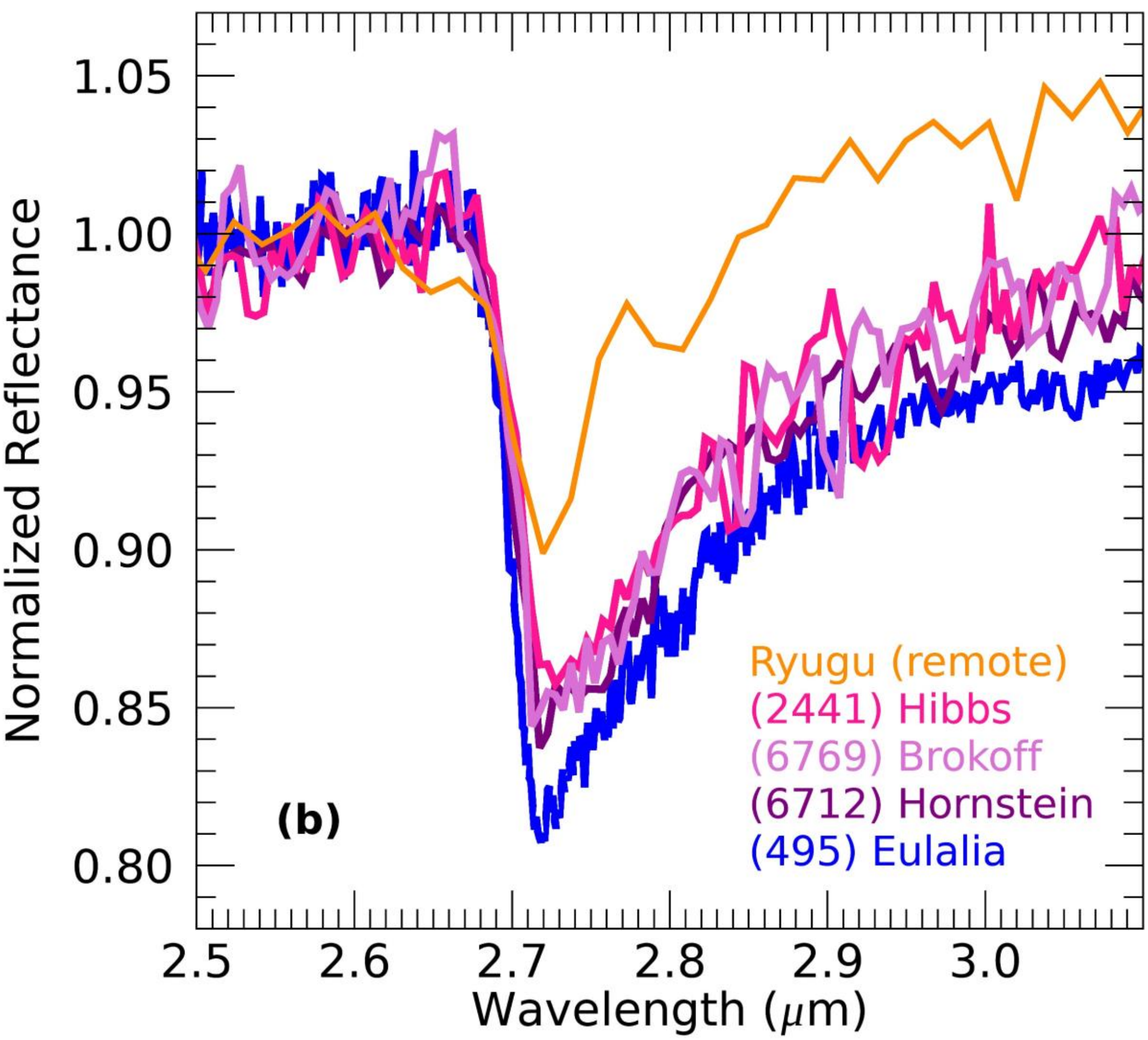
1.05
1.00
0.95
0.90
0.85
0.80
Normalized Reflectance
2.5
2.6
2.7
2.8
2.9
3.0
Wavelength (μm)
(b)
Ryugu (remote)
(2441) Hibbs
(6769) Brokoff
(6712) Hornstein
(495) Eulalia

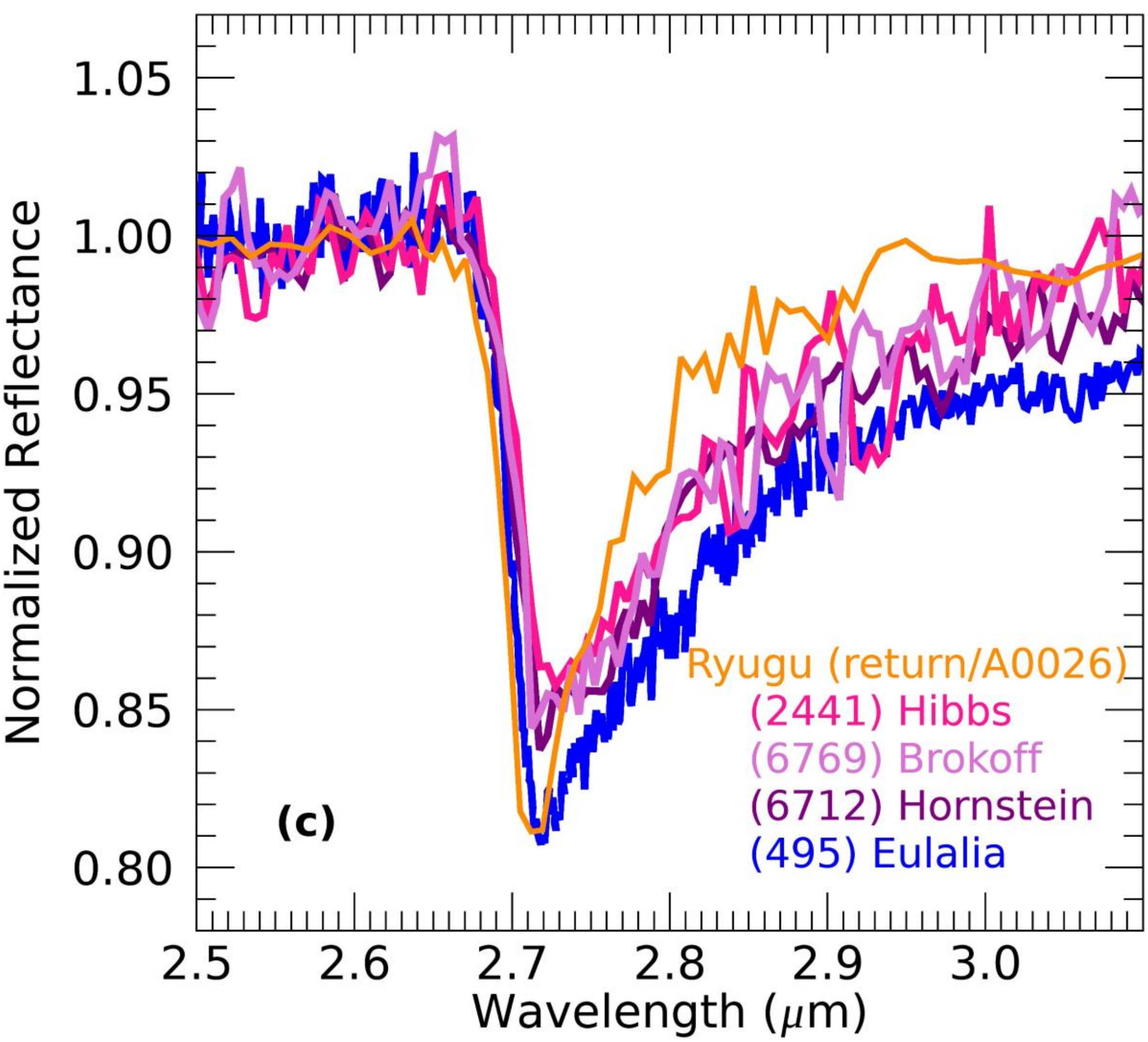
1.05
1.00
0.95
0.90
0.85
0.80
Normalized Reflectance
2.5
2.6
2.7
2.8
2.9
3.0
Wavelength (μm)
(c)
Ryugu (return/A0026)
(2441) Hibbs
(6769) Brokoff
(6712) Hornstein
(495) Eulalia

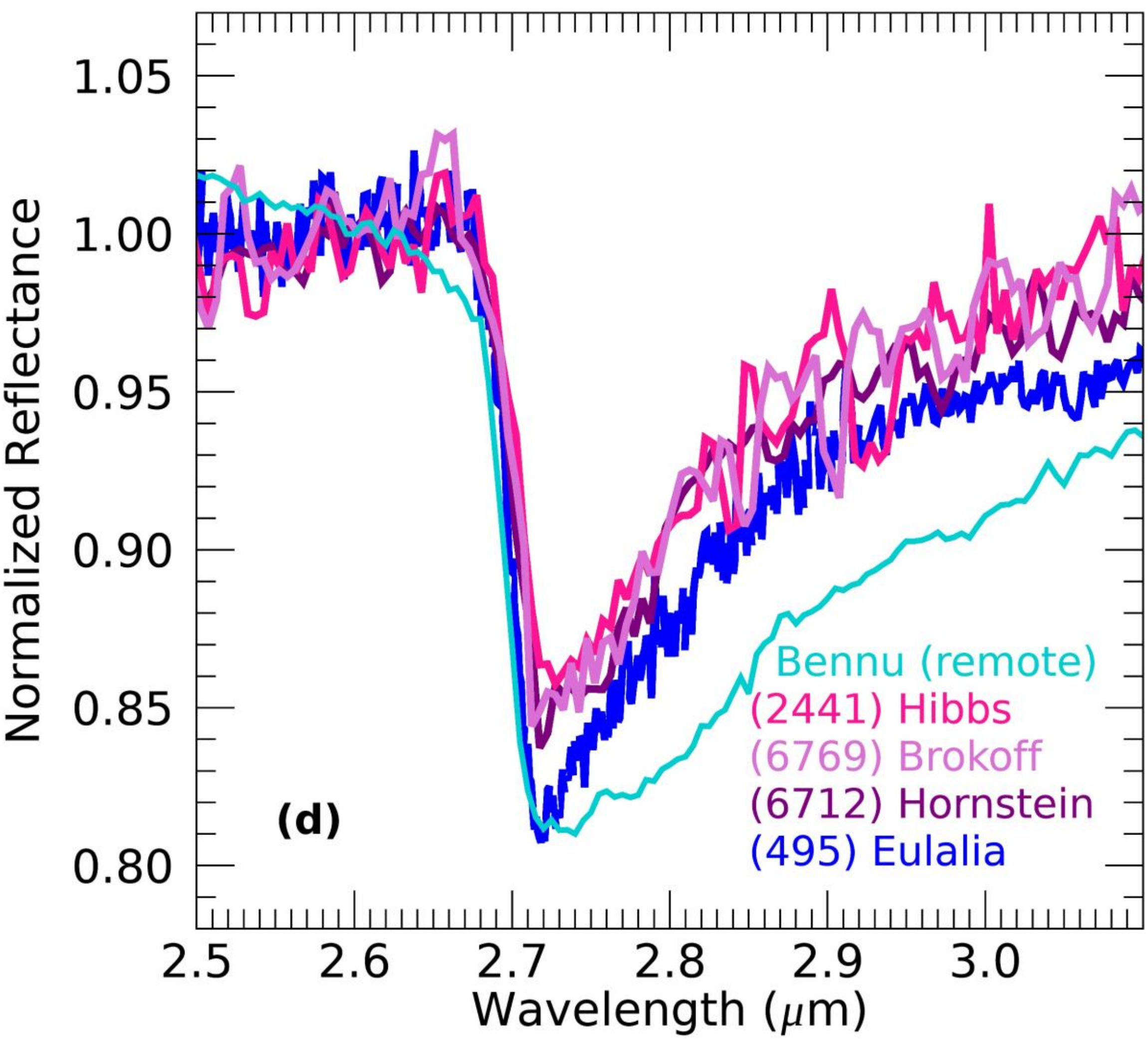
Normalized Reflectance
1.05
1.00
0.95
0.90
0.85
0.80
2.5
2.6
2.7
2.8
2.9
3.0
Wavelength (μm)
(d)
Bennu (remote)
(2441) Hibbs
(6769) Brokoff
(6712) Hornstein
(495) Eulalia

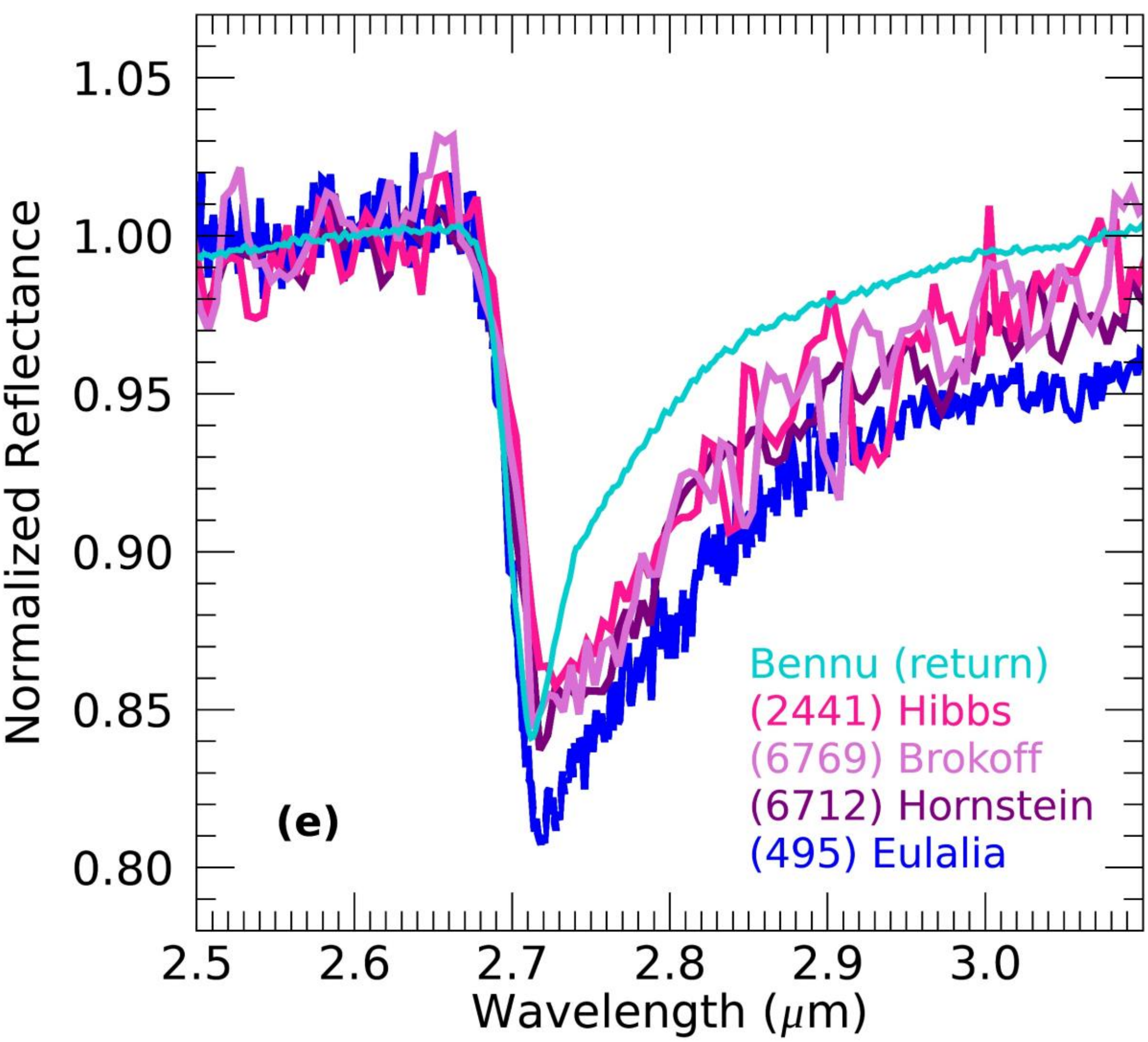

Normalized Reflectance
1.05
1.00
0.95
0.90
0.85
0.80
2.5
2.6
2.7
2.8
2.9
3.0
Wavelength (μm)
(e)
Bennu (return)
(2441) Hibbs
(6769) Brokoff
(6712) Hornstein
(495) Eulalia

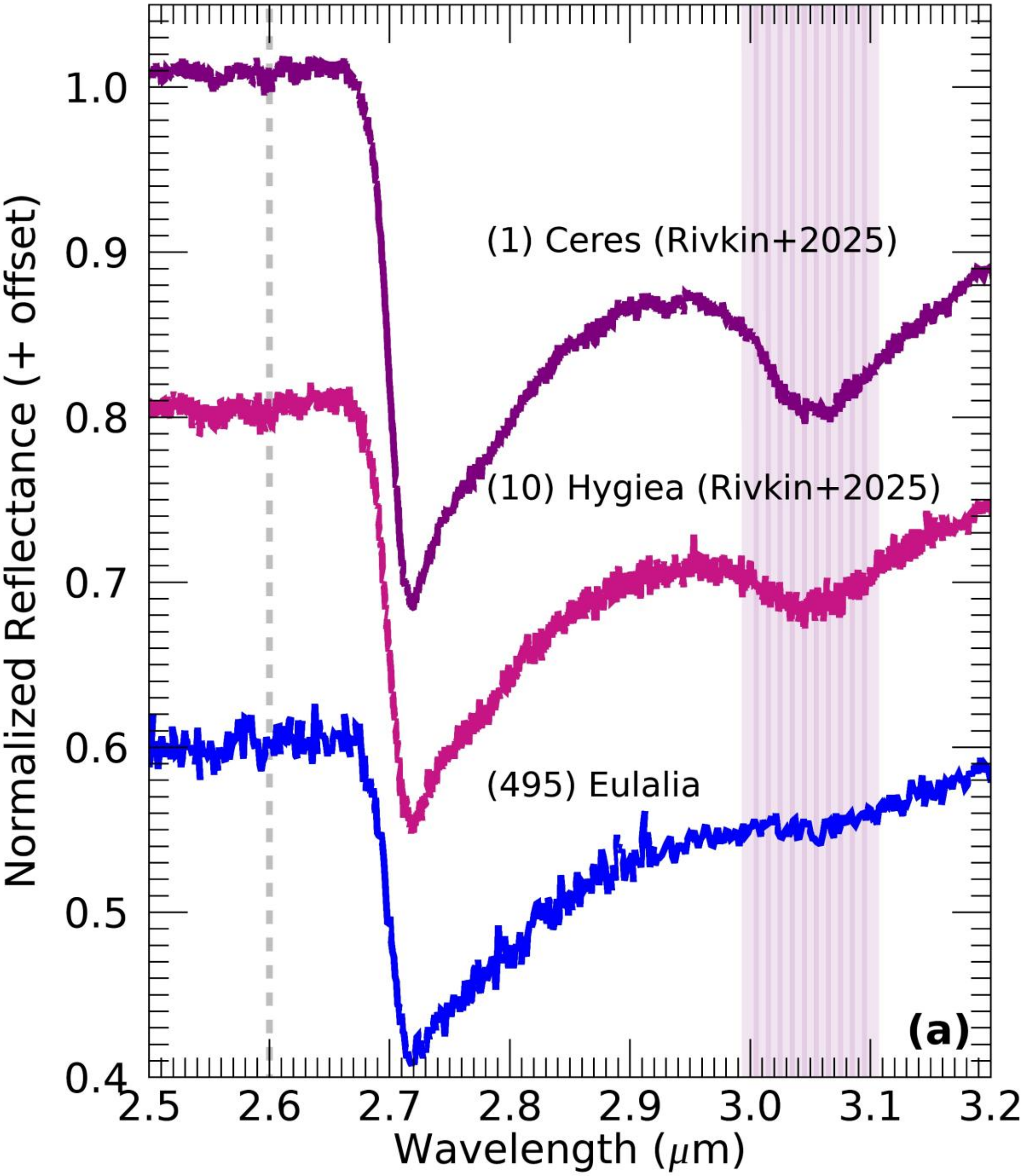

Normalized Reflectance (+ offset)
1.0
0.9
0.8
0.7
0.6
0.5
0.4
(1) Ceres (Rivkin+2025)
(10) Hygiea (Rivkin+2025)
(495) Eulalia
(a)
2.5
2.6
2.7
2.8
2.9
3.0
3.1
3.2
Wavelength (μm)

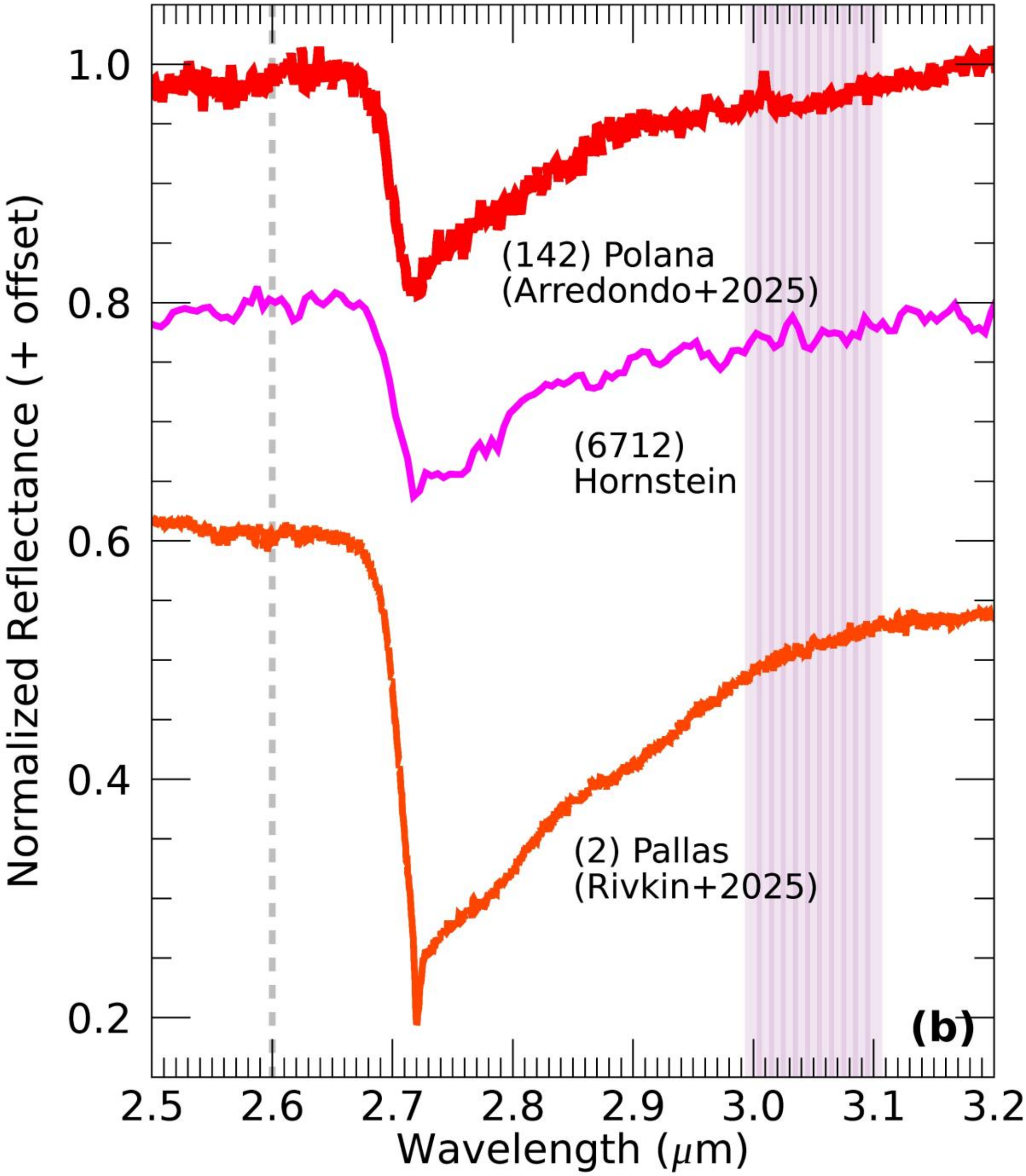

Normalized Reflectance (+ offset)
1.0
0.8
0.6
0.4
0.2
(142) Polana
(Arredondo+2025)
(6712)
Hornstein
(2) Pallas
(Rivkin+2025)
(b)
2.5
2.6
2.7
2.8
2.9
3.0
3.1
3.2
Wavelength (μm)

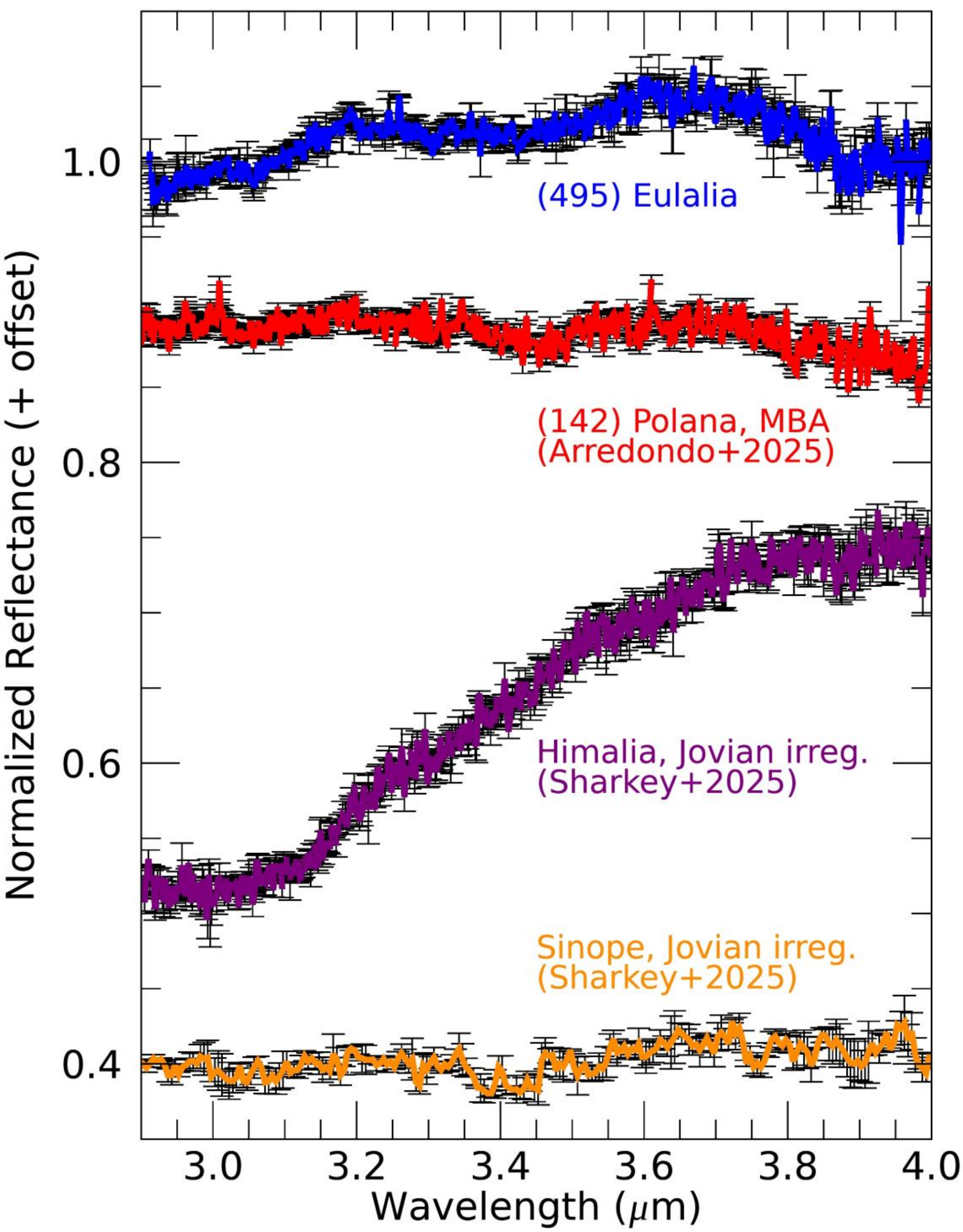
(495) Eulalia
(142) Polana, MBA
(Arredondo+2025)
Himalia, Jovian irreg.
(Sharkey+2025)
Sinope, Jovian irreg.
(Sharkey+2025)
Normalized Reflectance (+ offset)
Wavelength (μm)
1.0
0.8
0.6
0.4
3.0
3.2
3.4
3.6
3.8
4.0